\PassOptionsToPackage{prologue,dvipsnames,svgnames}{xcolor}
\documentclass[11pt]{article}
\usepackage{fullpage}
\usepackage[round]{natbib}

\usepackage{amsmath,amssymb}
\usepackage{bm}
\usepackage{amssymb}
\usepackage{mathrsfs}
\usepackage{graphicx}
\usepackage{subcaption}
\usepackage{pdfpages}
\usepackage{afterpage}
\usepackage{caption}
\usepackage{multirow}
\usepackage{rotating}
\usepackage[flushleft]{threeparttable}
\usepackage{multirow}
\usepackage{enumitem} 
\usepackage{algorithm2e}
\usepackage{svg}
\usepackage{stackengine}
\usepackage{float}
\RestyleAlgo{ruled}
\SetAlgorithmName{Procedure}{procedure}{List of Procedure}
\usepackage{tikz} 
\usetikzlibrary{matrix}

\usepackage{preamble/star}
\usepackage[group-separator={,},group-minimum-digits={3}]{siunitx}

\usepackage[colorlinks,linkcolor=blue,anchorcolor=blue,citecolor=blue,urlcolor=black]{hyperref}
\usepackage{cleveref}
\svgpath{{Figure/}}

\newcolumntype{P}[1]{>{\centering\arraybackslash}p{#1}}

\def\$#1\${\begin{align*}#1\end{align*}}

\newcommand{\Rom}[1]{\text{\uppercase\expandafter{\romannumeral #1\relax}}}

\renewcommand{\max}{\mathop{\mathrm{max}}}

\usepackage{color}

\begin{document}

\title{The ICML 2023 Ranking Experiment: Examining Author Self-Assessment in ML/AI Peer Review}

\author{Buxin Su\thanks{University of Pennsylvania. Email: \href{mailto:suw@wharton.upenn.edu}{suw@wharton.upenn.edu}.} \and Jiayao Zhang\footnotemark[1] \and Natalie Collina\footnotemark[1] \and Yuling Yan\thanks{Massachusetts Institute of Technology.} 
\and Didong Li\thanks{University of North Carolina at Chapel Hill.} \and Kyunghyun Cho\thanks{New York University.} \and Jianqing Fan\thanks{Princeton University.} \and Aaron Roth\footnotemark[1] 
\and Weijie Su\footnotemark[1]}

\date{August 2024; Revised September 2025}

\maketitle

\begin{abstract}

We conducted an experiment during the review process of the 2023 International Conference on Machine Learning (ICML), asking authors with multiple submissions to rank their papers based on perceived quality. In total, we received 1,342 rankings, each from a different author, covering 2,592 submissions. In this paper, we present an empirical analysis of how author-provided rankings could be leveraged to improve peer review processes at machine learning conferences. We focus on the Isotonic Mechanism, which calibrates raw review scores using the author-provided rankings. Our analysis shows that these ranking-calibrated scores outperform the raw review scores in estimating the ground truth ``expected review scores'' in terms of both squared and absolute error metrics. Furthermore, we propose several cautious, low-risk applications of the Isotonic Mechanism and author-provided rankings in peer review, including supporting senior area chairs in overseeing area chairs' recommendations, assisting in the selection of paper awards, and guiding the recruitment of emergency reviewers. 

\end{abstract}

\section{Introduction}\label{sec:1}

The peer review process is critical for advancing research by identifying high-quality, impactful work. However, concerns about the quality of peer review have grown across many fields~\citep{brezis2020arbitrariness,cheah2022should,liang2023random}, especially in machine learning (ML) and artificial intelligence (AI), where issues such as noisy or arbitrary reviews are prevalent~\citep{langford2015arbitrariness,lipton2019troubling,wang2020debiasing,russo2021some,yuan2022can}. For example, a randomized experiment at NeurIPS 2021---one of the three premier conferences in ML/AI, alongside ICML and ICLR---revealed that about half of the accepted papers would have been rejected upon a second round of reviews~\citep{cortes2021inconsistency,beygelzimer2023has}. One contributing factor to this decline is the exponential increase in submissions to ML/AI conferences. For instance, NeurIPS 2023 received 12,343 submissions, making it virtually impossible to recruit sufficient experienced reviewers capable of providing consistent, high-quality assessments for such a volume of submissions~\citep{sculley2018avoiding,stelmakh2021novice}.

In response, considerable effort has been devoted to improving the ML/AI conference peer review process, often aiming to enhance the accuracy of review scores~\citep{kobren2019paper,wang2019your}, which are the single most important factor in determining accept/reject decisions.\footnote{For example, the NeurIPS 2023 guidelines for area chairs state that any decision ``should be properly explained'' if any paper with an average score above (below) the threshold is rejected (accepted).} A common feature of these efforts focuses on the reviewer side, aiming either to incentivize reviewers, improve the matching of reviewers to submissions, or reduce reviewer bias~\citep{goldberg2025peer,jecmen2020mitigating,wang2020debiasing}. 

\begin{figure}[!htp]
    \centering
    \includegraphics[width=\textwidth]{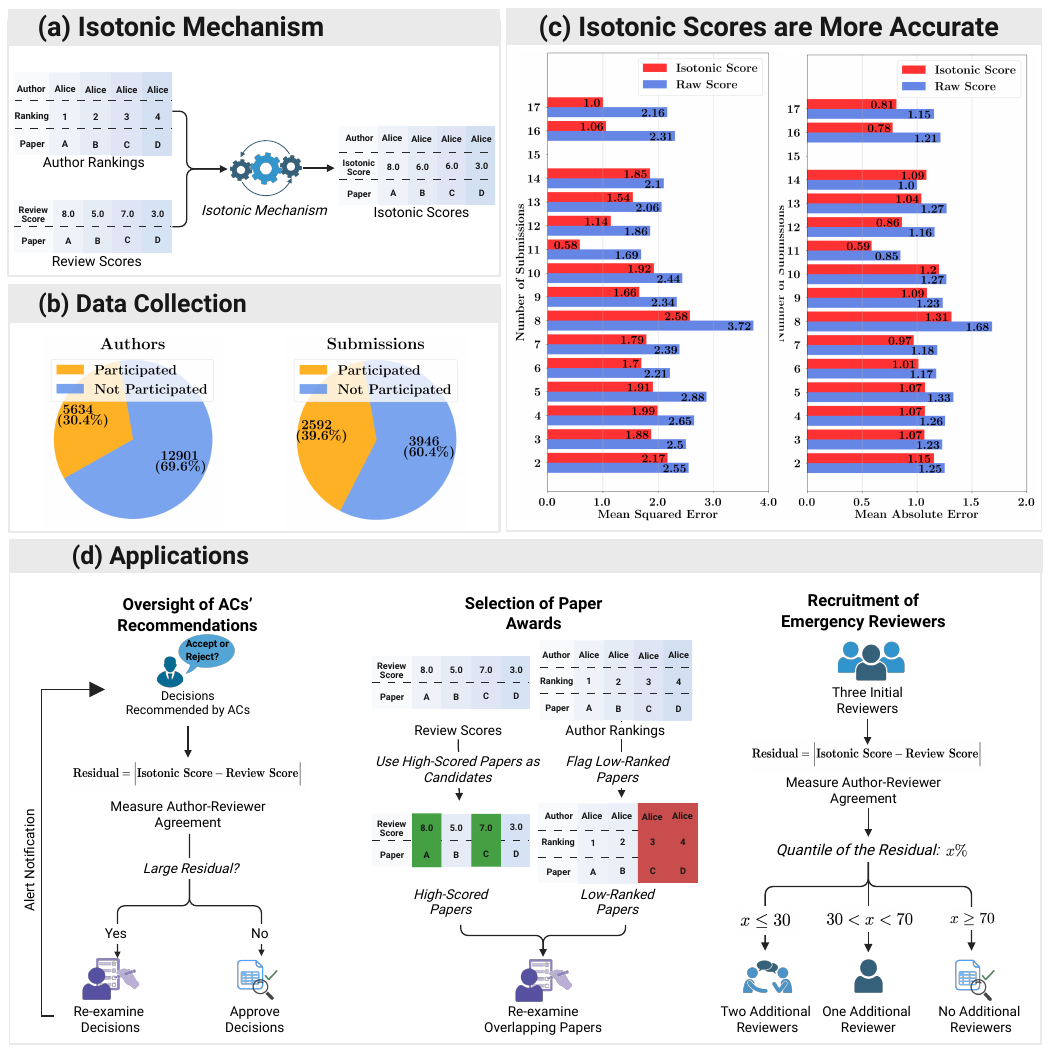}
    \caption{(a) Illustration of the Isotonic Mechanism (see Section~\ref{sec:mm}). (b) Some summary statistics of our survey experiment at ICML 2023 (see Section \ref{sec:basic} for more statistics). (c) Comparison between isotonic and review scores in terms of estimation accuracy (see Section~\ref{sec:main} for details). (d) Three potential applications of our mechanism (see Section~\ref{sec:other}).}
\label{fig:intro}
\end{figure}

In contrast to these reviewer-centric approaches, the recently proposed Isotonic Mechanism aims to yield more robust review scores by incorporating authors' assessments of their \textit{own} submissions to a conference~\citep{su2021you}. This mechanism requires authors with multiple submissions to rank their papers according to their perception of the relative quality of the papers and outputs calibrated scores that are modified from the original review scores to align with the author-provided rankings (see Figure~\ref{fig:intro}(a) for an illustration). This ranking-based calibration can be seen as a ``de-noising'' of the original review scores from the perspective of the authors~\citep{su2021you,su2022truthful}. The Isotonic Mechanism is particularly well-suited to ML/AI conferences, where it is commonplace for an author to submit multiple papers at the same time~\citep{iclr,rastogi2022authors}. A key advantage of this method is that it utilizes authors' insights to inform the review process with minimal additional effort, rather than increasing the burden on reviewers.

In this paper, we evaluate how effective the Isotonic Mechanism is in practice for peer review. To this end, we conducted a survey experiment at the 2023 International Conference on Machine Learning (ICML). In 2023, ICML received 6,538 submissions from 18,535 authors. On January 26, 2023, right after the ICML submission deadline, we sent a survey to all submitting authors who have OpenReview profiles to ask them to provide rankings of their submissions if they submitted at least two papers (see Figure~\ref{fig:intro}(b) for some summary statistics).

Specifically, we address the following questions by analyzing the ICML 2023 ranking data together with review scores and accept/reject decisions:\footnote{Our code is publicly available on GitHub (\url{https://github.com/BuxinSu/ICML_Ranking.git}).}
\begin{enumerate}
\item[(a)] {\em How do the outputs of the Isotonic Mechanism, referred to as isotonic scores, compare to original/raw review scores in terms of accurately reflecting submission quality?}

\item[(b)] {\em What near-term applications might there be for leveraging isotonic scores to enhance the peer review process?}

\item[(c)] {\em What are the limitations of the study, and which aspects of the mechanism should future experiments investigate?}
\end{enumerate}

To address the first question, we require a measure of ground truth submission quality. Since submissions typically receive multiple reviews, we use the average of the remaining scores as a proxy for the ground truth ``expected review score'' of a submission when evaluating the performance of an estimator applied to a single randomly selected review score. Figure \ref{fig:intro}(c) shows that the Isotonic Mechanism substantially reduces proxy estimation errors---specifically, mean squared error (MSE) and mean absolute error (MAE)---for the 2,592 submissions ranked by authors. Moreover, it suggests that the improvement becomes more substantial as the number of submissions of an author increases. This dependence implies that more author-provided rankings could lead to even more significant error reduction through the Isotonic Mechanism.\footnote{If all ICML 2023 authors were to provide their rankings, this would increase the average length of rankings. See 
Figure S.5 in the Supplementary Material.} 
Because the change in proxy estimation error is an unbiased estimator for the change in ground truth estimation error, this allows us to produce confidence intervals for the decrease in ground truth squared error, and we find substantial decreases, significant at the 99\% confidence level. More details are given in Section~\ref{sec:Isotonic_Scores}.

In Section \ref{sec:other}, we address the second question by proposing three potential applications of the Isotonic Mechanism, which are illustrated in Figure~\ref{fig:intro}(d). First, we suggest that senior area chairs (SACs) can use isotonic scores to help direct additional scrutiny of accept/reject recommendations made by area chairs (ACs). Second, we suggest using author-provided rankings to help direct attention during the selection of paper awards. This application occurs after the accept/reject decisions are made, which ensures that its impact is circumscribed, and serves to direct human attention rather than to make decisions directly. Third, we propose using the discrepancy between isotonic scores and raw review scores as an indicator of review quality. A significant discrepancy in comparison to other submissions could signal the need for an emergency reviewer to provide additional evaluation. We provide empirical evidence from the ICML 2023 data supporting the effectiveness of the latter two applications.

In Section \ref{sec:discuss}, we conclude our paper by discussing the third question, the limitations of our experimental results, and suggesting avenues for future research. While this method has shown empirical effectiveness at increasing the accuracy of noisy reviews, and is supported by theoretical underpinnings---including ``truthfulness'' guarantees---under certain conditions~\citep{su2022truthful}, using the isotonic mechanism to make important decisions may give rise to unforeseen consequences because of the possibility of strategic manipulation by authors in ways that are not captured by the stylized analysis. Therefore, our policy recommendations are deliberately circumscribed, and we advocate for more comprehensive investigations of the Isotonic Mechanism in future real-world experiments.

\subsection{Related Work}

By analyzing 1,313 reviews, \cite{pranic2021quality} found that while authors were most satisfied with reviews recommending acceptance, reviews suggesting revisions were of the highest quality. Two additional studies of particular relevance to our work are \cite{gardner2012peering} and \cite{rastogi2022authors}. In \cite{gardner2012peering}, authors were surveyed to rate their submissions to the Australasian Association for Engineering Education Annual Conference, and in \cite{rastogi2022authors}, NeurIPS 2021 authors were asked to estimate the acceptance probabilities of their submissions and compare them pairwise. Both studies revealed that authors tend to overestimate their submissions' chances of acceptance.

From a methodological standpoint, \cite{ugarov2023peer} proposed a mechanism that incentivizes reviewers through peer prediction. More recently, \cite{liang2023can} explored the use of large language models such as GPT-4 to generate initial reviews for research papers. For completeness, an emerging body of work considered methods to enhance peer review from the authors' perspectives \citep{aziz2019strategyproof,mattei2020peernomination,srinivasan2021auctions}. For example, \cite{srinivasan2021auctions} proposed a mechanism requiring authors to submit a bid for a review slot for each of their submitted papers.

\section{Experimental Design and Summary Statistics}
\label{sec:basic}

We first provide an overview of some basic statistics from ICML 2023. (a) Number of submissions: \num{6538}; (b) Number of authors: \num{18535};\footnote{Among the \num{18535} authors, 20 did not have OpenReview profiles. Our survey was sent via the OpenReview API to those \num{18515} authors who had OpenReview profiles.} (c) Number of submissions with at least one author having more than one submission: \num{5035} ($77.0\%$); (d) Number of authors with two or more submissions: \num{4505} (24.3\%); (e) Number of authors with at least \num{5} submissions: \num{508}; (f) Number of authors with at least \num{10} submissions: \num{74}; (g) Number of authors with at least \num{15} submissions: \num{26}; (h) Number of authors with at least \num{20} submissions: \num{7}.

\begin{figure*}[!htp]
    \centering
    \begin{subfigure}[b]{0.32\textwidth}
        \includegraphics[width=\textwidth]{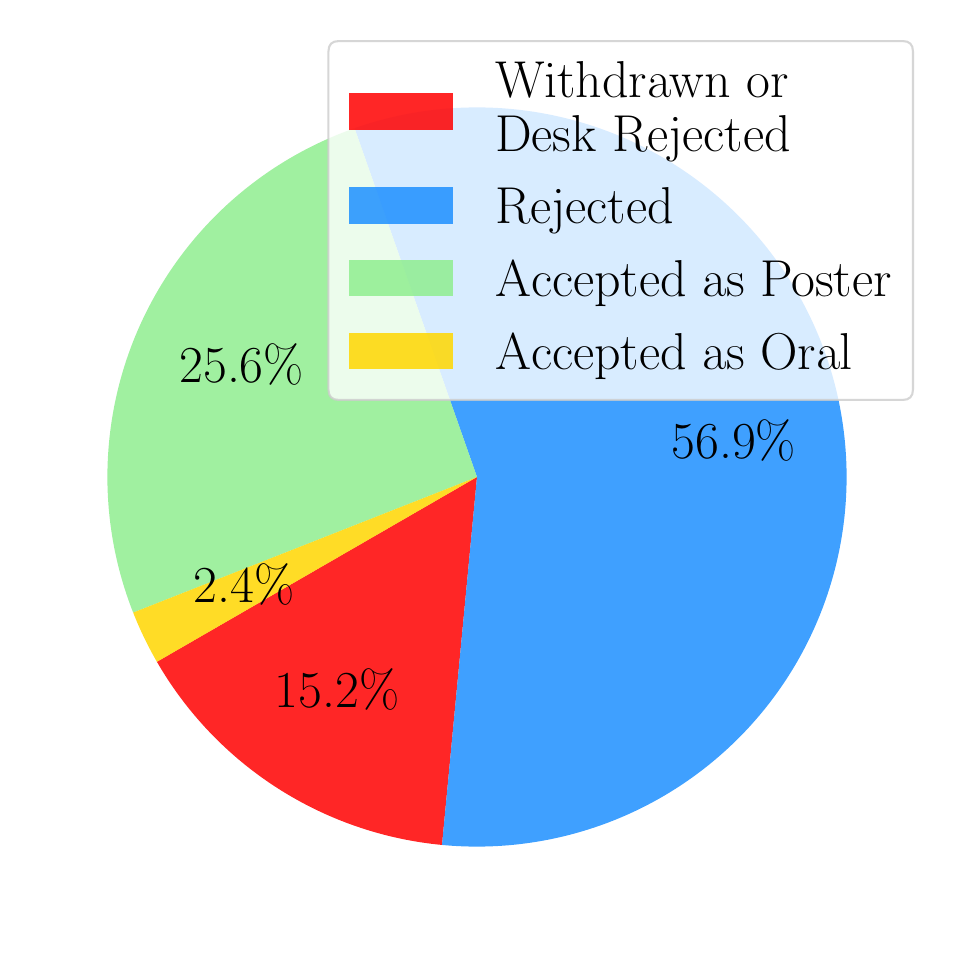}
    \end{subfigure}
    \begin{subfigure}[b]{0.33\textwidth}
        \includegraphics[width=\textwidth]{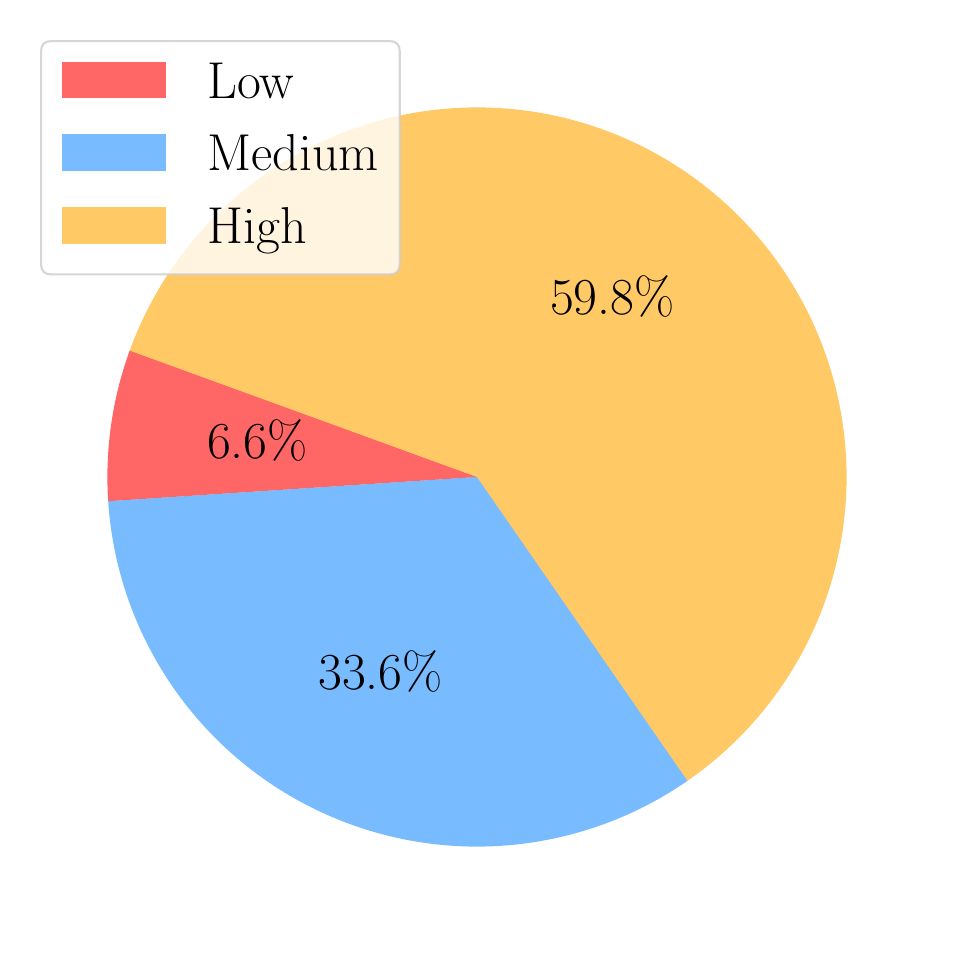}
    \end{subfigure}
    \begin{subfigure}[b]{0.32\textwidth}
        \includegraphics[width=\textwidth]{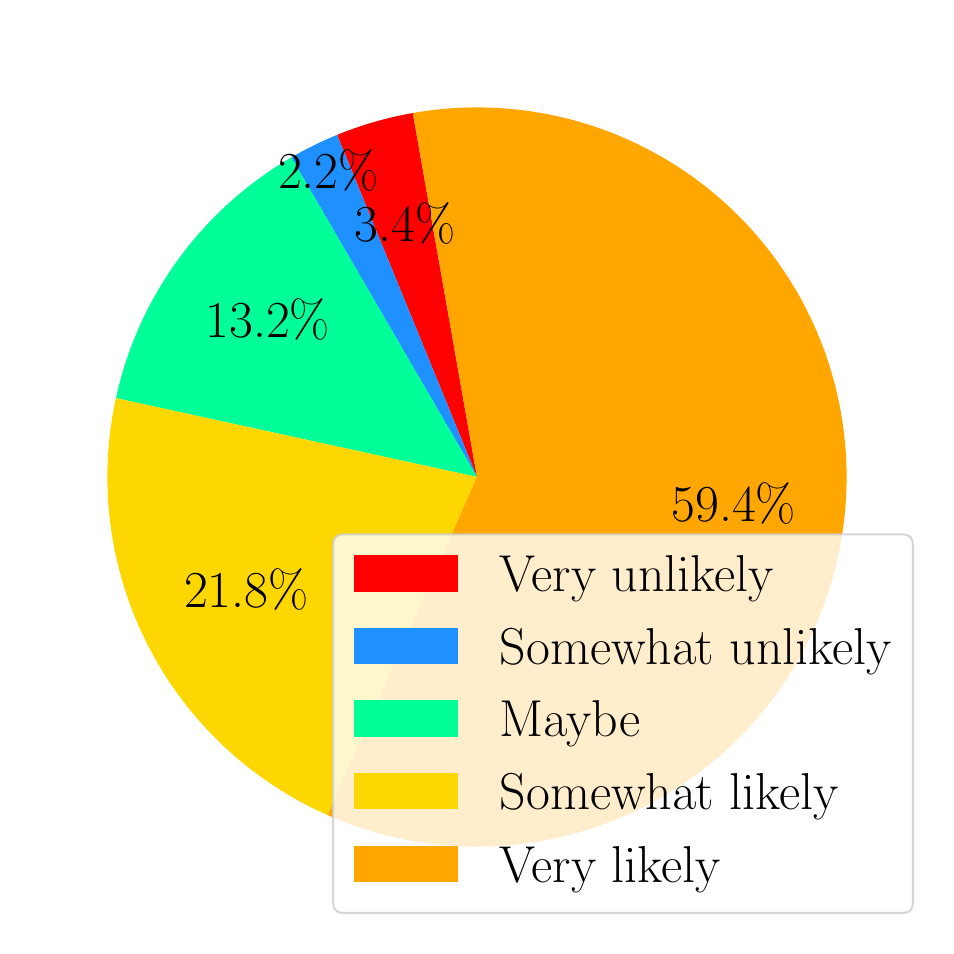}
    \end{subfigure}
    \caption{Distribution of ICML 2023 review outcomes (left), authors' responses to the question ``How confident are you about your ranking?'' (middle), and authors' responses to the question ``How likely would you be to provide the same ranking if it were to be used for decision-making?'' (right).}
    \label{fig:pie_chart}
\end{figure*}

Submissions to ICML 2023 were generally reviewed by three or four reviewers, who evaluated each submission on a scale from 1 (very strong reject) to 10 (award quality). The outcomes for the 6,538 submissions are summarized in the left pie chart of Figure \ref{fig:pie_chart}, which excludes the ``Outstanding Paper Award'' category, awarded to six submissions.



The mean of the average review scores\footnote{In this paper, the term ``average review score'' of a paper refers to the simple average of all scores provided by reviewers for that paper.} following the rebuttal period for the categories ``Rejected'', ``Accepted as Poster'', ``Accepted as Oral'', and ``Outstanding Paper Award'' is \num{4.32}, \num{5.93}, \num{6.82}, and \num{7.72}, respectively.

The survey-based experiment was conducted in OpenReview, which hosted the peer review process for ICML 2023, in conjunction with \href{https://openrank.cc/}{OpenRank.cc}, which we developed to implement the experiment. On January 26, 2023, immediately following the submission deadline, an official email was sent through OpenReview to all ICML authors requesting information about their submissions. Importantly, participants were informed that the survey data would not be used in the decision-making process for ICML 2023. 
Figure S.1
in the Supplementary Material shows the survey interface. Specifically, we solicited the following information:
\begin{itemize}
    \item Ranking: Authors with multiple submissions were asked to rank their papers based on their perceived quality, with allowances for ties in the rankings. Authors could order their papers by dragging them up or down at the OpenRank.cc interface.
    
    \item Additional questions: All authors, including those with only one submission, were asked to respond to some questions, such as their confidence in the provided rankings and their perceived probability of inconsistencies between their expectations and the review outcomes. All these questions are shown in 
    Figure S.1
    in the Supplementary Material.
        
\end{itemize}

Additionally, review scores and final decisions were retrieved from OpenReview. 

This study was conducted with the approval of the Institutional Review Boards (IRB) at the University of Pennsylvania. The experiment and subsequent analyses adhered to strict privacy and confidentiality standards. Specifically, the data were anonymized by excluding all personal identifying information, and analysis began only after the accept/reject decisions were announced. For further information regarding our privacy policy, please refer to \url{https://openrank.cc/legal/privacy}. Furthermore, it is our policy that all data collected from this experiment be completely deleted by December 31, 2024.

\paragraph{Summary statistics from the experiment.}

We first provide statistics about the rankings obtained from the survey: (a) Number of authors who completed the survey: \num{5634} ($30.4\%$); (b) Number of authors who had multiple submissions and provided rankings of their submissions: \num{1342}; (c) Number of submissions that were ranked by at least one author: \num{2592} ($39.6\%$); (d) Number of reviews received by these \num{2592} submissions: \num{7974} (3.08 reviews on average per submission);\footnote{This is the number of reviews received before the rebuttal, as of March 12, 2023. The average number of reviews per submission increased to 3.29 after the rebuttal period, as of April 22, 2023.} (e) The longest ranking list provided by an author: \num{17} submissions.

The dependence between the number of submissions by an author and their likelihood of completing the survey is shown in 
Figure S.5
in the Supplementary Material.
It appears that authors with more submissions were less likely to provide rankings.

Regarding the remaining questions in the survey, 59.8\% of the authors reported high confidence in their rankings, and 59.4\% would likely provide the same rankings if they were to be used in the decision-making process. More details are provided in the middle and right pie charts of Figure~\ref{fig:pie_chart}.

In response to the question, ``What is your estimated probability that your lowest-ranked paper will have a higher or equal average rating than your highest-ranked paper?'', over half of the authors estimated this probability to be at least 40\%. The average of these estimated probabilities is 36.6\%. The distribution of these probabilities is illustrated in 
Figure S.6
in the Supplementary Material. In contrast, the actual proportion of authors whose lowest-ranked papers received higher or equal scores prior to the rebuttal period, compared to their highest-ranked papers, is 42.2\%.

\paragraph{Preliminary analysis of rankings.}
Examining the relevance of author-provided rankings in predicting the quality of submissions is a key focus of our study. If these rankings were not predictive of review outcomes at all, incorporating them into the Isotonic Mechanism would be unlikely to enhance the efficacy of the review process. Yet, our preliminary analysis suggests that these rankings are indeed predictive, indicating their potential value to improve peer review.

To investigate this aspect, we grouped the highest-ranked paper by an author into one category and the lowest-ranked paper into another. The mean of average review scores received by the highest-ranked papers is 4.80 before the rebuttal period, while it is 4.50 for the lowest-ranked papers. Although this difference is not large,\footnote{A small difference is not surprising given that review scores tend to be noisy. Moreover, this regime is where the Isotonic Mechanism can produce significantly more accurate scores than the raw scores (see Theorem 3 in \cite{su2021you}).} it is statistically significant, with a $p$-value of $3.56 \times 10^{-20}$ under a paired one-sided $t$-test. A more detailed comparison between the two groups depending on the categories of decision is given in Table \ref{Table:ranking_final_decision_percentage},
which shows that highest-ranked submissions were more likely to obtain better review outcomes.

Moreover, this positive correlation extends beyond the rebuttal period, where highest-ranked papers were more likely to receive an increase in score, with an average increase of $0.23$, compared to $0.20$ for the lowest-ranked. However, our analysis found no statistically significant correlation between the rankings and the length of the rebuttals, as measured by word count. This is illustrated in 
Table S.1
in the Supplementary Material.

\begin{table}[!htbp] 
    \centering
    \begin{tabular}{c|c|c|c|c}
    \hline\hline
         & \begin{tabular}{@{}c@{}} ``Withdrawn or  \\ Desk Rejected'' \end{tabular} & ``Rejected'' & \begin{tabular}{@{}c@{}} ``Accepted as  \\ Poster'' \end{tabular} & \begin{tabular}{@{}c@{}} ``Accepted as  \\ Oral'' \end{tabular} \\
         \hline
         Highest-ranked & $9.02 \%$ & $53.20 \%$ & $33.31\%$ & $4.47\%$ \\
         \hline
         Lowest-ranked & $16.24\%$ & $57.68\%$ & $23.85\%$ & $2.24\%$ \\ 
         \hline
         $p$-value & $2.43 \times 10^{-8}$ & $2.20 \times 10^{-3}$ & $7.31 \times 10^{-8}$ & $1.87 \times 10^{-3}$
         \\
    \hline
    \end{tabular}
    \caption{Comparison of outcomes (``Withdrawn/Desk Rejected'', ``Rejected'', ``Accepted as Poster'', and ``Accepted as Oral'') for highest-ranked versus lowest-ranked submissions. Highest-ranked submissions received significantly better review outcomes. We use a Chi-square test to obtain $p$-values.}
    \label{Table:ranking_final_decision_percentage}
\end{table}

\section{Statistical Analysis of Isotonic Scores}
\label{sec:Isotonic_Scores}

In this section, we analyze the ICML 2023 ranking data, which comprises 1,342 rankings associated with 2,592 submissions. Our main finding is that the Isotonic Mechanism can reduce the MSE of review scores, and this improvement in estimation accuracy is not only statistically significant but also becomes more pronounced as the number of an author's submissions increases. This empirical finding aligns with the theoretical analysis of the method~\citep{su2021you,su2022truthful}.

\subsection{Method and Evaluation}\label{sec:mm}
The Isotonic Mechanism operates as follows~\citep{su2021you}. Consider an author who submits $n$ papers to a conference. The mechanism requires the author to rank these submissions in descending order of perceived quality. The ranking is denoted by $\pi$, which allows for ties. Given the (average) raw review scores $\by := (y_1, y_2, \ldots, y_n)$ for the $n$ submissions, the Isotonic Mechanism outputs the ranking-calibrated scores as the solution to the following optimization program:
\begin{equation}\nonumber
\begin{aligned}
&\min_{\br:=(r_1, \ldots, r_n)} & &\|\by - \br\|^2 \qquad \text{s.t.~} & &r_{\pi(1)} \ge \cdots \ge r_{\pi(n)},
\end{aligned}
\end{equation}
where $\|\cdot\|$ denotes $\ell_2$ norm or, equivalently, Euclidean distance. Formally, this convex optimization program is equivalent to isotonic regression~\citep{barlow1972isotonic}. For example, letting $\by = (8,7,4,3)$ and $(\pi(1), \pi(2), \pi(3), \pi(4)) = (1,3,2,4)$,\footnote{From this ranking, the author has the opinion that the last paper has the lowest quality and the first paper has the highest quality.} the isotonic scores are $(8,5.5,5.5,3)$.

An essential aspect of this method lies in the assumption that the author is knowledgeable about the quality of their submissions. The Isotonic Mechanism is perhaps the simplest blend of authors' and reviewers' perspectives. Furthermore, under certain conditions, authors are incentivized to truthfully report their rankings when these modified scores are used for decision-making~\citep{su2021you}. With truthful author-provided rankings, the isotonic scores are more accurate than the raw scores in estimating the ground truth quality of submissions. Extensions of this mechanism for broader practical settings are discussed in \cite{yan2023isotonic} and \cite{wu2023isotonic}.

\begin{figure}[!htp]
  \centering
  \includegraphics[width = 0.9\textwidth]{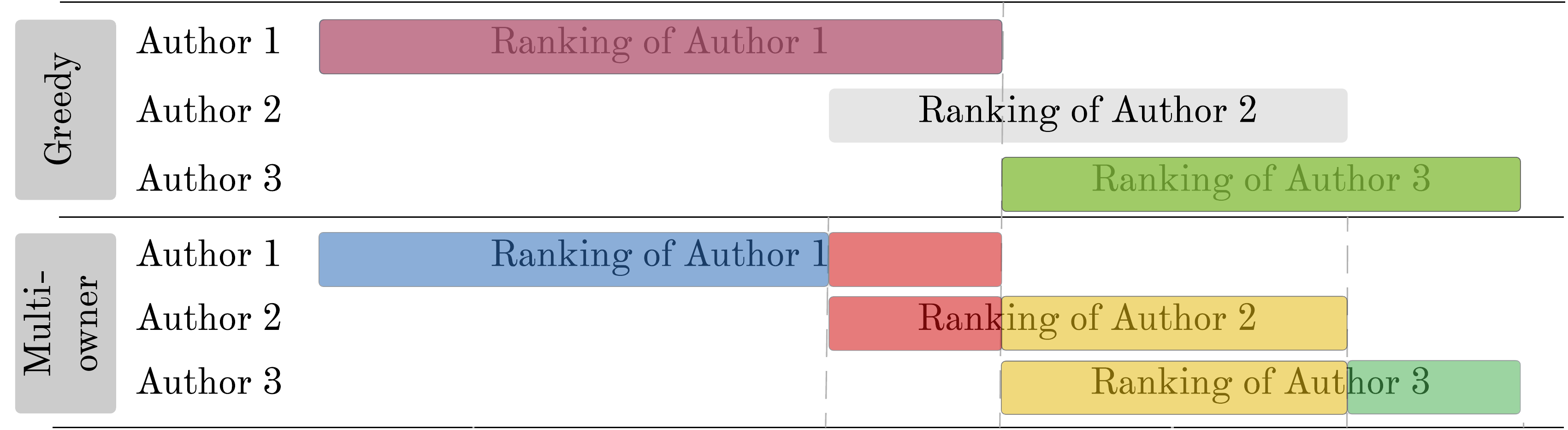}
  \caption{Illustration of the greedy and multi-owner strategies for the Isotonic Mechanism in the setting of multiple authors. Author-submission pairs highlighted in gray (Author 2's submissions in the greedy strategy) are excluded from consideration in the mechanism. In the multi-owner strategy, any paper in the red block has its score averaged over the two isotonic scores from Author 1 and Author 2.}
  \label{fig:iso_procedure}
\end{figure}

To adapt the Isotonic Mechanism to the practical setting where most papers each have multiple authors, we consider three strategies:

\begin{itemize}
    
    \item \textbf{Simple-averaging strategy.} The first strategy runs the mechanism for each author who provides a ranking. For a given submission, different authors often yield different modified scores. The isotonic score for the submission is calculated as the simple average of these different modified scores.
    
    \item \textbf{Greedy strategy.} This strategy starts by running the Isotonic Mechanism for the author who provides the longest ranking, and then removes this author and all submissions of the author from further consideration. This process is repeated for the remaining authors and their submissions until each submission has exactly one isotonic score.\footnote{If the length of the ranking is 1, the isotonic score will be identical to the raw review score.}
    
    \item \textbf{Multi-owner strategy.} The first step of this strategy is to partition all submissions into disjoint blocks such that in each block every submission shares a common set of authors. In each block,  run the Isotonic Mechanism taking as input a ranking within the block from each author to yield modified scores. The last step is to average the modified scores separately for each block. See more details about this approach in \cite{wu2023isotonic}.
\end{itemize}

In both the greedy and multi-owner strategies, each run of the Isotonic Mechanism operates on a sub-ranking that involves a subset of the submissions from an author. However, it is important to note that these sub-rankings can be derived from complete rankings. Therefore, authors can provide complete rankings, regardless of the strategy being implemented. Notably, under certain conditions, the greedy and multi-owner strategies ensure the truthful reporting of rankings by authors~\citep{wu2023isotonic}. In contrast, the simple-averaging approach does not generally guarantee this.

\paragraph{Evaluation metrics.} Evaluating the performance of isotonic scores compared to raw scores presents a challenge due to the unknown ground truth quality of submissions. To address this challenge, we leverage the fact that a submission typically receives multiple reviews, resulting in multiple review scores. For simplicity, consider $y$ and $y'$ as two independent scores of the same submission, both assumed to be unbiased estimators of the ground truth.\footnote{Scores may exhibit biases conditional on certain factors~\citep{wang2020debiasing}. In our context, ``unbiasedness'' is understood in a marginal sense, achieved through random selection of scores without conditioning on variables such as confidence level or reviewer seniority. Moreover, it is more appropriate to interpret ``ground truth'' as the ``ground truth score'' rather than the intrinsic merit of a paper. Practically, the ground truth score could be considered as the average score given by a very large number (say, 1,000) of reviewers.} Let $\hat y$ denote any estimator of the ground truth using only the data $y$. The performance of $\hat y$ is measured using either $(\hat{y} - y')^2$ or $|\hat{y} - y'|$, which we refer to as the ``proxy'' MSE and MAE, respectively. Note that the conventional MSE and MAE of $\hat y$ are defined as $\mathbb{E} (\hat{y} - \text{ground truth})^2$ and $\mathbb{E} |\hat{y} - \text{ground truth}|$, respectively, which are not observable. In contrast, $(\hat{y} - y')^2$ and $|\hat{y} - y'|$ can be precisely calculated from the raw scores.

Both proxy MSE and MAE are upward-biased estimators of their conventional counterparts. This can be seen, as the expectation of the proxy MSE is expressed as
\begin{equation}\nonumber
\mathbb{E} (\hat{y} - y')^2 = \mathbb{E} \left( \hat{y} - \mathbb{E} y' \right)^2 + \mathrm{Var}(y') = \mathrm{MSE}(\hat{y}) + \mathrm{Var}(y').
\end{equation}
In essence, the bias of the proxy MSE is equal to the variance of the ``noisy target'' $y'$. For the proxy MAE, note that it satisfies $\mathbb{E}|\hat{y} - y'| \ge  \mathbb{E}|\hat{y} - \mathbb{E} y'| = \mathrm{MAE}(\hat y)$. Here, the inequality follows from Jensen's inequality when applied to the convex function $|c - x|$ for any constant $\hat y = c$.

Despite this bias, the proxy MSE retains the ability to compare two estimators in expectation: for any two estimators $\hat{y}$ and $\tilde{y}$, their proxy MSE's difference satisfies
\begin{align} \label{eqn:unbias}
\begin{split}
\mathbb{E} (\hat{y} - y')^2 - \mathbb{E} (\tilde{y} - y')^2 &= \mathrm{MSE}(\hat{y}) + \mathrm{Var}(y') - \mathrm{MSE}(\tilde{y}) - \mathrm{Var}(y')\\
&= \mathrm{MSE}(\hat{y}) - \mathrm{MSE}(\tilde{y}).
\end{split}
\end{align}
Therefore, if $\hat{y}$ outperforms $\tilde y$ in terms of MSE, then $\hat{y}$ will also have a smaller proxy MSE than $\tilde y$ in expectation, and vice versa.

When analyzing the ICML 2023 ranking data, we randomly selected one review score per submission as the data $y$ for estimating the submission's quality using either the Isotonic Mechanism or the raw-score estimator. The average of the remaining review scores serves as the noisy target $y'$. For this purpose, a submission must have at least two review scores. This is applicable to 2,530 out of the 2,592 ranked submissions.

\subsection{Results}
\label{sec:main}
\begin{figure}[!htbp]
    \centering
    \begin{subfigure}[b]{0.47\textwidth}
        \footnotesize
        \stackunder[0pt]{\includegraphics[height = 0.62\textwidth]{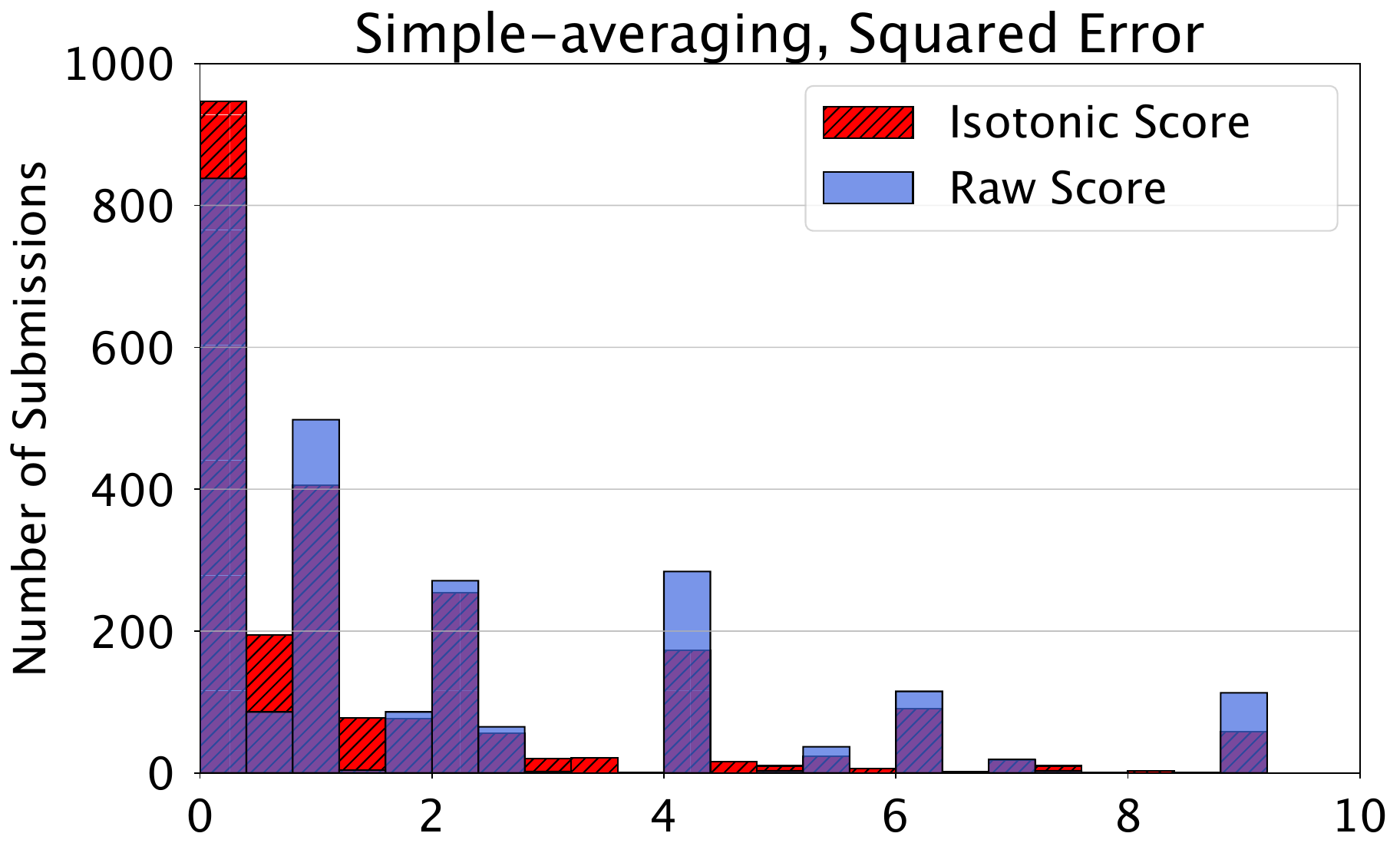}
        }{Proxy MSE}
            \end{subfigure}
    \begin{subfigure}[b]{0.47\textwidth}
        \footnotesize
        \stackunder[0pt]{\includegraphics[height = 0.62\textwidth]{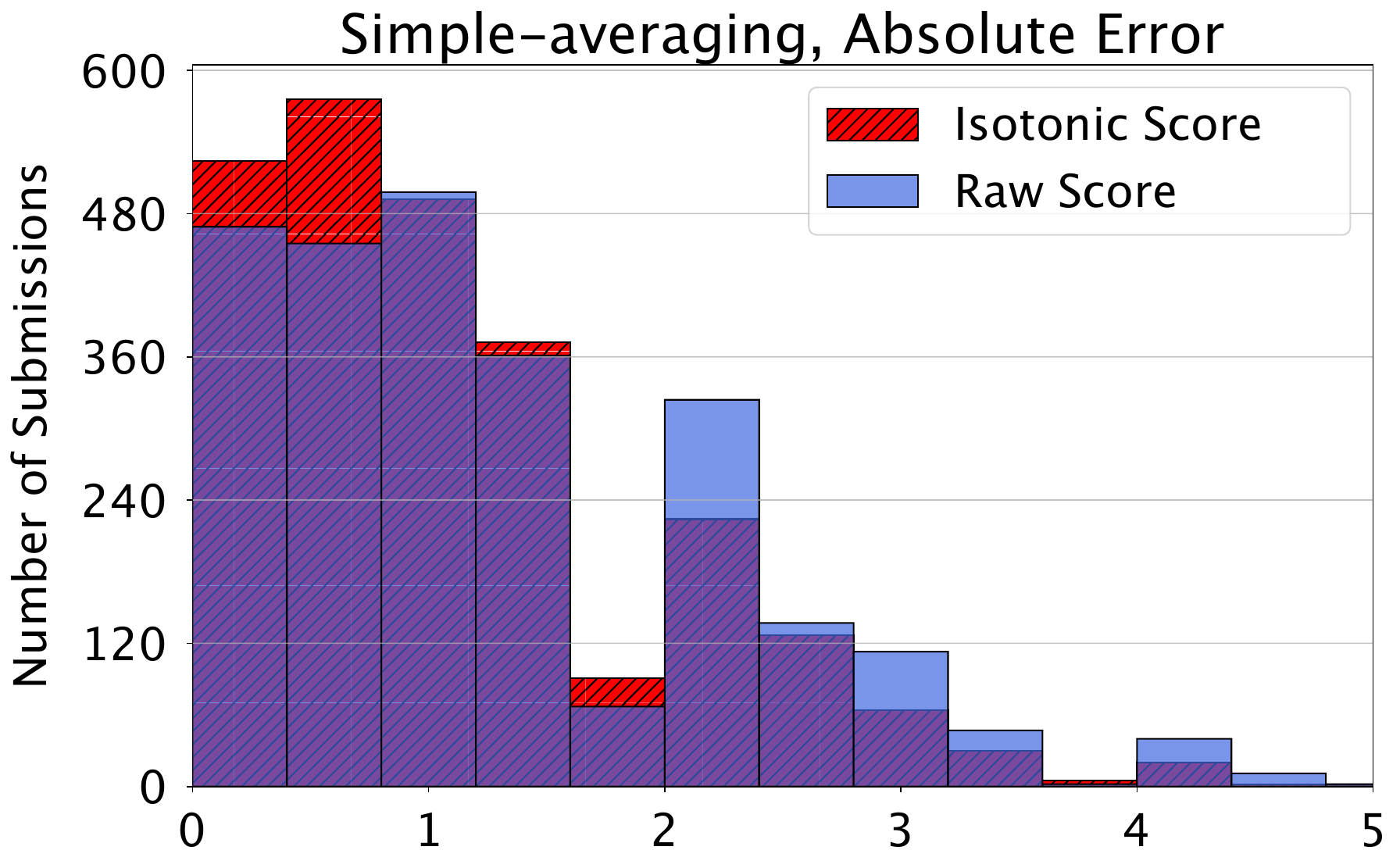}
        }{Proxy MAE}
          \end{subfigure}

    \vspace{2mm}

    \begin{subfigure}[b]{0.47\textwidth}
        \footnotesize
        \stackunder[0pt]{\includegraphics[height = 0.62\textwidth]{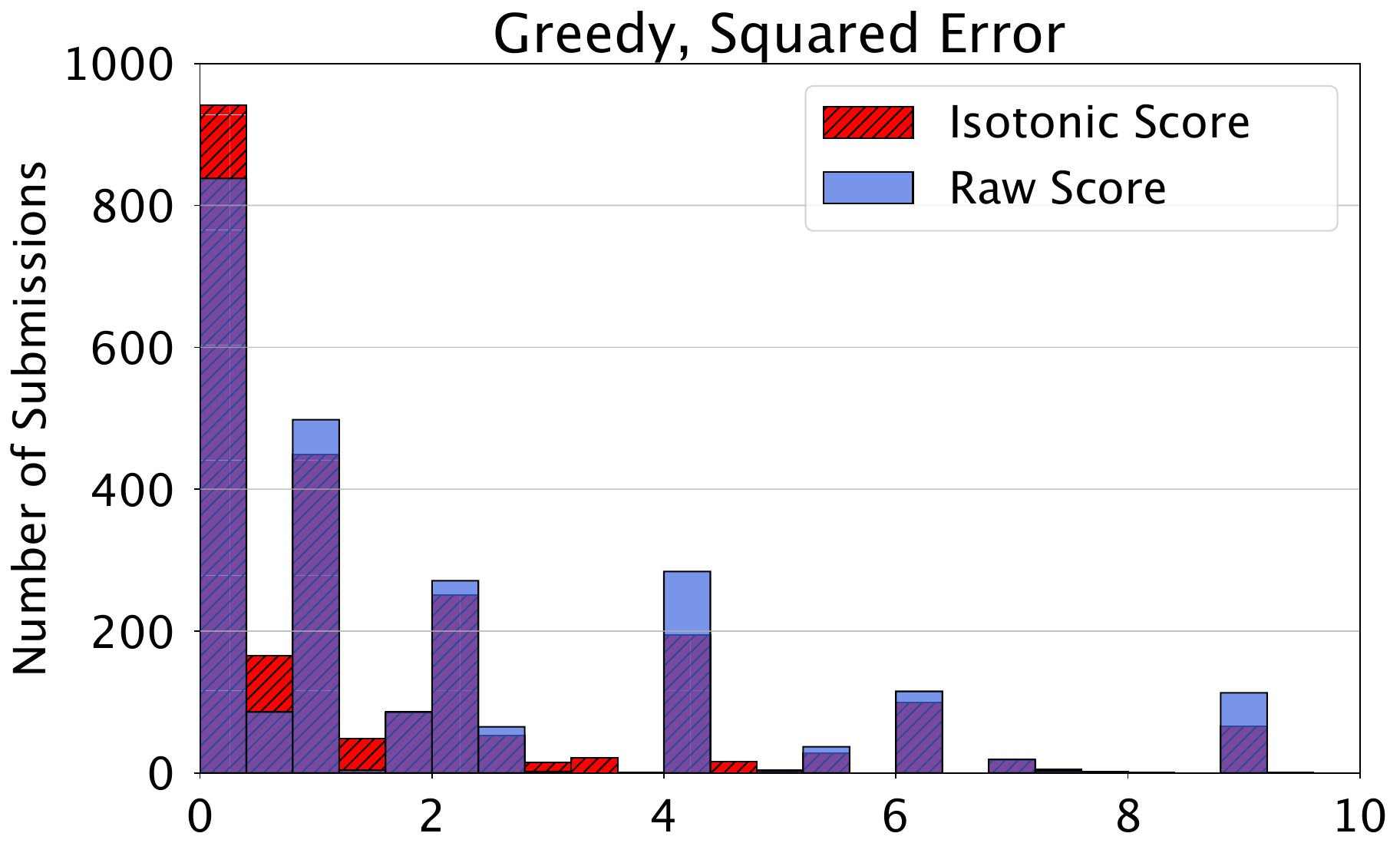}
        }{Proxy MSE}
            \end{subfigure}
    \begin{subfigure}[b]{0.47\textwidth}
        \footnotesize
        \stackunder[0pt]{\includegraphics[height = 0.62\textwidth]{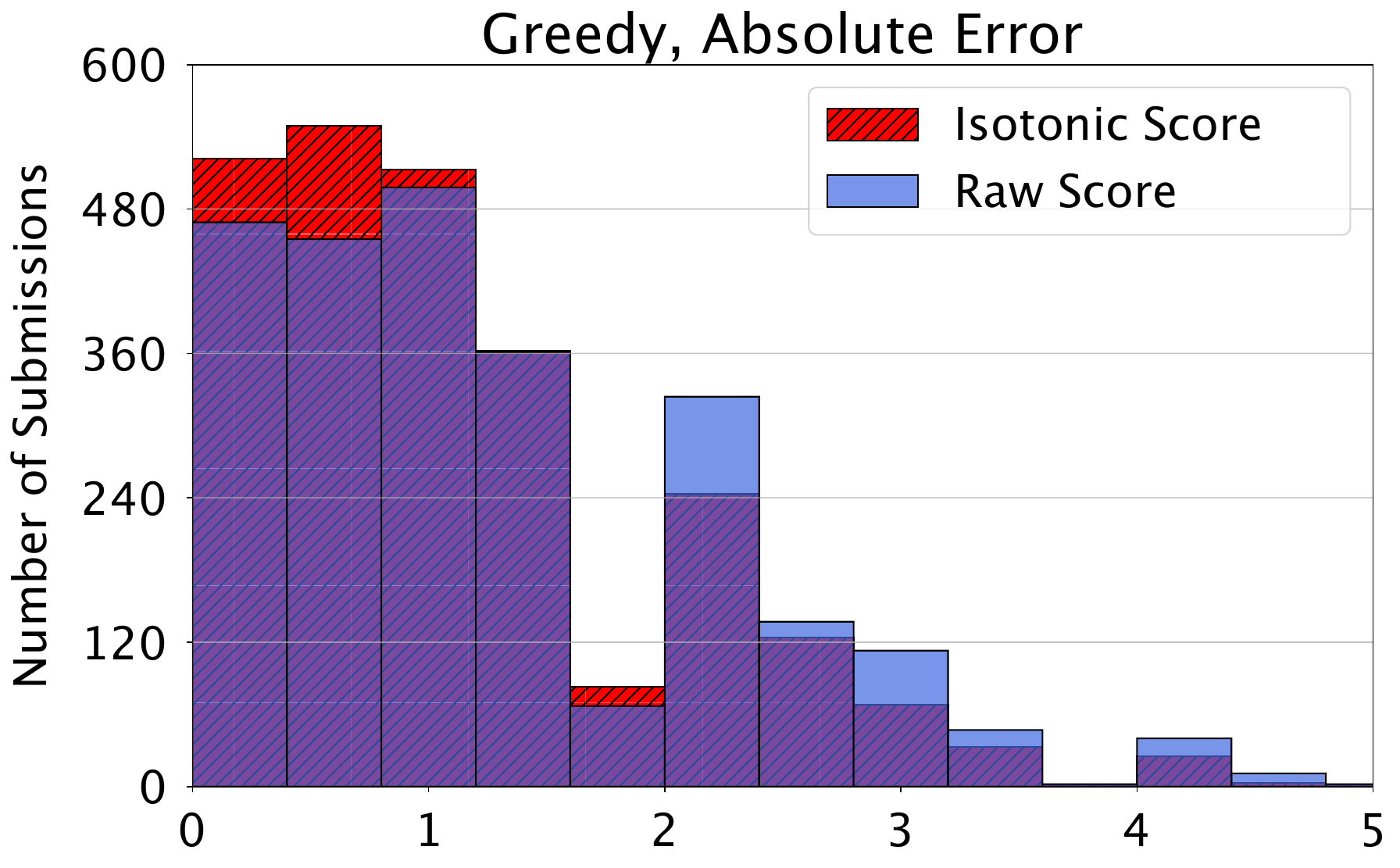}
        }{Proxy MAE}
            \end{subfigure}

    \vspace{2mm}
        
    \begin{subfigure}[b]{0.47\textwidth}
        \footnotesize
        \stackunder[0pt]{\includegraphics[height = 0.62\textwidth]{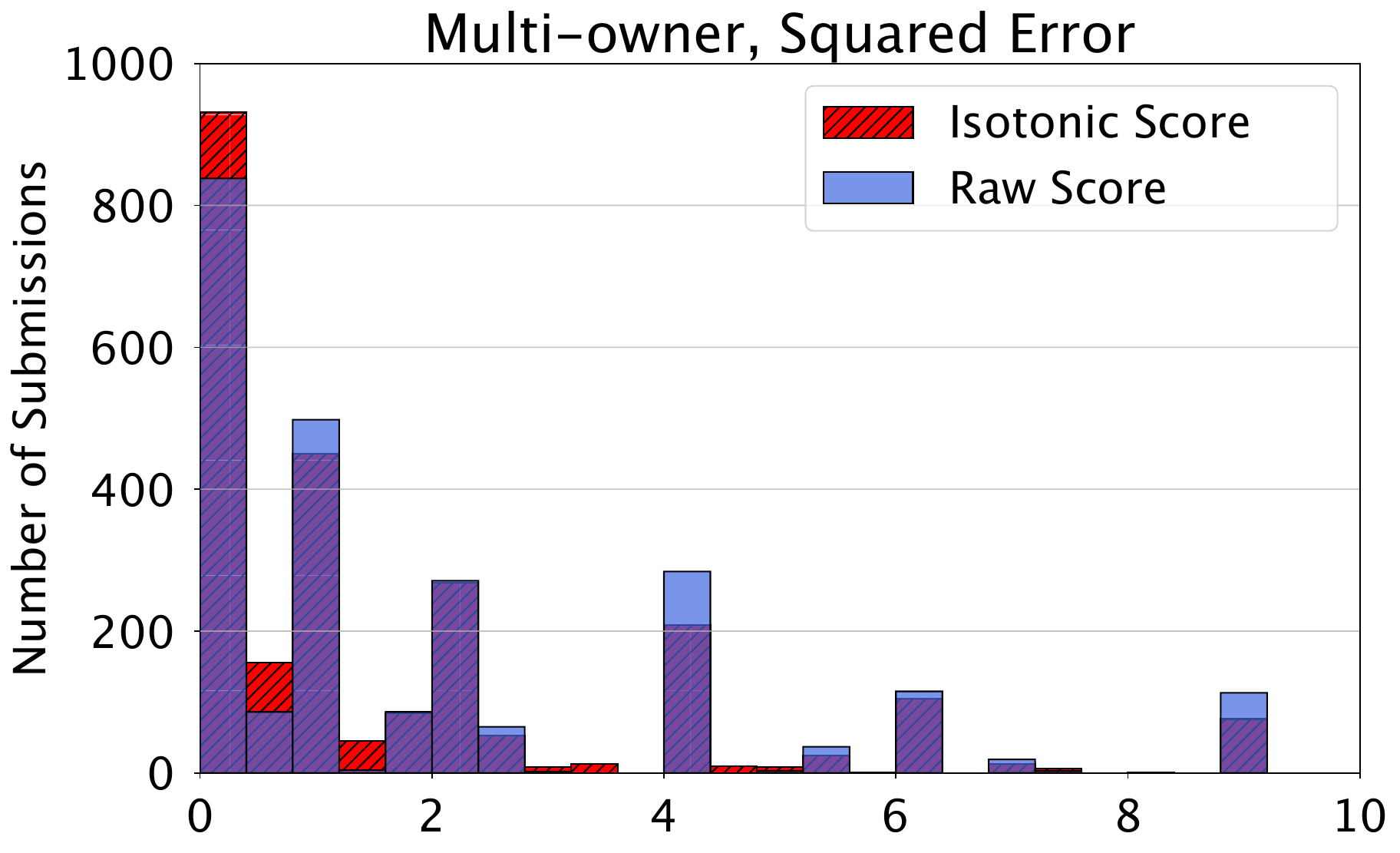}
        }{Proxy MSE}
            \end{subfigure}
    \begin{subfigure}[b]{0.47\textwidth}
        \footnotesize
        \stackunder[0pt]{\includegraphics[height = 0.62\textwidth]{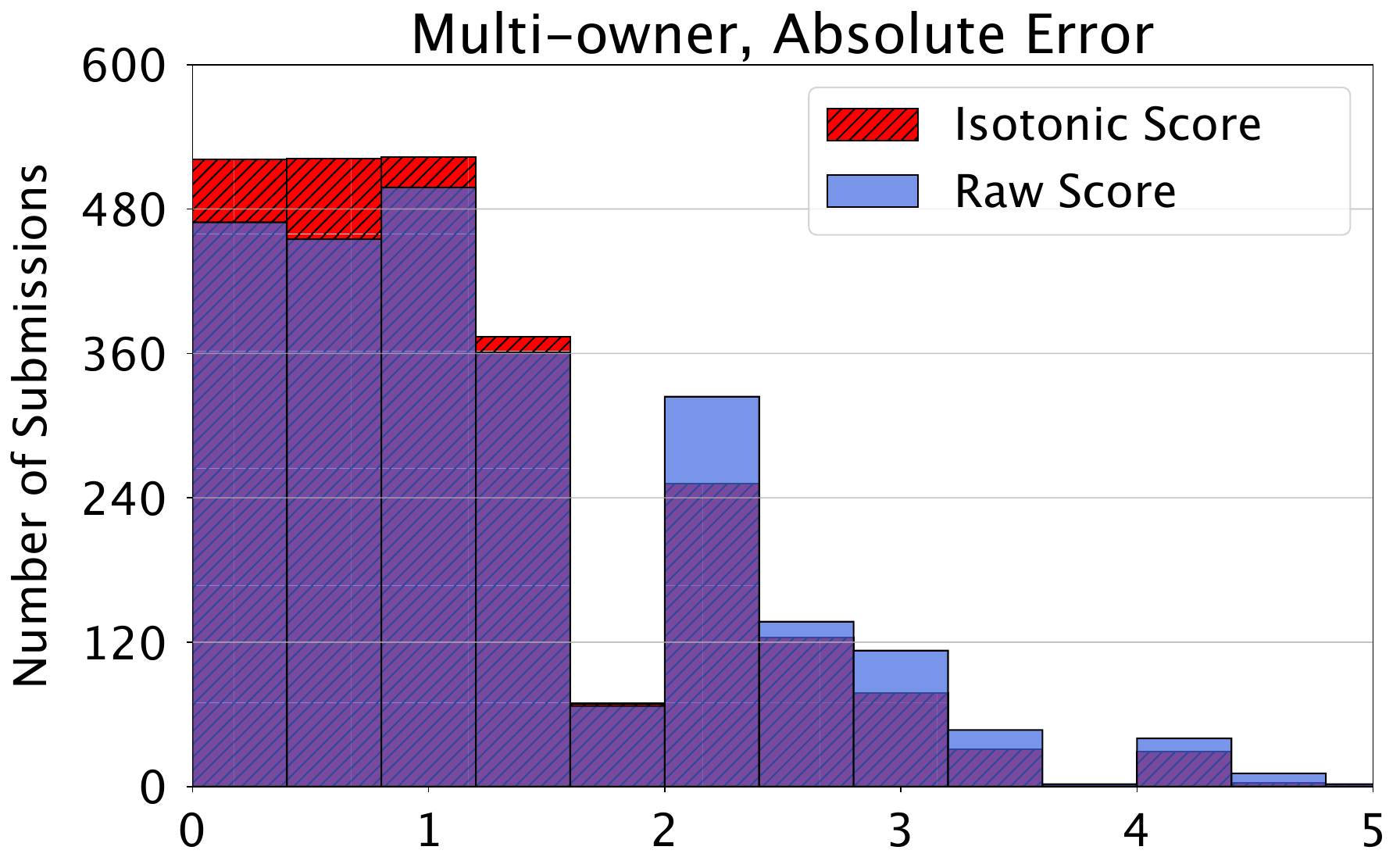}
        }{Proxy MAE}
            \end{subfigure}
    \caption{Histograms comparing the distributions of isotonic and raw scores in terms of proxy MSE (left) and proxy MAE (right). The distribution of isotonic scores is more heavily weighted towards smaller errors.}
    \label{fig:histogram_residual}
\end{figure}

To compare the isotonic and raw scores, Figure~\ref{fig:histogram_residual} presents histograms of the distributions of proxy MSE and MAE for both isotonic and raw scores across the \num{2530} submissions. The distributions show that the isotonic scores yield lower proxy errors than the raw scores, with the density of submissions shifting toward smaller errors. This consistency appears across all three strategies of the Isotonic Mechanism, suggesting an overall improvement in the estimation of submission quality. Furthermore, 
Figure S.8
in the Supplementary Material corroborates the findings presented in 
Figure~\ref{fig:histogram_residual}.

Table \ref{tab:greedy_multi_isotonic_mse} demonstrates that the Isotonic Mechanism using any of the three strategies reduces both the overall proxy MSE and MAE compared to raw scores. Specifically, the greedy strategy achieves a $21.3\%$ reduction in the proxy MSE and $11.7\%$ in the proxy MAE. Furthermore, evidence suggests that the reduction in conventional MSE is likely to exceed $21.3\%$ when employing isotonic scores, as elaborated in 
Section C
in the Supplementary Material.

\begin{table}[!htp]
\centering
\renewcommand{\arraystretch}{1.05}
\resizebox{\textwidth}{!}{
\begin{tabular}{l||c|c|c|c|c|c}

\hline \hline
\multirow{2}{*}{} & \multicolumn{3}{c|}{Proxy MSE } & \multicolumn{3}{c}{Proxy MAE } \\

\cline{2-7}
& \begin{tabular}{@{}c@{}} Error \end{tabular} & \begin{tabular}{@{}c@{}} Improvement \end{tabular} & 
\begin{tabular}{@{}c@{}} $p$-value \end{tabular} & \begin{tabular}{@{}c@{}} Error \end{tabular} & \begin{tabular}{@{}c@{}} Improvement \end{tabular} & 
\begin{tabular}{@{}c@{}} $p$-value \end{tabular} 
\\

\hline
\begin{tabular}{@{}c@{}} Raw Score \end{tabular} & $2.57$ & \texttt{NA} & \texttt{NA} & $1.26$ & \texttt{NA} & \texttt{NA}\\ 

\hline
\begin{tabular}{@{}c@{}} Simple-averaging Strategy \end{tabular} & $1.97$ & $23.48 \%$ & $1.09 \times 10^{-43}$ & $1.10$ & $12.75 \%$ & $1.02 \times 10^{-39}$\\ 

\hline
\begin{tabular}{@{}c@{}} Greedy Strategy \end{tabular} & $2.02$ & $21.30 \%$ & $1.46 \times 10^{-36}$ & $1.11$ & $11.71 \%$ & $2.06 \times 10^{-32}$\\ 

\hline
\begin{tabular}{@{}c@{}} Multi-owner Strategy \end{tabular} & $2.07$ & $19.38 \%$ & $5.58 \times 10^{-34}$ & $1.12$ & $10.62 \%$ & $5.75 \times 10^{-31}$\\ 

\hline
\end{tabular}}
\caption{Reduction of proxy MSE and MAE using the Isotonic Mechanism with various strategies. A paired one-sided $t$-test shows that the reduction in proxy errors is statistically highly significant. Note that the results use a randomly selected raw-score estimator. We repeat this random selection process six times and aggregate the results in 
Tables S.5 and S.6 
in the Supplementary Material, which remain highly consistent with those shown in Table~\ref{tab:greedy_multi_isotonic_mse}, suggesting that the randomness in the selection process has little impact on the observed improvements.}
\label{tab:greedy_multi_isotonic_mse}
\end{table}

Denote by $\hat{y}^{\textnormal{Iso}}$ and $\hat{y}^{\textnormal{Ave}}$ the isotonic score and raw score, respectively. As shown in \eqref{eqn:unbias}, $(\hat{y}^{\textnormal{Ave}} - y')^2 - (\hat{y}^{\textnormal{Iso}} - y')^2$ is an unbiased estimator of $\mathrm{MSE}(\hat{y}^{\textnormal{Ave}}) - \mathrm{MSE}(\hat{y}^{\textnormal{Iso}})$. This observation allows us to construct confidence intervals for the average reduction in ground truth MSE---that is, $\mathrm{MSE}(\hat{y}^{\textnormal{Ave}}) - \mathrm{MSE}(\hat{y}^{\textnormal{Iso}})$ averaged over all ranked submissions---and we present the results in Table~\ref{tab:mse_inference}. At the 95\% confidence level, $\mathrm{MSE}(\hat{y}_i^{\textnormal{Iso}})$ on average is smaller than $\mathrm{MSE}(\hat{y}_i^{\textnormal{Ave}})$ by 0.4 or more.

\begin{table}[!htp]
\centering
    \begin{tabular}{c|c|c|c}
    \hline\hline
         Confidence Level & \begin{tabular}{@{}c@{}} Simple-averaging Strategy \end{tabular} & \begin{tabular}{@{}c@{}} Greedy Strategy \end{tabular} & \begin{tabular}{@{}c@{}} Multi-owner Strategy \end{tabular} \\
         \hline
         \begin{tabular}{@{}c@{}} $95\%$ \end{tabular} & $[0.52, 0.69]$ & $[0.46, 0.63]$ & $[0.42, 0.58]$ \\
         \hline
         \begin{tabular}{@{}c@{}} $99\%$ \end{tabular} & $[0.49, 0.71]$ & $[0.44, 0.66]$ & $[0.39, 0.60]$ \\
         \hline
    \end{tabular}
\caption{95\% and 99\% confidence intervals for the average reduction of MSE by using isotonic scores compared to raw scores. This makes use of the fact that the reduction in proxy MSE is an unbiased estimator of ground truth MSE reduction, as shown in \eqref{eqn:unbias}.
} 
\label{tab:mse_inference}
\end{table}

\begin{figure}[!htbp]
    \centering
    \resizebox{0.313\textwidth}{!}{\begin{minipage}[b]{0.40\textwidth}
        \stackunder[0pt]{ \includegraphics[width=\textwidth]{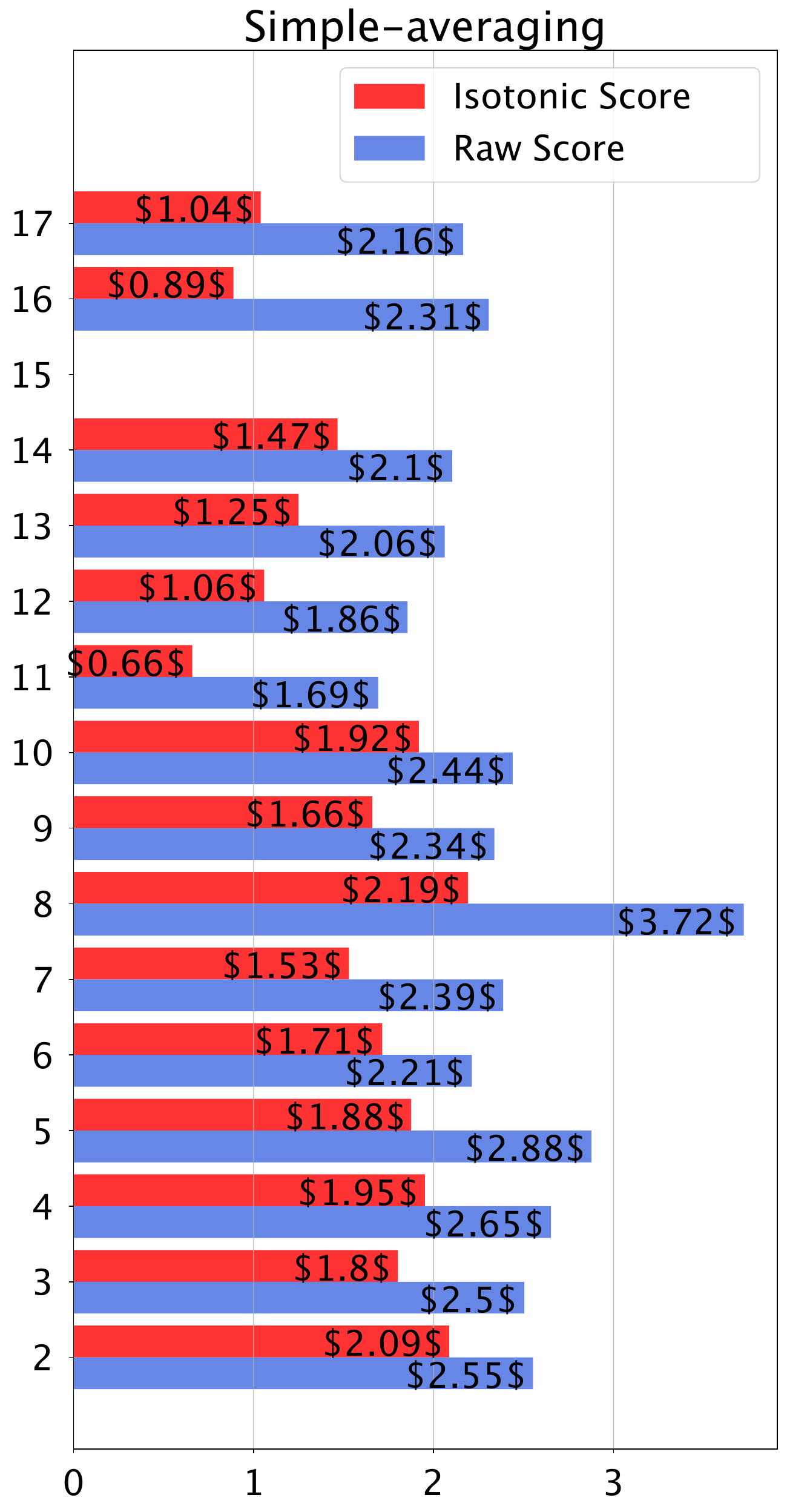}
        }{Mean Squared Error}
    \end{minipage}}
    \resizebox{0.313\textwidth}{!}{\begin{minipage}[b]{0.40\textwidth}
        \stackunder[0pt]{ \includegraphics[width=\textwidth]{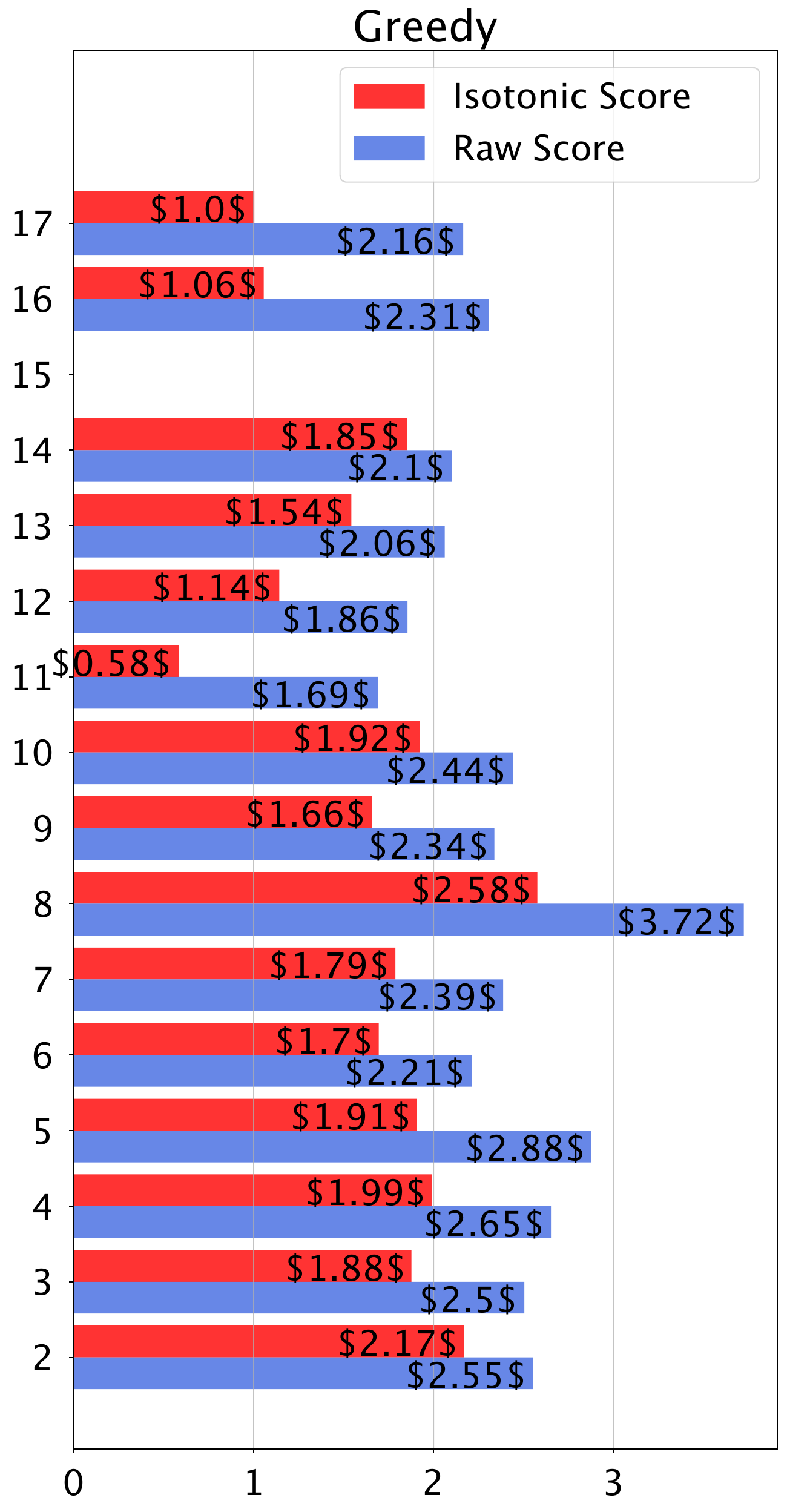}
        }{Mean Squared Error}
    \end{minipage}}
    \resizebox{0.313\textwidth}{!}{\begin{minipage}[b]{0.40\textwidth}
        \stackunder[0pt]{ \includegraphics[width=\textwidth]{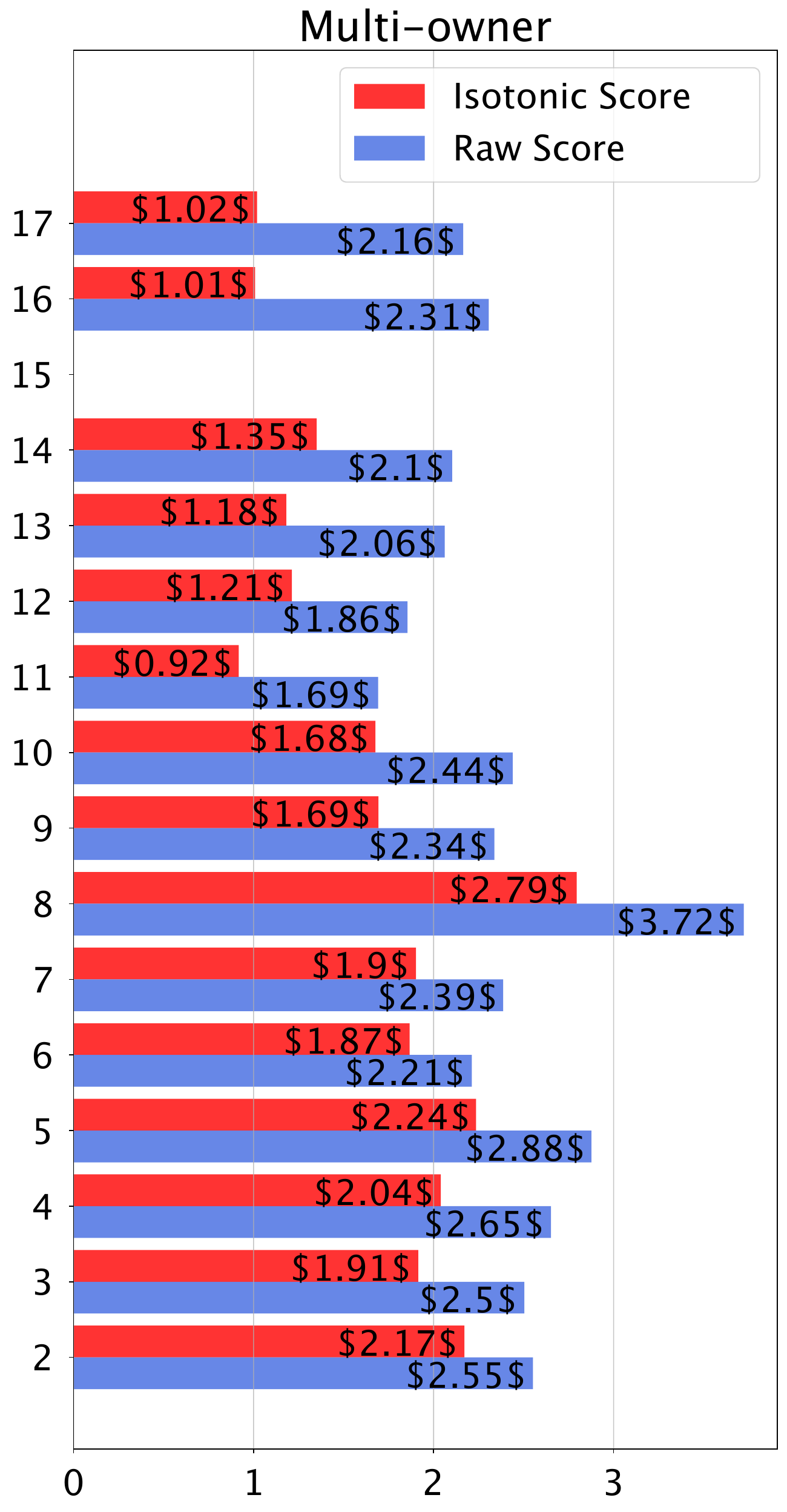}
        }{Mean Squared Error}
    \end{minipage}}
    \resizebox{0.313\textwidth}{!}{\begin{minipage}[b]{0.40\textwidth}
        \stackunder[0pt]{\includegraphics[width=\textwidth]{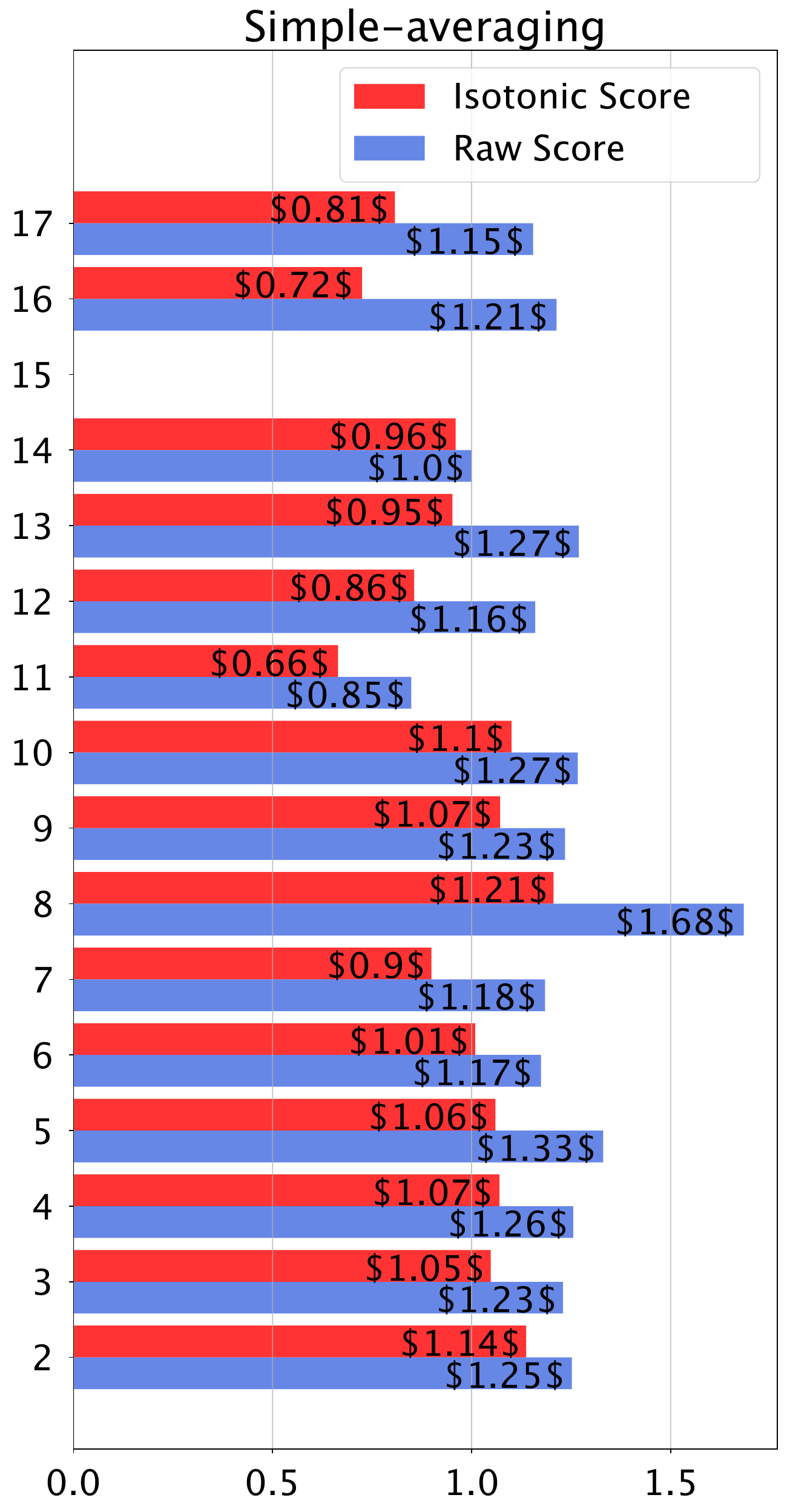}
        }{Mean Absolute Error}
    \end{minipage}}
    \resizebox{0.313\textwidth}{!}{\begin{minipage}[b]{0.40\textwidth}
        \stackunder[0pt]{ \includegraphics[width=\textwidth]{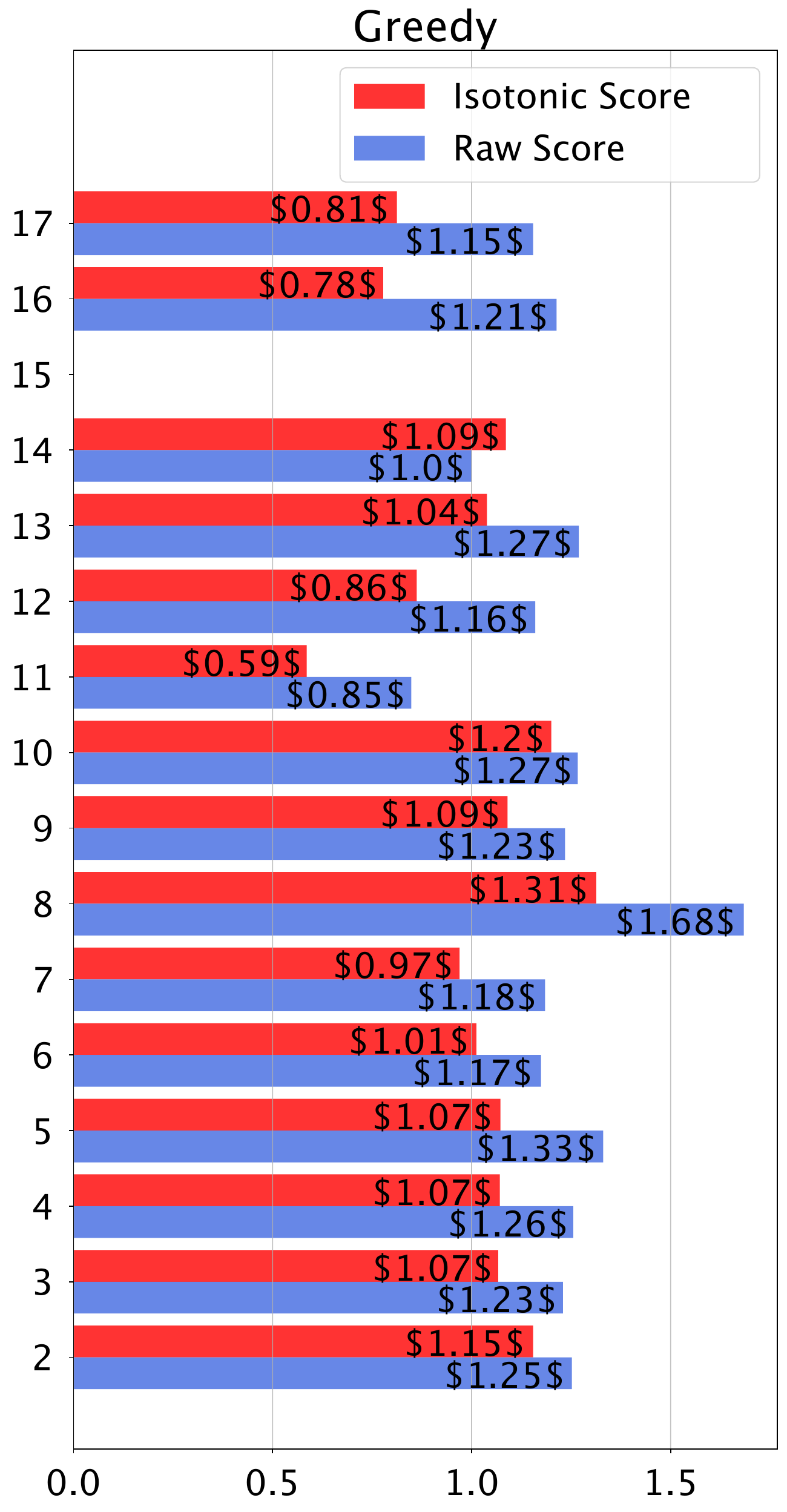}
        }{Mean Absolute Error}
    \end{minipage}}
    \resizebox{0.313\textwidth}{!}{\begin{minipage}[b]{0.40\textwidth}
        \stackunder[0pt]{ \includegraphics[width=\textwidth]{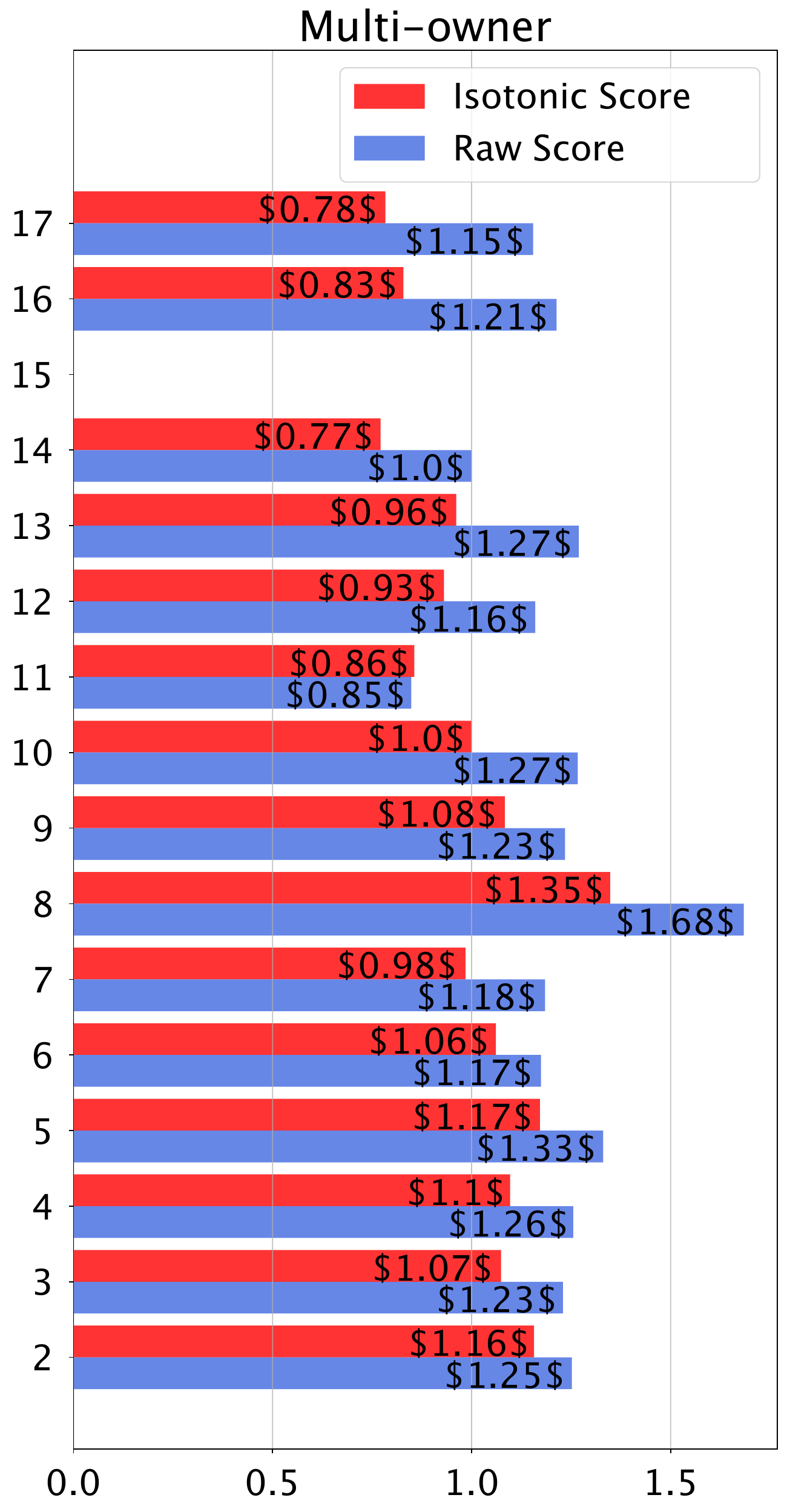}
        }{Mean Absolute Error}
    \end{minipage}}
    \caption{Comparison between isotonic and raw scores in terms of proxy MSE and MAE averaged over ICML 2023 authors who submitted the same-length rankings. 
    \Cref{fig:intro}(c) corresponds to the middle column.}
    \label{fig:no_submission_proxy}
\end{figure}

In investigating the impact of an author's number of submissions on the Isotonic Mechanism's performance, Figure \ref{fig:no_submission_proxy} 
averages the proxy MSE and MAE for isotonic and raw scores across authors with the \textit{same} number of submissions. The results indicate a tangible and statistically significant improvement in estimation accuracy using the Isotonic Mechanism, irrespective of the number of submissions. Overall, this improvement becomes more pronounced with an increase in submission quantity, as shown in 
Figure S.4
in the Supplementary Material.

These findings align with the theoretical analysis in \cite{su2022truthful}, which proves that the Isotonic Mechanism achieves greater performance with larger numbers of submissions. However, shorter rankings are typically more common in our dataset, as authors with more submissions in ICML 2023 tended to have lower response rates. Taken together, our results imply that the advantages of the Isotonic Mechanism would be more pronounced if all authors in ICML 2023 provided their complete rankings.

\section{Applications}
\label{sec:other}

Section \ref{sec:Isotonic_Scores} provides strong evidence that the isotonic scores are more accurate estimates of ground truth compared to the raw review scores, as measured by both MSE and MAE. Despite these results, however, we do not advocate for using them to make a major change in how paper acceptance decisions are made at major ML/AI conferences at the moment. Instead, we suggest a more modest and cautious approach for using isotonic scores in the review process, while at the same time conducting additional empirical evaluations of the Isotonic Mechanism to attempt to better understand the consequences of using it in the review process. 

Towards this end we have identified several specific applications of the Isotonic Mechanism and author-provided rankings that appear to be beneficial without significant negative consequences. These applications primarily target scenarios where authors are aware of important nuances regarding the scientific value of their papers, which might be overlooked by reviewers or ACs. These applications share the common feature that the isotonic scores or author-provided rankings are accessible only to certain high-level roles within the peer review hierarchy, such as SACs and above, and thus give a separation between the isotonic scores and the majority of accept/reject decisions.

\subsection{Oversight of ACs' Recommendations}

The isotonic scores can be used to flag submissions in need of more scrutiny by SACs. In this application, the isotonic scores are made visible to SACs and those in higher roles, who can then use these scores to more effectively oversee the recommendations made by ACs. For instance, significant discrepancies between isotonic scores and ACs' recommendations could serve as red flags, prompting SACs to scrutinize and discuss these cases with the ACs.

The use of isotonic scores in this application presents low risk because ACs, who make the initial accept/reject recommendations (the majority of which are also the final decisions), do not have access to these scores. To further mitigate risk, when an SAC identifies a red flag for a submission's review, the SAC could request that the AC conduct a further review of the submission without specifying that the request is due to a large discrepancy between the accept/reject recommendation and the authors' own opinions.

\subsection{Selection of Paper Awards}
\label{sec:rec}

ML/AI conferences select certain papers to receive awards. The process typically begins with the formation of a shortlist, including papers with high average scores or those nominated by ACs. A committee then carefully reviews these shortlisted papers to identify the award recipients. The committee is intended to carefully weigh each paper on its merits, rather than relying solely on reviewer scores. Given the significant attention that outstanding paper awards attract, it is crucial to ensure that awarded papers are of high quality and free of substantive flaws. This rigor is vital for upholding the reputation of published ML/AI research and maintaining public confidence in the work highlighted by major conferences.


Author-provided rankings could be given as an additional useful piece of information for the committee involved in the selection of paper awards. As evidence that this information might be useful, three out of the six papers awarded as Outstanding Papers at ICML 2023 were ranked by one of their authors and, notably, were all ranked first by their authors. Furthermore, of the 84 submissions that received oral presentations (a distinction that is given to the top few percent of papers) and had rankings from their authors, 69.1\% were ranked first by at least one of their authors. These statistics highlight a strong correlation between the authors' rankings and the recognition the papers received.

In the selection of papers for awards, the rankings could be made visible to some program chairs (PCs) who are not on the selection committee.\footnote{Here, the rankings instead of the isotonic scores are used for a reason that will be elaborated in Section \ref{sec:emergency}.
} The committee relies on their expertise in the selection of the paper awards without knowledge of the author-provided rankings. Once the recommendation is made by the selection committee, the PCs could then scrutinize and raise flags if a recommended paper receives low rankings from its authors, in which case the committee may need to gather additional evidence before considering it for an award.

The award selection takes place following the accept/reject decisions. This phase does not impact most authors, thereby minimizing the potential for unforeseen outcomes when using author-provided rankings.

\subsection{Recruitment of Emergency Reviewers}
\label{sec:emergency}

In ML/AI conferences, it is common practice to recruit emergency reviewers in response to indicators of low review quality, often triggered by low-confidence reviews or significant disagreement among reviewers for a submission. For instance, NeurIPS 2023 recommended recruiting an additional emergency reviewer for each low-confidence review in addition to the four regular reviewers. An effective mechanism for assigning emergency reviewers is an economical way of utilizing the limited pool of qualified reviewers \citep{peng2018cvpr, stelmakh2021novice} by adaptively assigning them to papers based on the quality of the initial round of reviews.

Determining review quality is an inherently noisy process. Incorporating authors' elicited rankings into this determination when assigning emergency reviewers could both improve its accuracy and enhance the community's trust in the credibility of the peer review processes. It is crucial to convey to authors that their concerns are taken seriously, especially when they disagree with the reviews. This can be achieved by leveraging isotonic scores, based on the premise that discrepancies between raw review scores and isotonic scores, which we refer to as isotonic residuals,\footnote{In contrast to the setup in Section \ref{sec:main},
here we run the Isotonic Mechanism on the average of \textit{all} review scores for each submission.} might signal concerns about review quality from the authors' viewpoint.

\begin{figure}[!htbp]
    \centering
    \begin{subfigure}[b]{0.46\textwidth}
        \small
        \stackunder[0pt]{ 
        \includegraphics[height=0.6\textwidth]{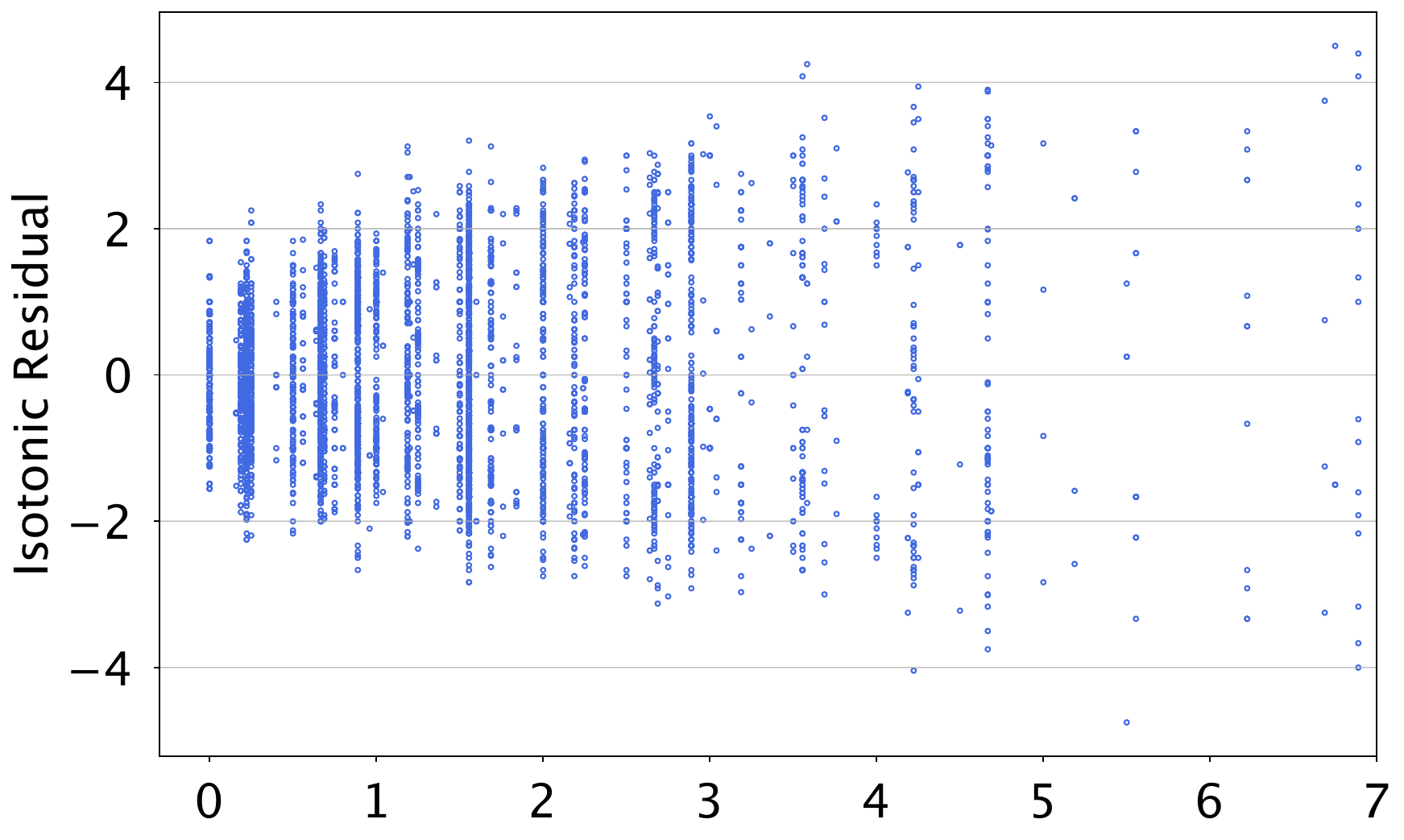}
        }{Variance}
    \end{subfigure}
    \begin{subfigure}[b]{0.46\textwidth}
        \small
        \stackunder[0pt]{ 
        \includegraphics[height=0.6\textwidth]{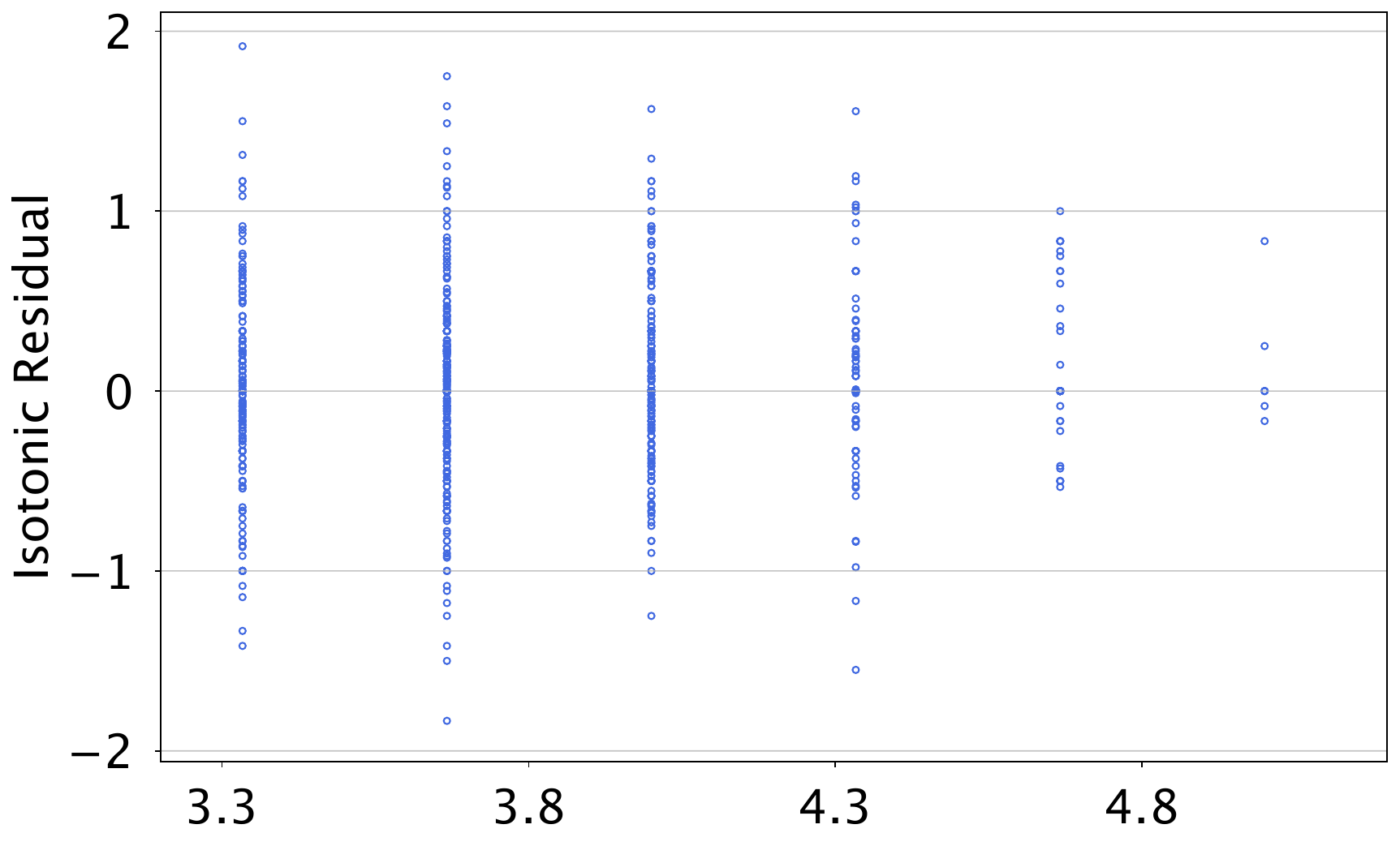}
        }{Confidence}
    \end{subfigure}
    \caption{Scatter plots showing isotonic residuals using the simple-averaging strategy, plotted against review confidence levels and against raw score variance, for submissions with a confidence level of 3 or above.}
    \label{fig:variance_residual}
\end{figure}

To provide empirical support, we examine the relationship between isotonic residuals and both the variance of review scores and review confidence levels, using data from ICML 2023. High score variance or low confidence levels are often used as indicators for recruiting emergency reviewers \citep{shah2018design}. Figure \ref{fig:variance_residual} 
illustrates that isotonic residuals in absolute value have a strong negative correlation with confidence levels and a positive correlation with score variance. Furthermore, little dependence is found between score variance and confidence levels, with a correlation of only $2.05 \times 10^{-2}$. This suggests that isotonic residuals might offer a more comprehensive measure of review quality.

We also evaluate the effectiveness of isotonic residuals using data from a second survey, which asked authors to identify submissions where the review outcomes differed most from their expectations. Table \ref{Table:unexpected} 
shows that isotonic residuals are the most predictive of submissions with the most ``unexpected'' review outcomes. This finding is expected as isotonic scores, by definition, reflect authors' expectations.

\begin{table}[!htp] 
    \centering
    \begin{tabular}{c|c|c|c}
    \hline\hline
         & \begin{tabular}{@{}c@{}} Isotonic Residual \end{tabular} & \begin{tabular}{@{}c@{}} Score Variance \end{tabular} & \begin{tabular}{@{}c@{}} Score Confidence \end{tabular} \\
         \hline
         \begin{tabular}{@{}c@{}} Prediction Accuracy \end{tabular} & $254/322 = 78.9\%$ & $162/322 = 50.3\%$ & $136/322 = 42.2\%$ \\
        \hline
    \end{tabular}
    \caption{Prediction accuracy of the most ``unexpected'' review outcomes, determined using the largest mean isotonic residual in absolute value, the greatest variance of review scores, and the lowest average confidence levels}
    \label{Table:unexpected}
\end{table}

Given this empirical evidence, we propose using large isotonic residuals as an indicator of the need for emergency reviewers to provide additional expert opinions. In implementing this mechanism, it is crucial to ensure that submissions receive roughly the same number of reviewers on average, regardless of whether a submission is included in this mechanism or not.\footnote{A submission might not be included because no author has multiple submissions. However, this applies to a minority of submissions in large ML/AI conferences. For example, 77.0\% of the ICML 2023 submissions have at least one author with more than one submission.} One approach to achieving this balance is to assign three initial reviewers to papers participating in the mechanism, while four reviewers to those not included. For papers in the participating group, we assign two emergency reviewers for isotonic residual magnitudes in the top 30\%, and one for magnitudes between the 30th and 70th percentiles.\footnote{The numbers 30\% and 70\% are arbitrary as long as they ensure that the expected number of emergency reviewers for a paper is one.} Consequently, on average, a paper has four reviewers, regardless of its group. To ensure a cautious approach, the quantile of the isotonic residual in absolute value will be made available to ACs and above, but not the raw isotonic score itself. Without knowing whether the isotonic residual is positive or negative, ACs cannot ascertain whether authors hold a high or low opinion of their submissions, thereby minimizing the likelihood of bias influencing the ACs' decisions.

It is important to note that the isotonic score differs from the raw review score only when at least one author has multiple submissions to the conference. Consequently, the Isotonic Mechanism cannot uniformly increase review accuracy across all submissions, which is also the rationale for suggesting the use of rankings rather than isotonic scores in the selection of paper awards, as discussed in Section \ref{sec:rec}.
As demonstrated, isotonic scores are better estimates of the ground truth ``expected review scores'' than the raw review scores, and the accuracy improves with the number of papers an author has submitted and ranked \citep{su2021you}. In particular, authors with fewer submissions are more likely to have higher variance in their isotonic scores. Thus, caution is warranted when comparing isotonic residuals of submissions across authors. Nevertheless, despite the fact that the Isotonic Mechanism does not improve accuracy uniformly for all submissions, as it is accuracy improving, it seems prudent to (cautiously) use the information to improve the review process---using the isotonic residuals to recruit emergency reviewers is one such example of cautious use.

\section{Discussion}
\label{sec:discuss}

This paper has analyzed the ICML 2023 ranking data collected from \num{1342} submitting authors who participated in our experiment to empirically evaluate the Isotonic Mechanism. Our findings suggest that this mechanism can effectively mitigate noise in review scores. Additionally, we have proposed three cautious applications of the Isotonic Mechanism and author-provided rankings to improve peer review processes.

Although the Isotonic Mechanism has been shown to be dominant-strategy truthful under certain conditions \citep{su2022truthful}, authors may still engage in strategic behaviors in real-world deployments. In interpreting these results, it is crucial to recognize that the rankings were provided under the assumption that they would not influence decision-making processes. Nevertheless, 59.4\% of authors in our survey indicated that they would submit the same rankings, even if they were to be used for decision-making. Examples of strategic behaviors not covered in the game-theoretic analysis include: authors showing a preference for papers where they are the first author, or professors ranking a student's paper higher than warranted to enhance the student's job market prospects, or authors seeking to get a weak paper published early while deferring stronger papers to the next conference series. The potential for strategic behaviors presents an important consideration for the deployment of the Isotonic Mechanism in consequential decision-making, and warrants further investigation. The optimal way to examine such behaviors would be through a randomized experiment at an ML/AI conference, where some authors' rankings are used in decision-making, while others are not. Another concern is that the variance reduction effect of the Isotonic Mechanism increases with the number of papers an author submits (and has no effect for authors who submit only a single paper). This complicates comparisons of scores across authors with differing submission counts and is another aspect that requires further study.

The analysis could be improved by addressing the potential for non-response bias, which stems from the 30.4\% response rate in our experiment. For example, the mean of the average review scores of all submissions to ICML 2023 is $4.53$ before the rebuttal period, whereas it is $4.66$ for the 2,592 ranked submissions. The response rate is largely influenced by the number of submissions per author, with more prolific authors being less likely to provide rankings. However, it is these authors for whom the Isotonic Mechanism could be most effective. Consequently, this bias may lead to a conservative estimate of the mechanism's effectiveness. In future experiments, incentivizing authors to participate could increase the response rate. One possible incentive would be to offer earlier notification of decisions to authors who provide rankings.

\begin{table}[!htp]
\centering
    \begin{tabular}{c|c|c|c|c|c|c}
    \hline\hline
         Percentile & 5\% & 10\% & 15\% & 20\% & 25\% & 30\% \\
         \hline
         \begin{tabular}{@{}c@{}} Overlapping Fraction \end{tabular} & $70.63 \%$ & $69.17 \%$ & $73.09 \%$ & $78.06 \%$ & $77.53 \%$ & $83.27 \%$ \\
         \hline
    \end{tabular}
    \caption{Fractions of overlapping top-scored submissions between isotonic scores (using greedy strategy) and raw scores, for different percentiles among the \num{2530} ranked submissions.}
    \label{tab:overlap}
\end{table}

Another potentially useful approach to improving the quality of reviews is to aggregate rankings provided by reviewers. Consider a conference where each reviewer is asked to rank the submissions they have reviewed. Consider each submission as a node and draw an edge between two nodes if they share a common reviewer. This creates a submission-reviewer network, which divides the submissions into several connected graphs. For each connected graph, the Plackett--Luce model can be employed to estimate the preference score of each paper using the spectral method described in \cite{fan2024spectral}. This approach yields ranking-based scores, which refine raw review scores through the aggregation of preferences from all reviewers. These scores can also be integrated with those generated by the Isotonic Mechanism. Notably, preference scores can be consistently estimated even when reviewers only identify their top choices or provide partial rankings \citep{fan2024ranking}.

To conclude, we outline several directions for future work. From a practical perspective, our team has been working for years with the goal of applying the Isotonic Mechanism into decision-making processes at some ML/AI conferences. We have already collaborated with the OpenReview team to incorporate ranking collection into the platform and conducted experiments at ICML for three consecutive years from 2023 to 2025. Specifically, 5,665 authors participated in our ICML 2024 experiment, with 2,184 submissions ranked by their authors; our ICML 2025 experiment excluded authors with only one submission, and we collected 3,749 rankings covering 3,171 submissions. Nevertheless, it remains valuable to test this mechanism across other large-scale conferences, such as NeurIPS, ICLR, KDD, and AAAI, which would help us adapt the mechanism to conference-specific contexts and inform future refinements.


From a methodological perspective, the Isotonic Mechanism in its current form does not account for reviewers' confidence levels. Developing weighted review scores that are incorporated into the mechanism could potentially improve its performance. Additionally, exploring how the mechanism could leverage rankings provided by reviewers \citep{fan2024spectral} represents another valuable research avenue. The practical challenge of coauthors holding differing opinions on their submissions, as evidenced by the 30.7\% of coauthor pairs providing inconsistent comparisons (which drops to 26.1\% for submissions with substantial review score differences), necessitates the development of a variant of the mechanism that accommodates these ranking inconsistencies \citep{wu2023isotonic}. A related research direction is the incorporation of uncertainty into the rankings within the mechanism. Another potential improvement to the Isotonic Mechanism is the removal of potentially low-quality rankings, with preliminary evidence suggesting that this could enhance accuracy, as shown in 
Table S.4
in the Supplementary Material. In practice, whether a paper is accepted largely depends on whether its score exceeds a threshold, such as the top 20\% of all submissions. Table~\ref{tab:overlap} illustrates how the top-scored papers would change when using raw scores versus isotonic scores (see the analysis of overlap with submissions receiving ``Accepted as Oral'' in 
Section C
of the Supplementary Material). Notably, the overlapping fraction increases as the cutoff percentage increases. A relevant question for future study is to compare these two lists of top-scored papers in terms of the citations they have accumulated over time.


{\small
\subsection*{Acknowledgments}
We would like to thank Melisa Bok, Emma Brunskill, Barbara Engelhardt, Xiao-Li Meng, Nihar Shah, Sherry Xue, and James Zou for helpful discussions in early stages of this project. We also thank the editor, associate editor, and reviewers for constructive comments that helped us significantly improve the presentation of the paper. We are grateful to the ICML Board for providing the opportunity for this experiment. This research was supported in part by NSF grant CCF-1934876 and Wharton AI for Business.

\bibliographystyle{abbrvnat}
\bibliography{ref}
}

\clearpage
\appendix
\section*{Supplementary Material}



\section{Additional Details of the Survey Experiment}

\begin{figure}[!htp]
  \centering
  \includegraphics[width=0.8\linewidth]{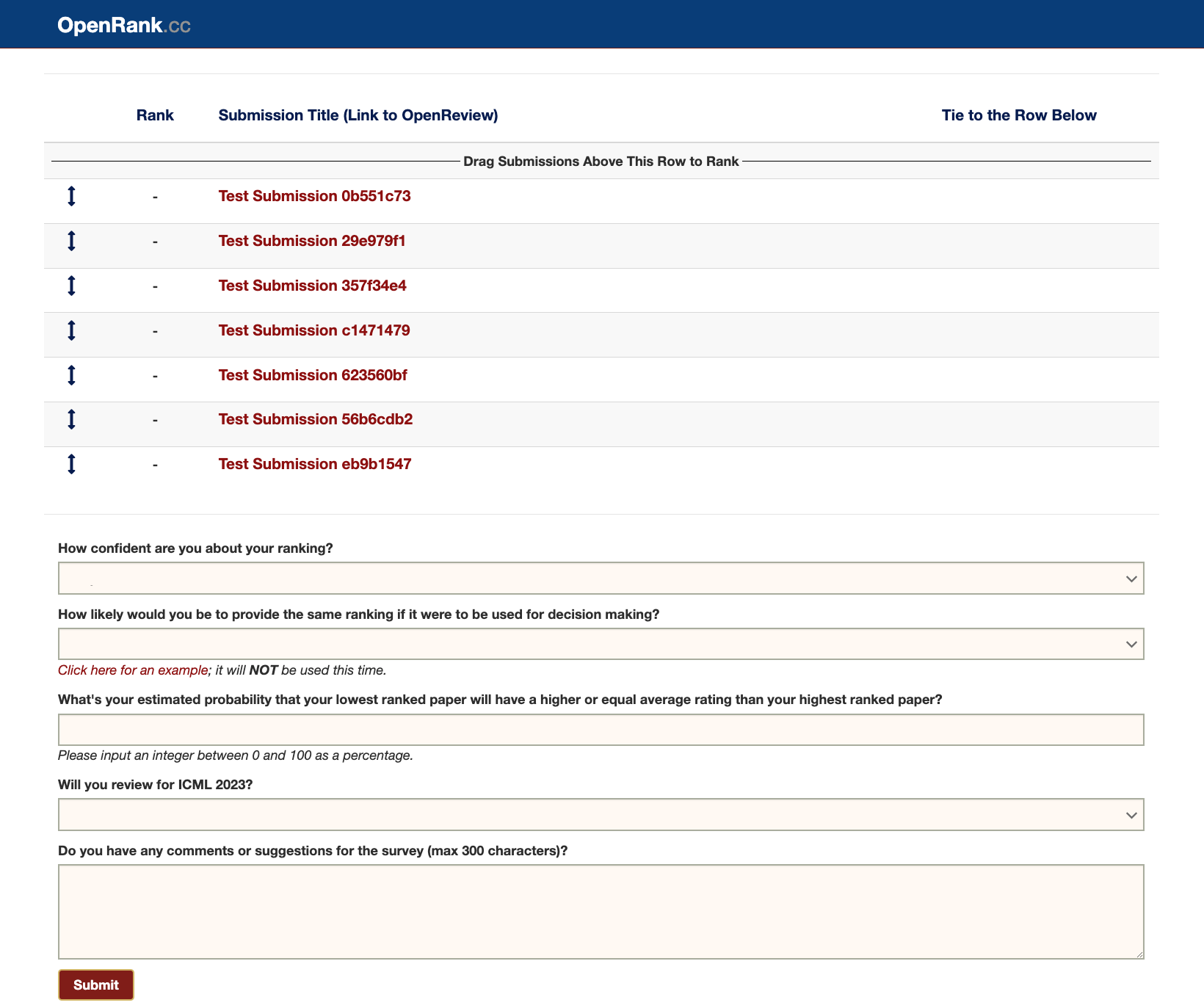}
  \caption{
  Screenshot of the first author survey.
  This survey was sent out on January 26, 2023, and
  closed on February 10, 2023. Among the \num{1342} authors with multiple submissions who provided valid rankings, they took an average of around \num{4} days to submit the survey: $25\%$ of them submitted the survey within $6.21$ hours and $90\%$ of them submitted the survey within $8.79$ days.
  } \label{fig:survey_procedure:1}
\end{figure}

\begin{figure}[!htp]
  \centering
  \includegraphics[width=0.9\linewidth]{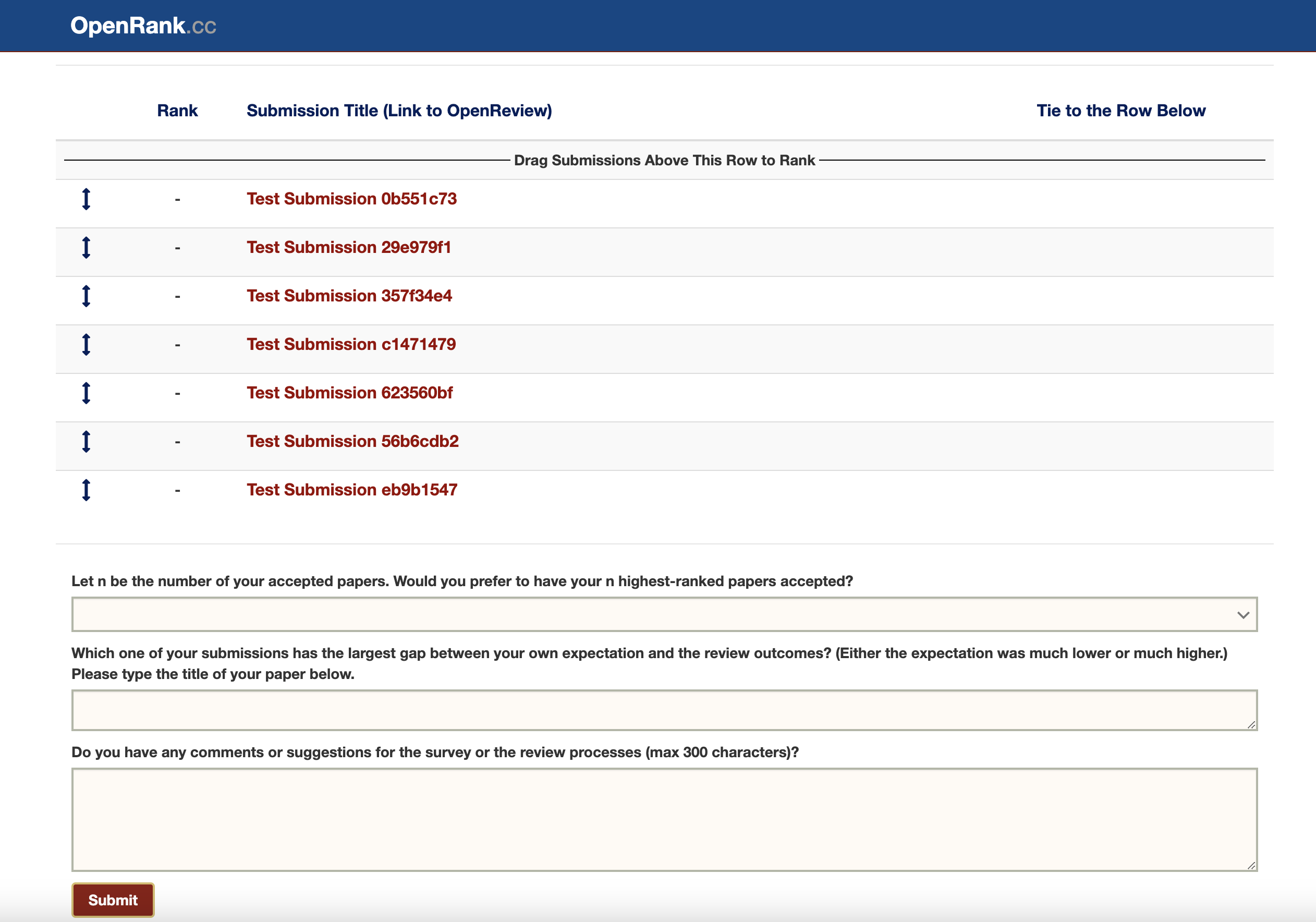}
  \caption{
Screenshot of the second author survey. This survey was sent out on April 24, 2023, and closed on May 15, 2023. There were 1,328 authors with multiple submissions at the time of both surveys (not counting those who withdrew submissions and became single-submission authors at the second survey); 103 ($7.76\%$) authors among them changed the ranking of any submissions compared with the first survey. The high participation rate may be explained by the fact that the second survey automatically presented the same rankings the authors provided in the first survey. The same rankings would be submitted if the authors clicked the ``Submit'' button without any other actions.
  }
  \label{fig:survey_procedure}
\end{figure}

\begin{figure*}[!htp]
    \centering
    \includegraphics[width=0.8\textwidth]{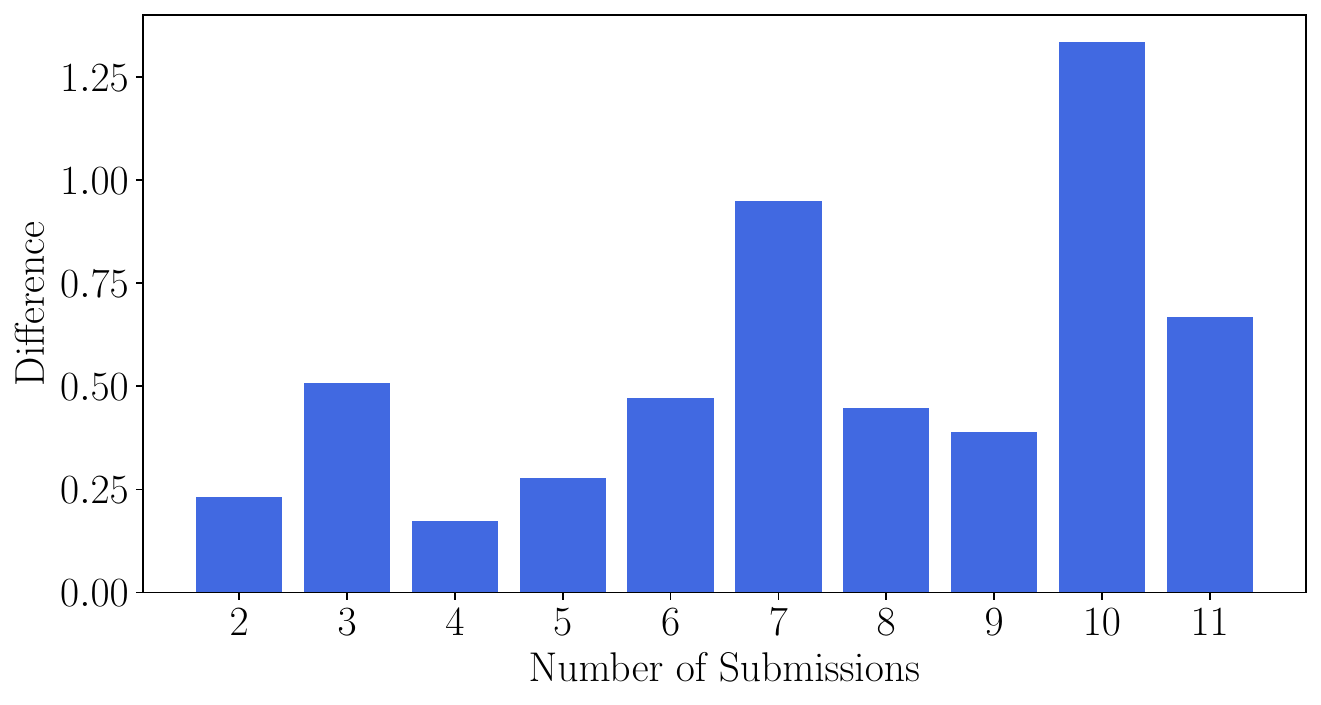} 
    \caption{The score difference between the highest-ranked and lowest-ranked paper scores, averaged over ICML 2023 authors who submitted rankings of the same length.}
    \label{appfig:gap_no_submission}
\end{figure*}

\begin{figure}[!htp]
  \centering
  \includegraphics[width=0.8\textwidth]{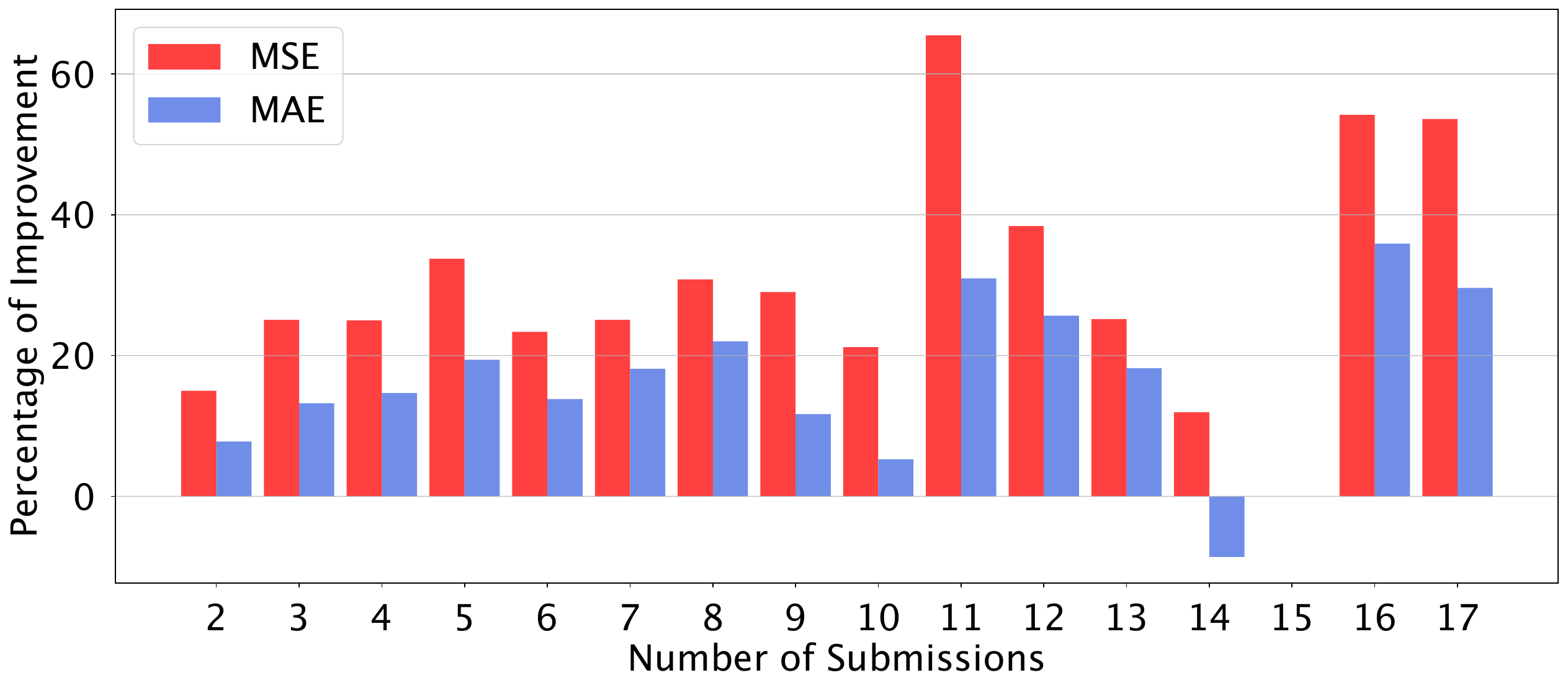}
  \caption{Percentage improvement in proxy MSE and MAE, in the same setting as 
  Figure 1(c) 
  and the middle column of 
  Figure 5. For MSE, the average percentage improvement for numbers of submissions between 2 and 10 is 25\% and it is 41\% for submissions more than 11. For MAE, the average percentage improvement for numbers of submissions between 2 and 10 is 14\% and it is 22\% for submissions more than 11.
  }
  \label{fig:improvement}
\end{figure}

\begin{figure}[!htp]
  \centering
  \small
  \resizebox{0.9\textwidth}{!}{\includegraphics[width=\textwidth]{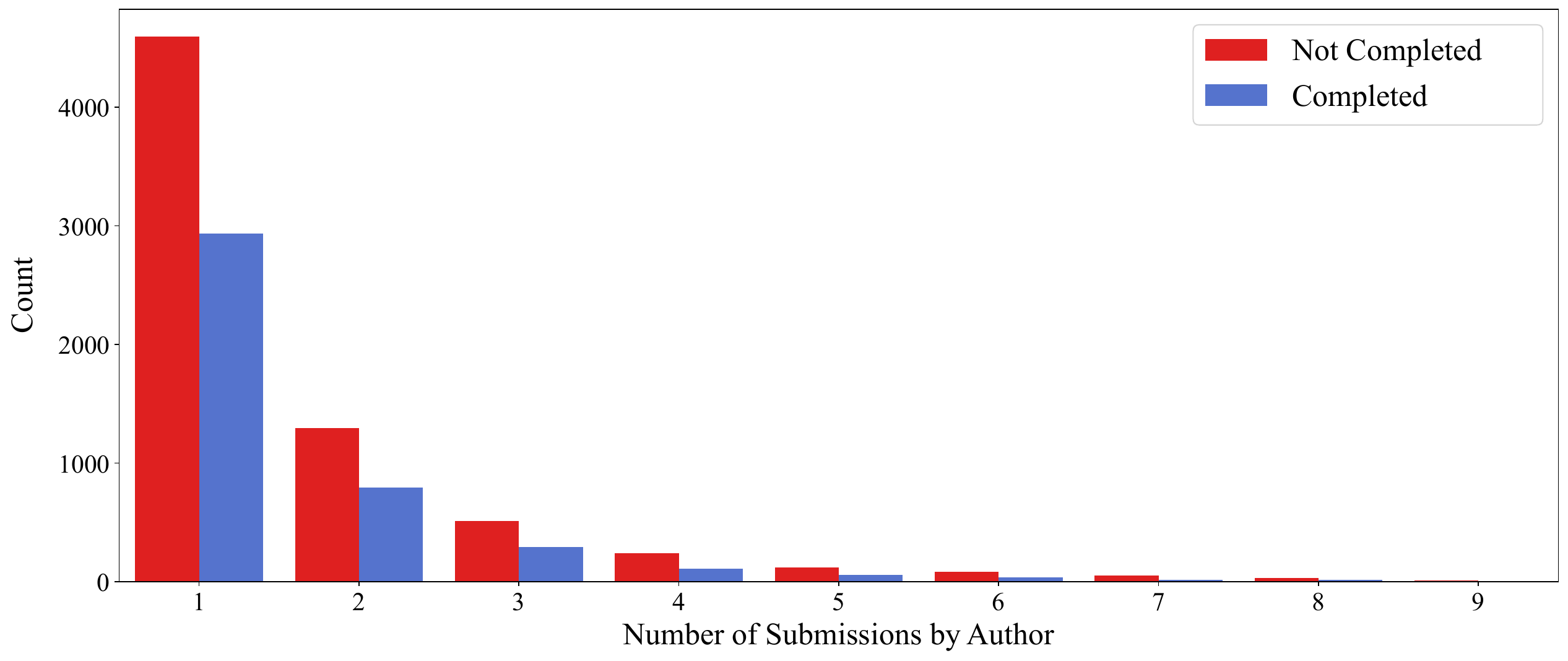}}
  \caption{The histogram of author participation in the first survey (complete the survey or not) grouped by the number of submissions the author has, showing only the authors with less than 10 submissions (where each group contains at least 10 authors). Among these nine groups of authors, the authors with three submissions have the highest response rate (35.5\%) and the authors with seven submissions have the lowest response rate (23.9\%).
  }
  \label{fig:attempt_completed}
\end{figure}

\begin{figure}[!htp]
  \centering
  \includegraphics[ width=0.4\textwidth]{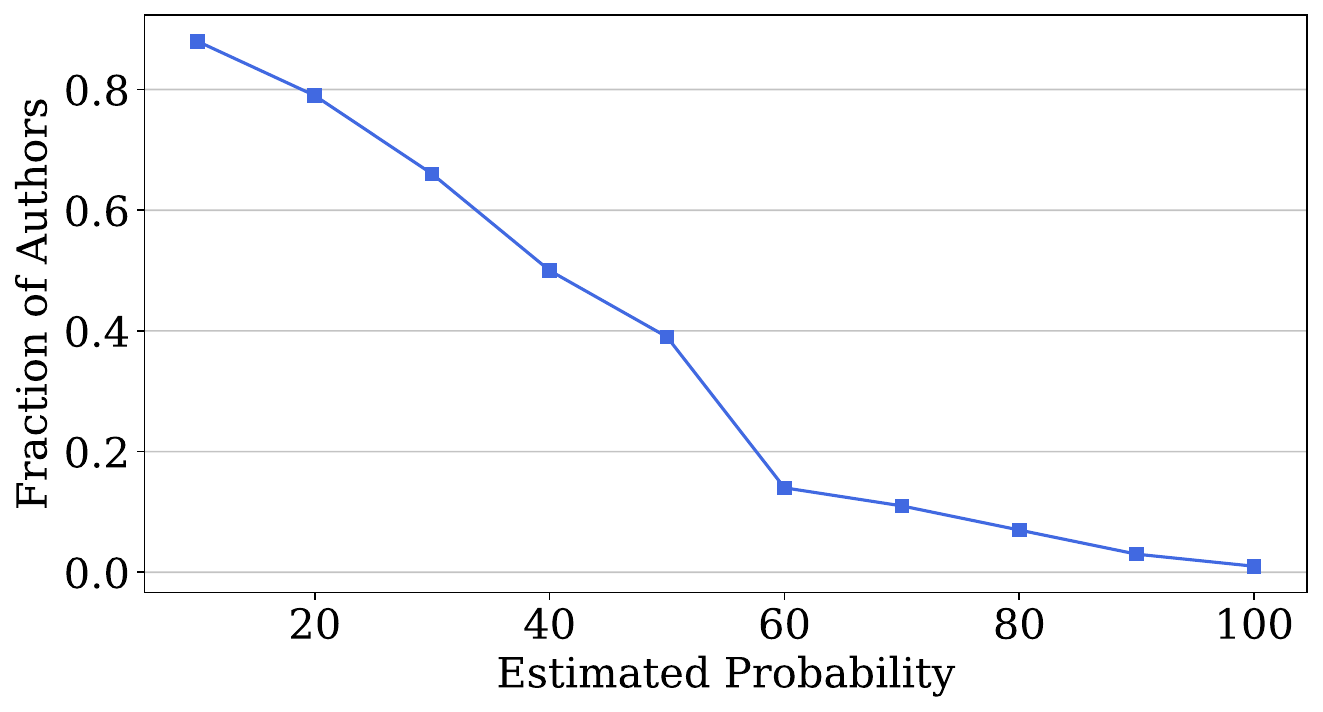}
  \caption{For a given threshold (estimated probability) on the $x$-axis, the $y$-axis represents the proportion of the \num{1342} authors who estimated that the probability of their lowest-ranked paper receiving an average score higher than or equal to that of the highest-ranked paper is at least the threshold.}
  \label{fig:est_prob}
\end{figure}

\begin{itemize}
\item
Figures \ref{fig:survey_procedure:1} and \ref{fig:survey_procedure} present screenshots of the first and second surveys, respectively.

\item 
Figure \ref{appfig:gap_no_submission} shows how the score difference between an author's highest-ranked and lowest-ranked papers depends on the number of submissions the author has.

\item 
Figure~\ref{fig:improvement} illustrates the percentage improvement achieved by the Isotonic Mechanism using the greedy strategy.

\item
Figure \ref{fig:attempt_completed} shows how the response rate depends on the number of submissions an author has.

\item 
Figure \ref{fig:est_prob} demonstrates the estimated probability that an author’s lowest-ranked paper receives
an average score greater than or equal to their highest-ranked paper.

\end{itemize}

\section{Rebuttal}\label{sec:rebuttal}

While the highest-ranked papers are more likely to have their reviewer scores increased during the rebuttal period, as demonstrated in 
Section 2,
Table \ref{table:rebuttal_attitude} 
reveals that the highest-ranked submissions do not exhibit a significantly higher word count or a greater number of replies in the rebuttal period compared to the lowest-ranked submissions.

\begin{table}[!htp]
\centering
\renewcommand{\arraystretch}{1.2}
\begin{tabular}{l||c|c|c|c}
\hline\hline
\multicolumn{5}{c}{\begin{tabular}{@{}c@{}} Word count and number of replies during rebuttal \end{tabular}}
\\ 

\hline
\multirow{2}{*}{} & \multicolumn{2}{c|}{Accepted submissions} & \multicolumn{2}{c}{Rejected submissions} \\ 

\cline{2-5}
& \begin{tabular}{@{}c@{}} Word count \end{tabular} & \begin{tabular}{@{}c@{}} Number of replies \end{tabular} & \begin{tabular}{@{}c@{}} Word count \end{tabular} & \begin{tabular}{@{}c@{}} Number of replies \end{tabular}  \\ 

\hline \hline
\begin{tabular}{@{}c@{}} Highest-ranked \end{tabular} & $256.08$ & $1.79$ & $314.35$ & $1.95$  \\ 

\hline
\begin{tabular}{@{}c@{}}  Lowest-ranked \end{tabular} & $229.21$ & $1.63$ & $279.51$ & $1.81$ \\ 

\hline
\begin{tabular}{@{}c@{}}  $p$-value \end{tabular} & $1.04 \times 10^{-1}$ & $5.13 \times 10^{-2}$ & $5.97 \times 10^{-2}$ & $6.08 \times 10^{-2}$ \\ 

\hline
\end{tabular}
\caption{We provide average word count and number of replies in the rebuttal period, grouped by highest-ranked submissions, lowest-ranked submissions, and the final decisions.}
\label{table:rebuttal_attitude}
\end{table}

\begin{figure}[!htbp]
    \centering
    \resizebox{0.313\textwidth}{!}{\begin{minipage}[b]{0.40\textwidth}

        \stackunder[0pt]{ \includegraphics[width=\textwidth]{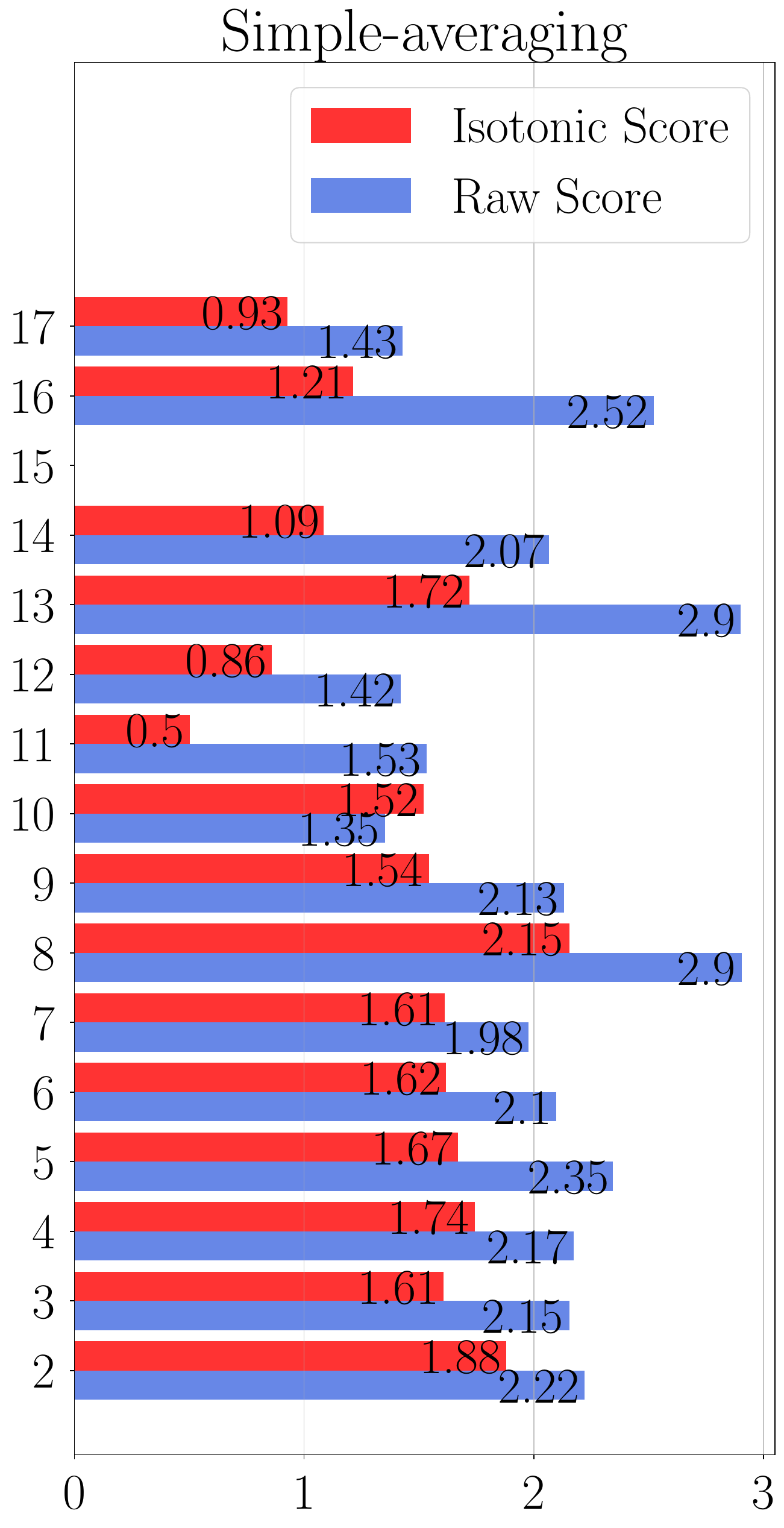}
        }{Mean Squared Error}
    \end{minipage}}
    \resizebox{0.313\textwidth}{!}{\begin{minipage}[b]{0.40\textwidth}

        \stackunder[0pt]{ \includegraphics[width=\textwidth]{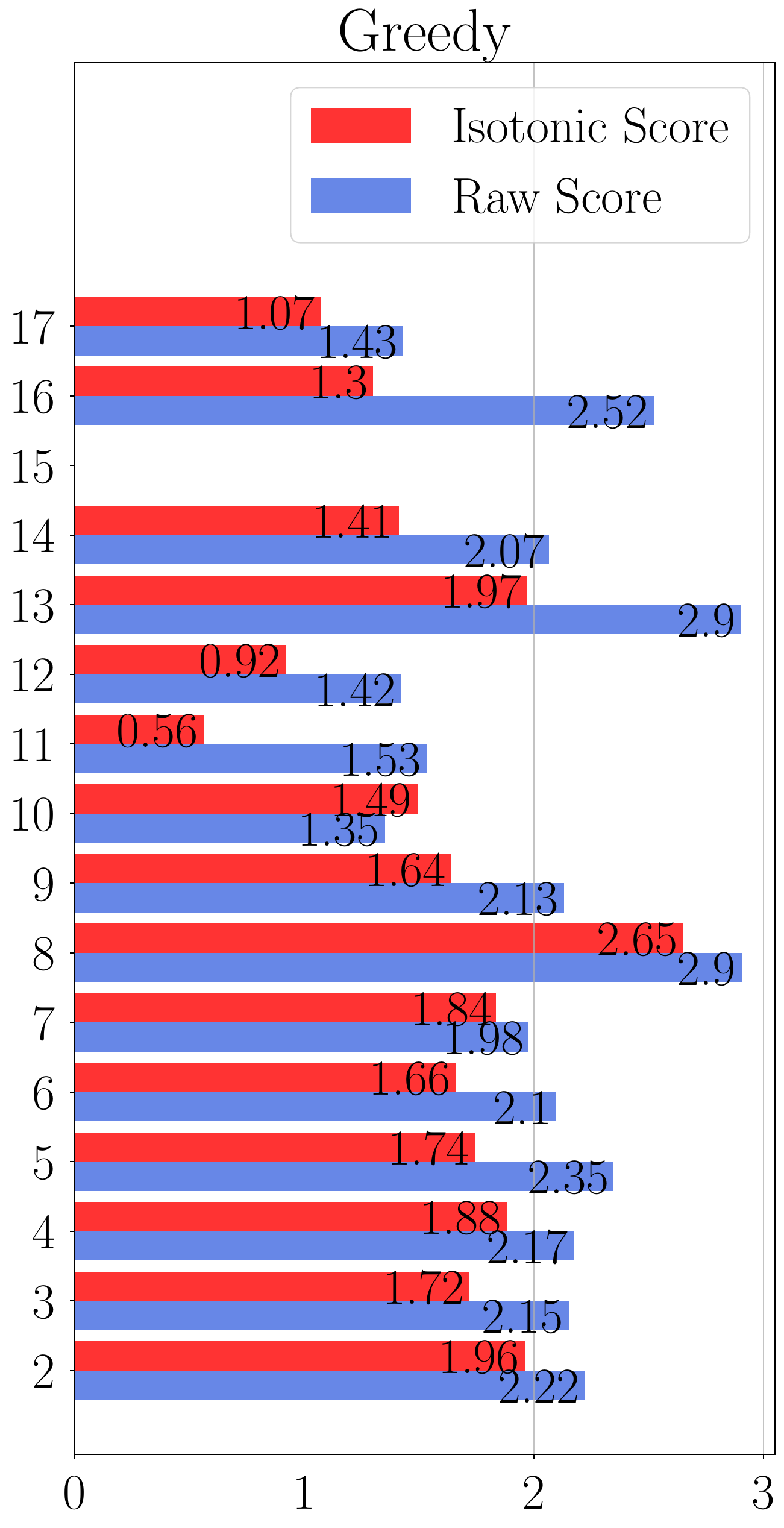}
        }{Mean Squared Error}
    \end{minipage}}
    \resizebox{0.313\textwidth}{!}{\begin{minipage}[b]{0.40\textwidth}

        \stackunder[0pt]{ \includegraphics[width=\textwidth]{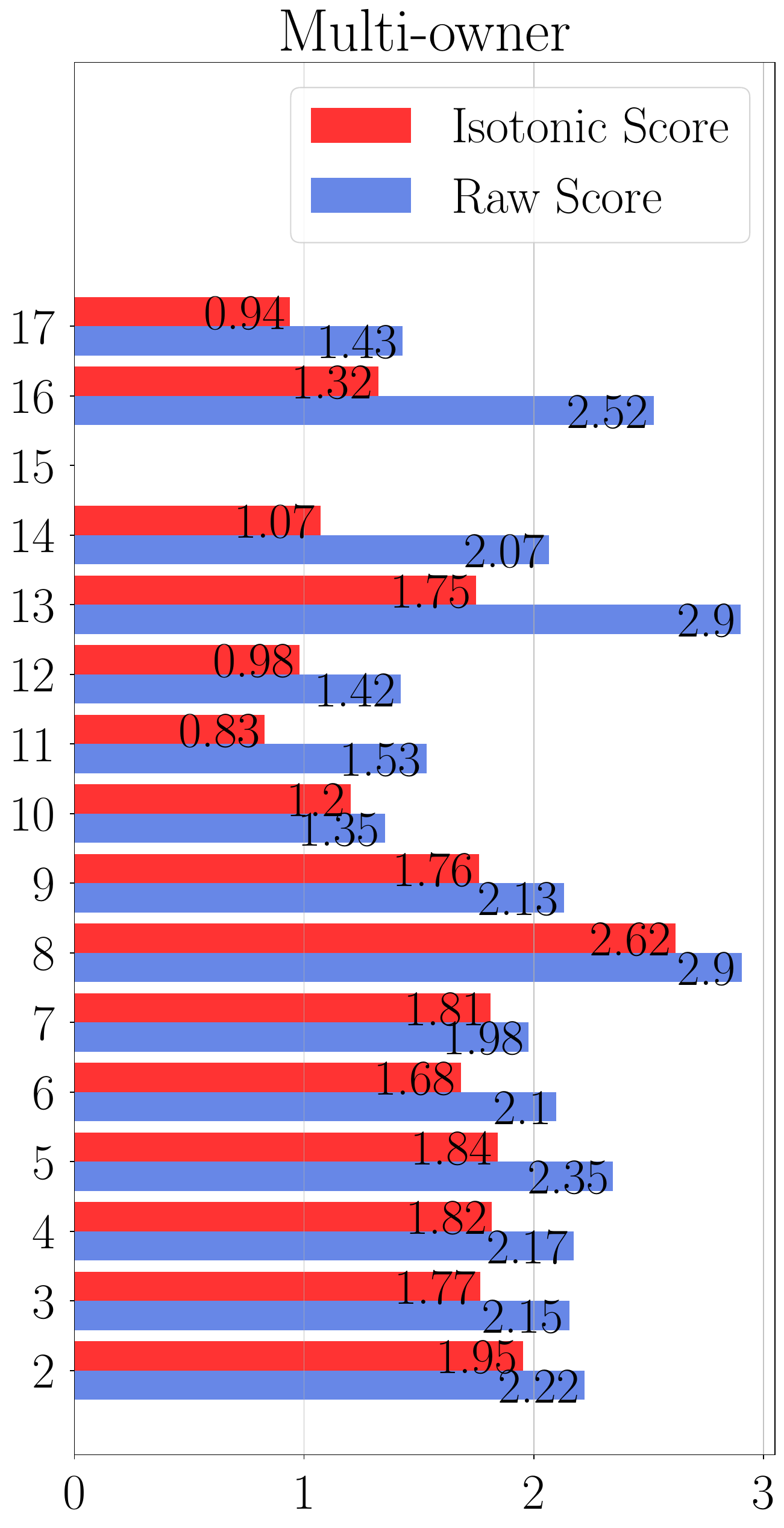}
        }{Mean Squared Error}
    \end{minipage}}
    \resizebox{0.313\textwidth}{!}{\begin{minipage}[b]{0.40\textwidth}

        \stackunder[0pt]{\includegraphics[width=\textwidth]{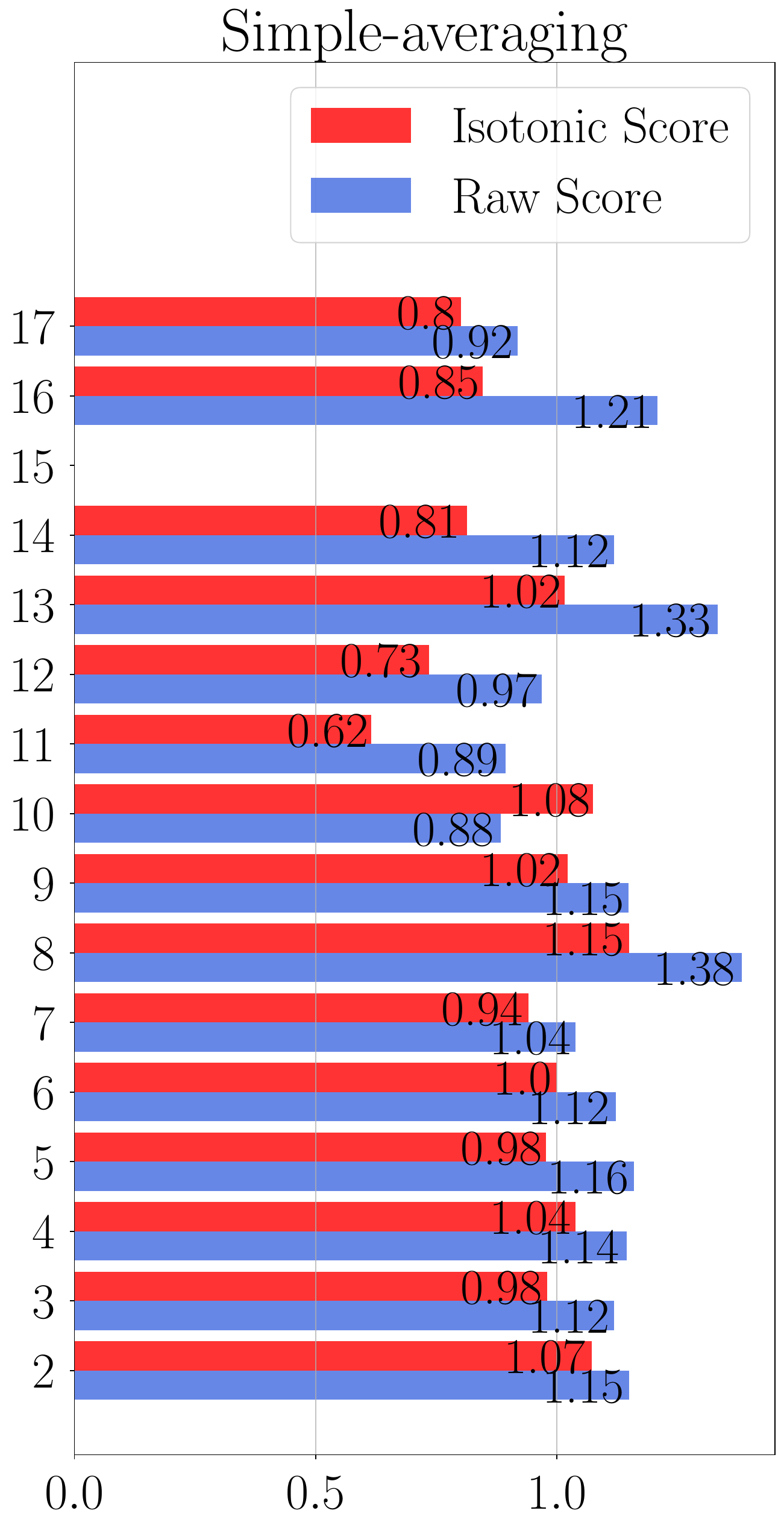}
        }{Mean Absolute Error}
    \end{minipage}}
    \resizebox{0.313\textwidth}{!}{\begin{minipage}[b]{0.40\textwidth}

        \stackunder[0pt]{ \includegraphics[width=\textwidth]{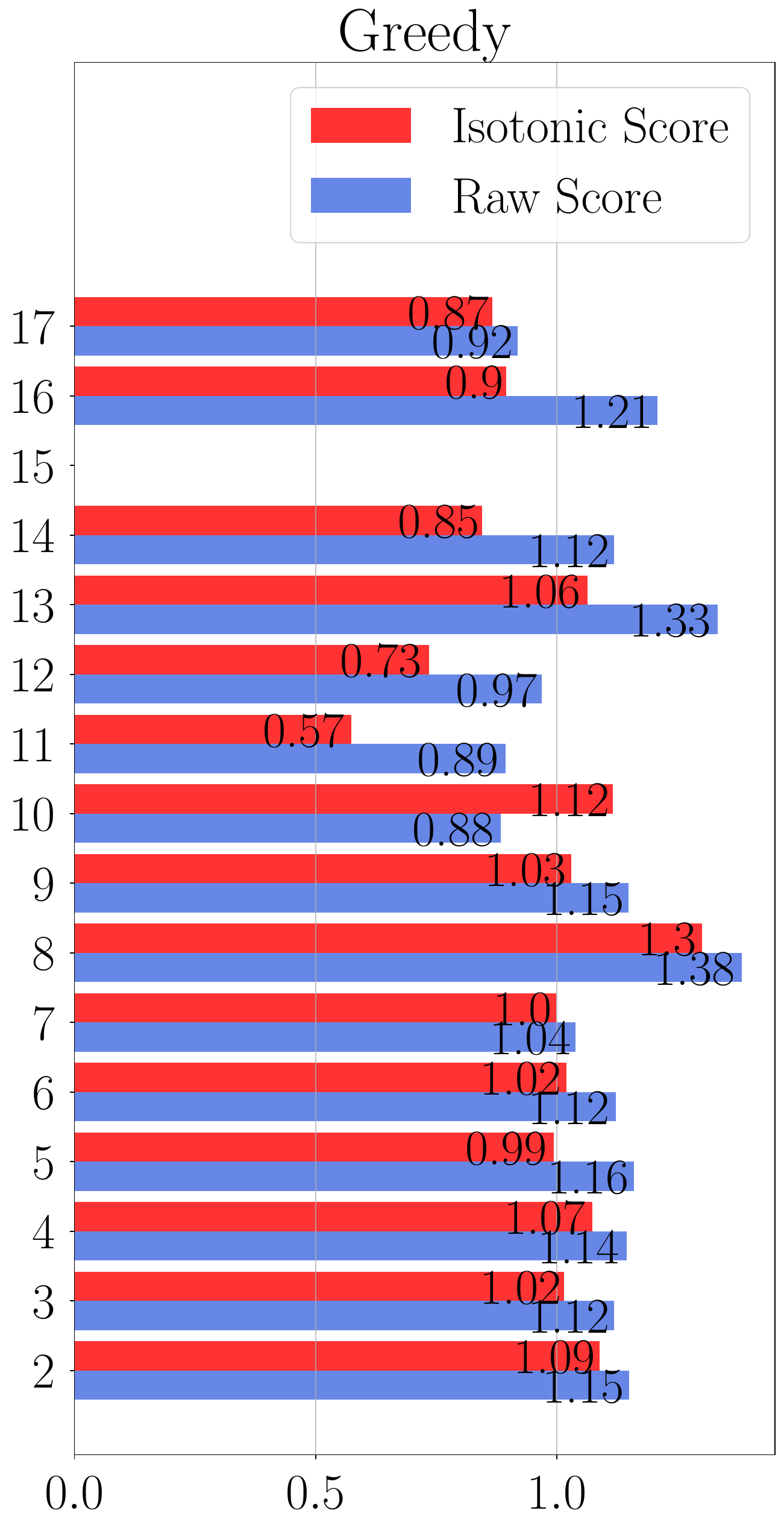}
        }{Mean Absolute Error}
    \end{minipage}}
    \resizebox{0.313\textwidth}{!}{\begin{minipage}[b]{0.40\textwidth}
        \stackunder[0pt]{ \includegraphics[width=\textwidth]{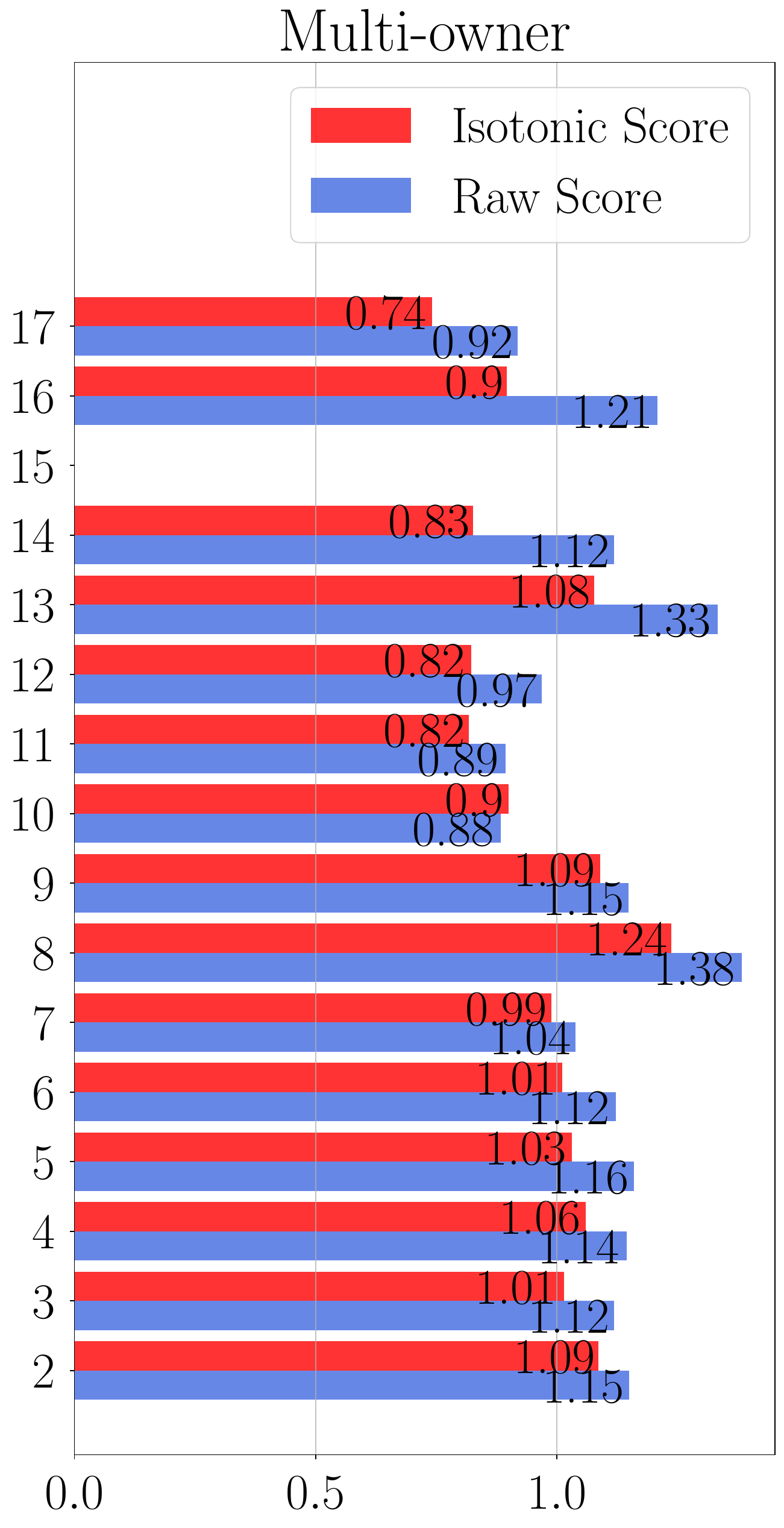}
        }{Mean Absolute Error}
    \end{minipage}}
    \caption{Comparison between isotonic and raw scores in terms of proxy MSE and MAE (post-rebuttal) averaged over ICML 2023 authors who submitted the same-length rankings.}
    \label{fig:no_submission_proxy_post_rebuttal}
\end{figure}

Table \ref{tab:greedy_multi_isotonic_mse_post_rebuttal} shows the efficacy of the multiple strategies for the Isotonic Mechanism in reducing both proxy MSE and MAE based on post-rebuttal review scores. 
In particular, the greedy strategy for the Isotonic Mechanism leads to a reduction in the proxy MSE by $16.2\%$ and the proxy MAE by $8.1\%$. 
Similar to 
Figure 5, Figure \ref{fig:no_submission_proxy_post_rebuttal} demonstrates that the Isotonic Mechanism achieves a more substantial decrease in both proxy MSE and MAE as the number of submissions by an author grows.

\begin{table}[H]
\centering
\renewcommand{\arraystretch}{1.05}
\resizebox{\textwidth}{!}{
\begin{tabular}{l||c|c|c|c|c|c}

\hline \hline
\multirow{2}{*}{} & \multicolumn{3}{c|}{Proxy MSE } & \multicolumn{3}{c}{Proxy MAE } \\

\cline{2-7}
& \begin{tabular}{@{}c@{}} Error \end{tabular} & \begin{tabular}{@{}c@{}} Improvement \end{tabular} & 
\begin{tabular}{@{}c@{}} $p$-value \end{tabular} & \begin{tabular}{@{}c@{}} Error \end{tabular} & \begin{tabular}{@{}c@{}} Improvement \end{tabular} & 
\begin{tabular}{@{}c@{}} $p$-value \end{tabular} 
\\ 

\hline
\begin{tabular}{@{}c@{}} Raw Score \end{tabular} & $2.20$ & \texttt{NA} & \texttt{NA} & $1.09$ & \texttt{NA} & \texttt{NA}\\ 

\hline
\begin{tabular}{@{}c@{}} Simple-averaging Strategy \end{tabular} & $1.78$ & $19.10 \%$ & $4.65 \times 10^{-28}$ & $1.03$ & $9.46 \%$ & $7.14 \times 10^{-21}$\\ 

\hline
\begin{tabular}{@{}c@{}} Greedy Strategy \end{tabular} & $1.84$ & $16.16 \%$ & $4.64 \times 10^{-19}$ & $1.05$ & $8.08 \%$ & $2.21 \times 10^{-14}$\\ 

\hline
\begin{tabular}{@{}c@{}} Multi-owner Strategy \end{tabular} & $1.86$ & $15.39 \%$ & $1.45 \times 10^{-20}$ & $1.05$ & $7.74 \%$ & $9.10 \times 10^{-16}$\\ 

\hline
\end{tabular}}
\caption{Reduction of proxy MSE and MAE (post-rebuttal) using the Isotonic Mechanism with various strategies. A paired one-sided $t$-test shows that the reduction in proxy errors is statistically highly significant.}
\label{tab:greedy_multi_isotonic_mse_post_rebuttal}
\end{table}

\section{Additional Details for the Main Text}
\label{app:de_sec}

\begin{figure}[!htbp]
    \centering
    \begin{subfigure}[b]{0.48\textwidth}
        \footnotesize
        
        \includegraphics[ height = 0.63\textwidth]{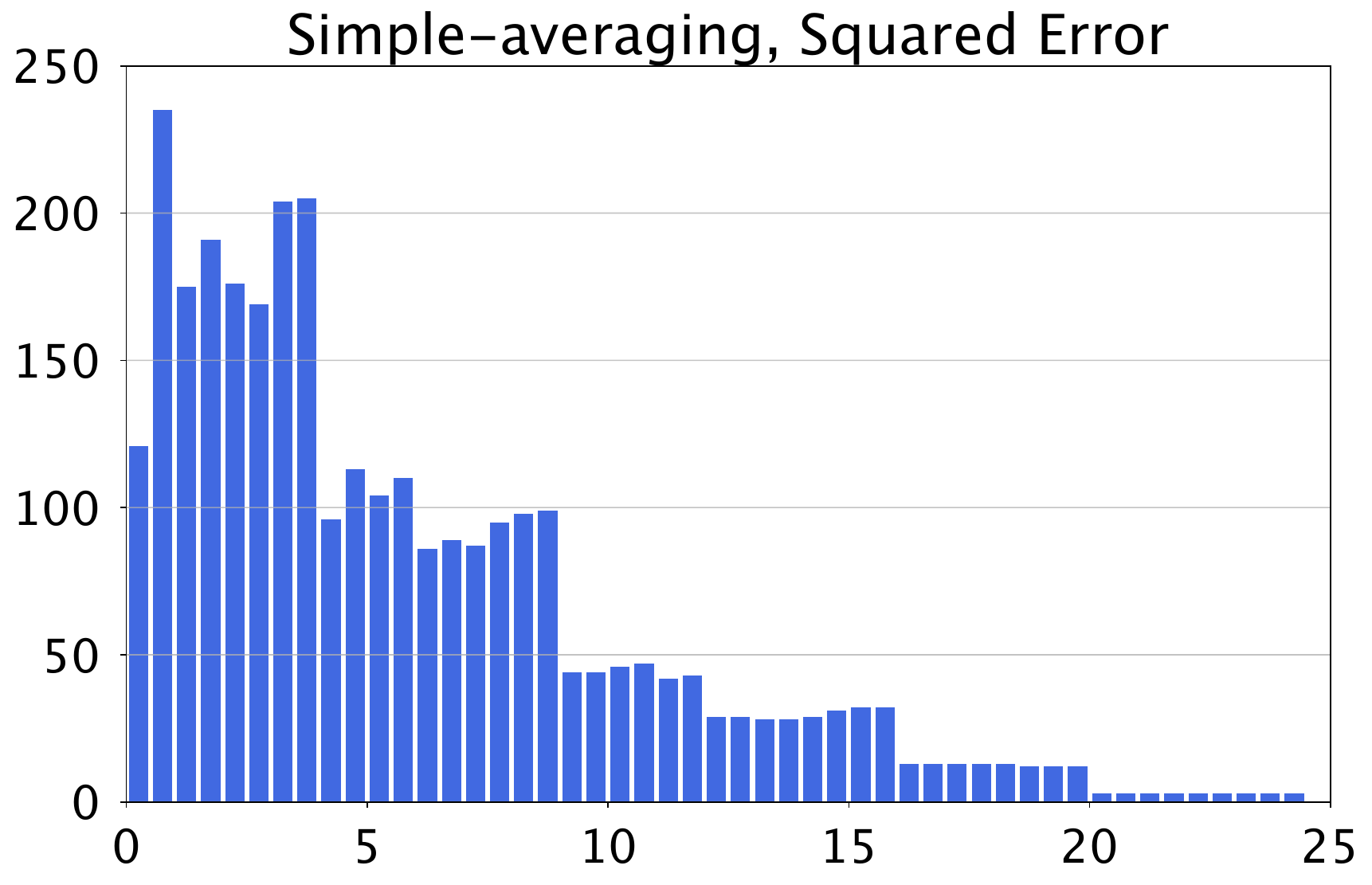}
            \end{subfigure}
    \begin{subfigure}[b]{0.48\textwidth}
        \footnotesize
        
        \includegraphics[ height = 0.63\textwidth]{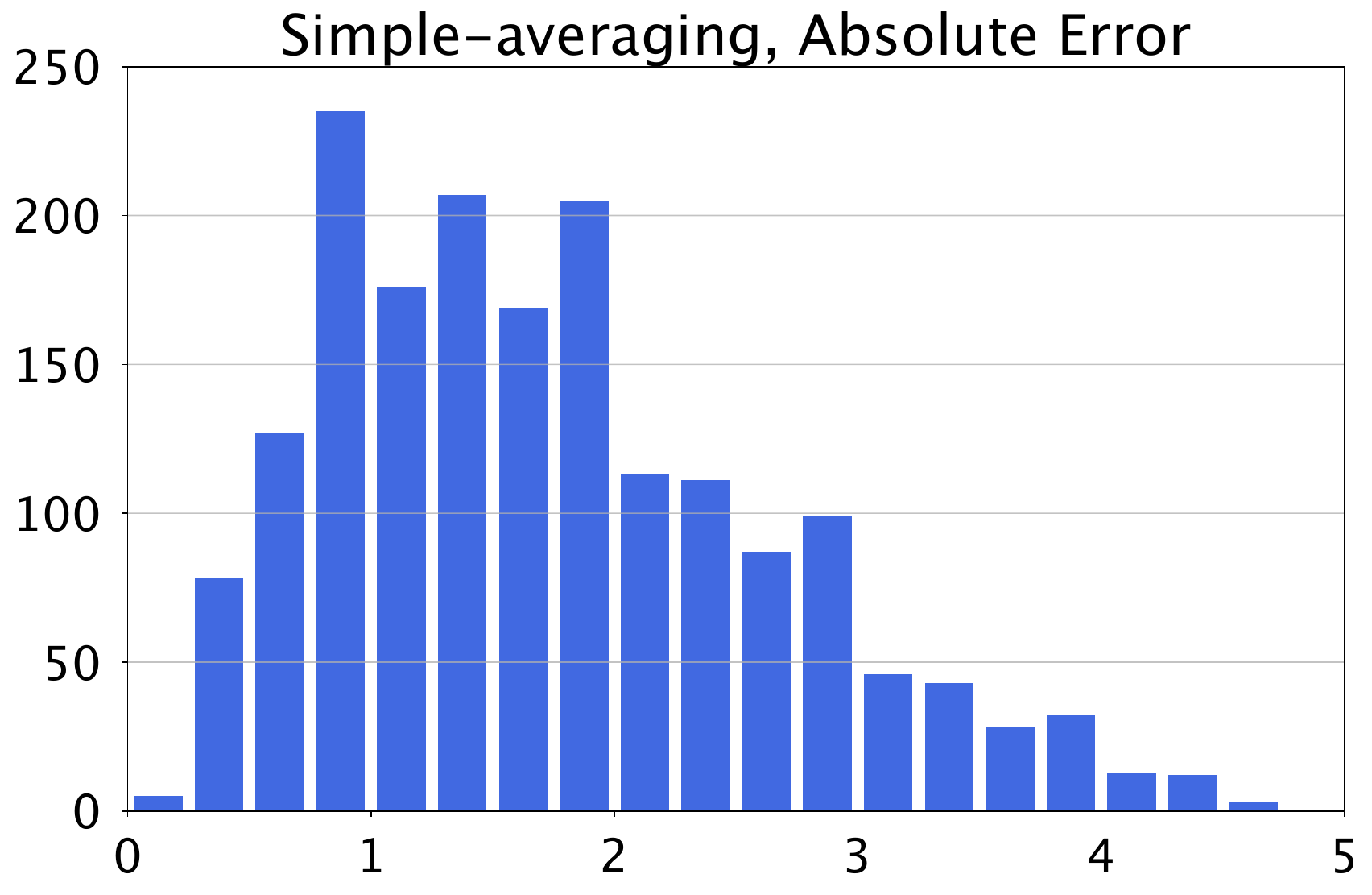}
          \end{subfigure}
        
    \vspace{2mm}
        
    \begin{subfigure}[b]{0.48\textwidth}
        \footnotesize
        
        \includegraphics[ height = 0.63\textwidth]{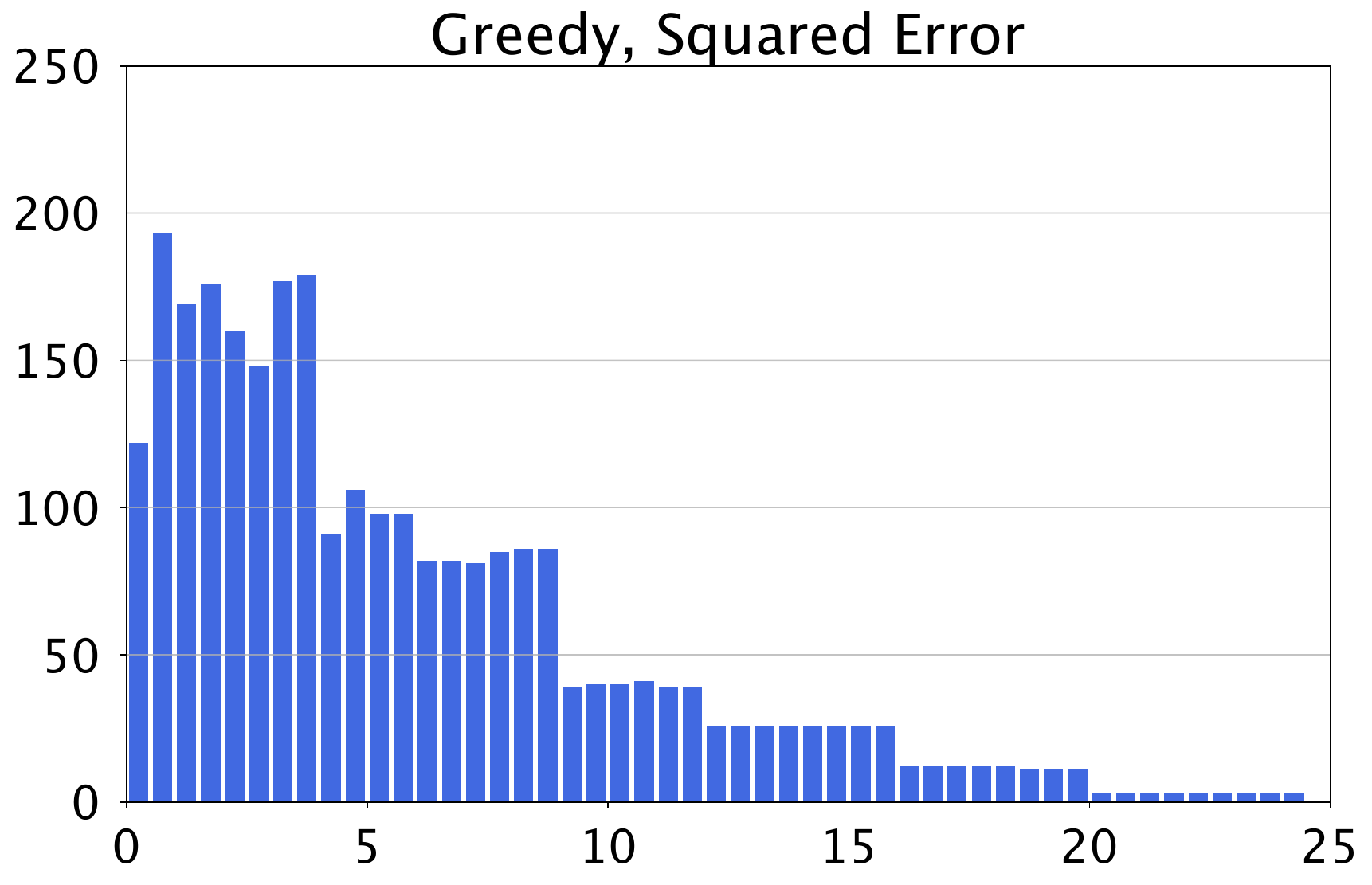}
            \end{subfigure}
    \begin{subfigure}[b]{0.48\textwidth}
        \footnotesize
        
        \includegraphics[ height = 0.63\textwidth]{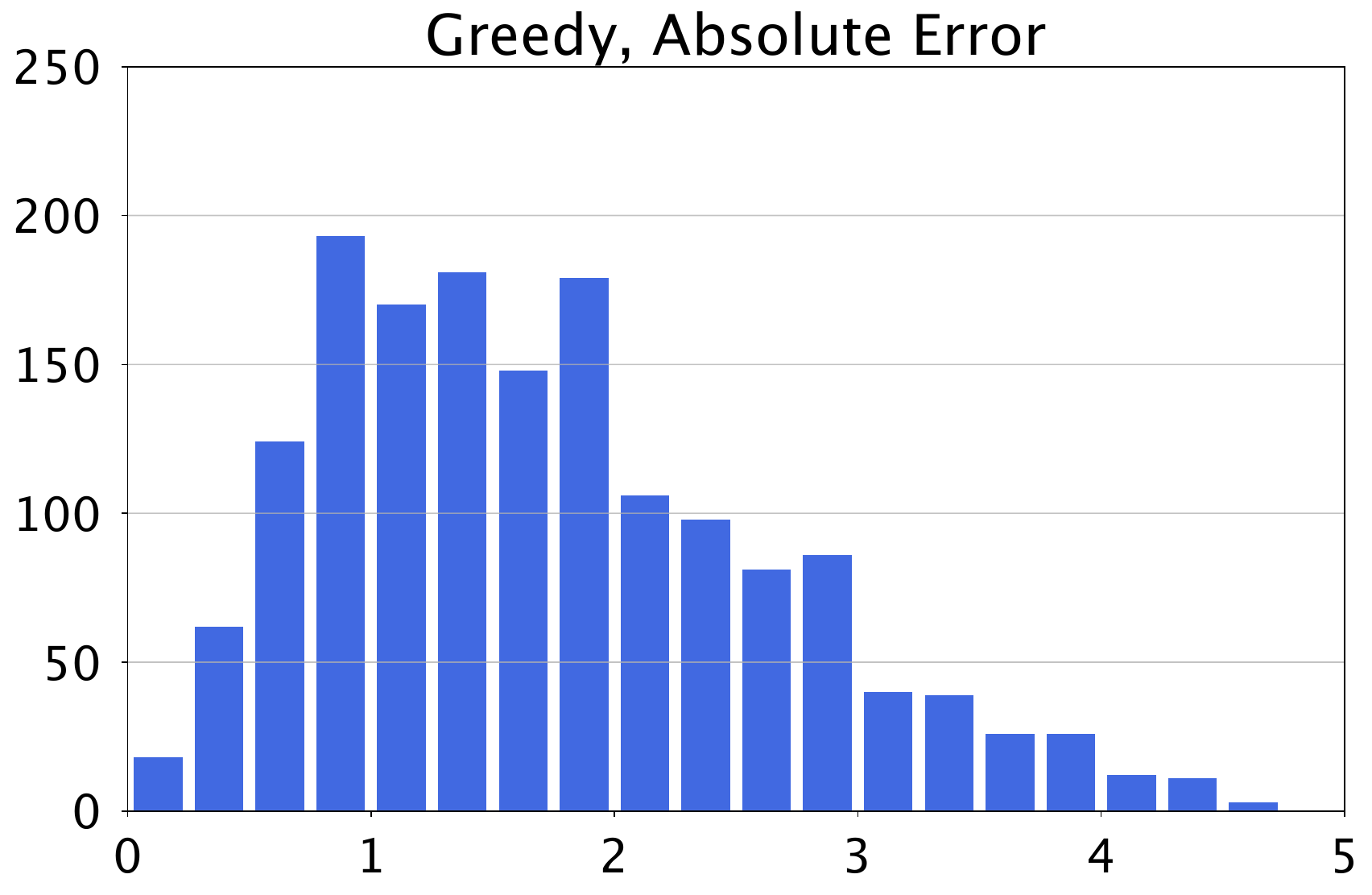}
            \end{subfigure}
        
    \vspace{2mm}
        
    \begin{subfigure}[b]{0.48\textwidth}
        \footnotesize
        
        \includegraphics[ height = 0.63\textwidth]{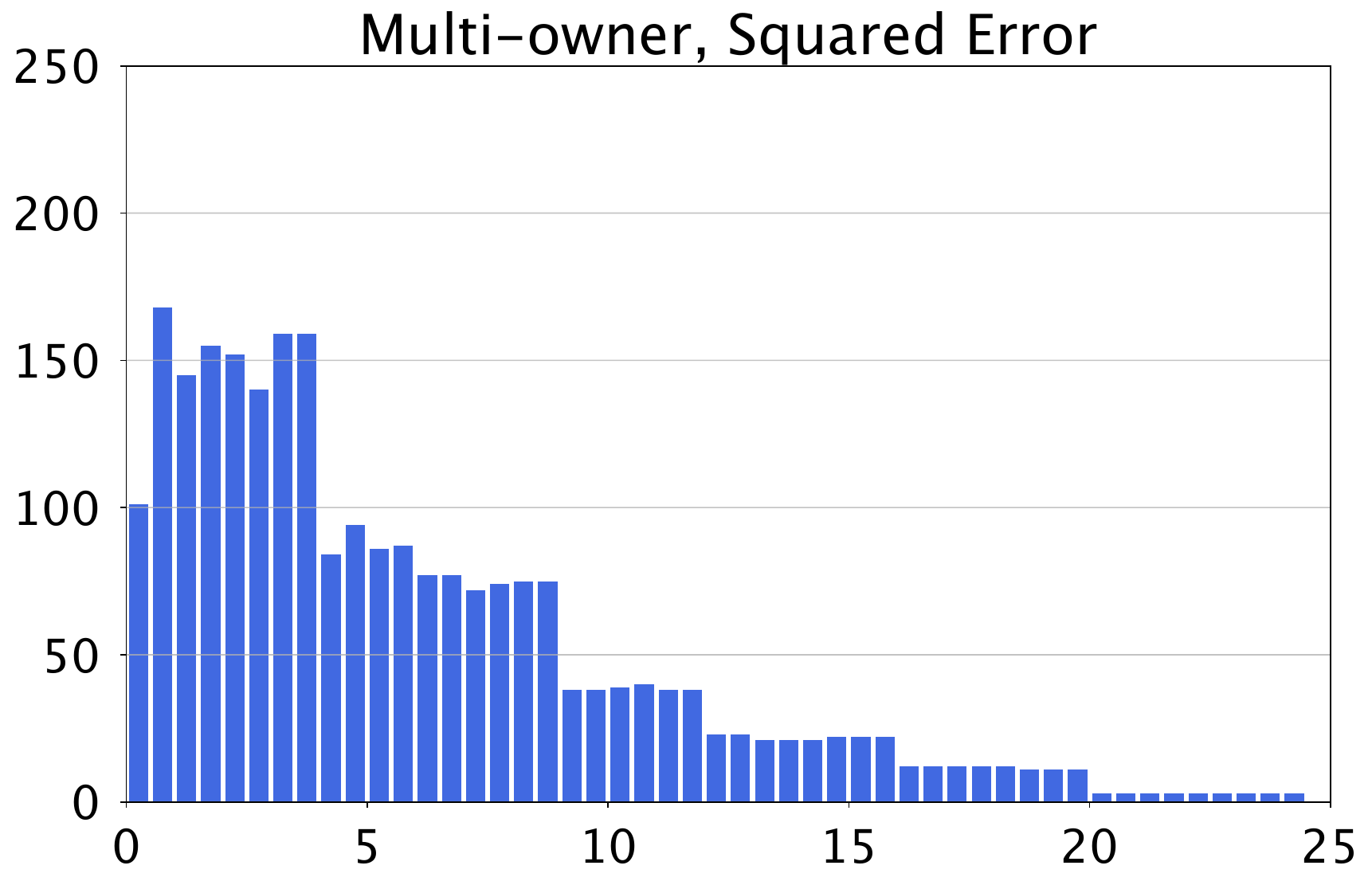}
            \end{subfigure}
    \begin{subfigure}[b]{0.48\textwidth}
        \footnotesize
        
        \includegraphics[ height = 0.63\textwidth]{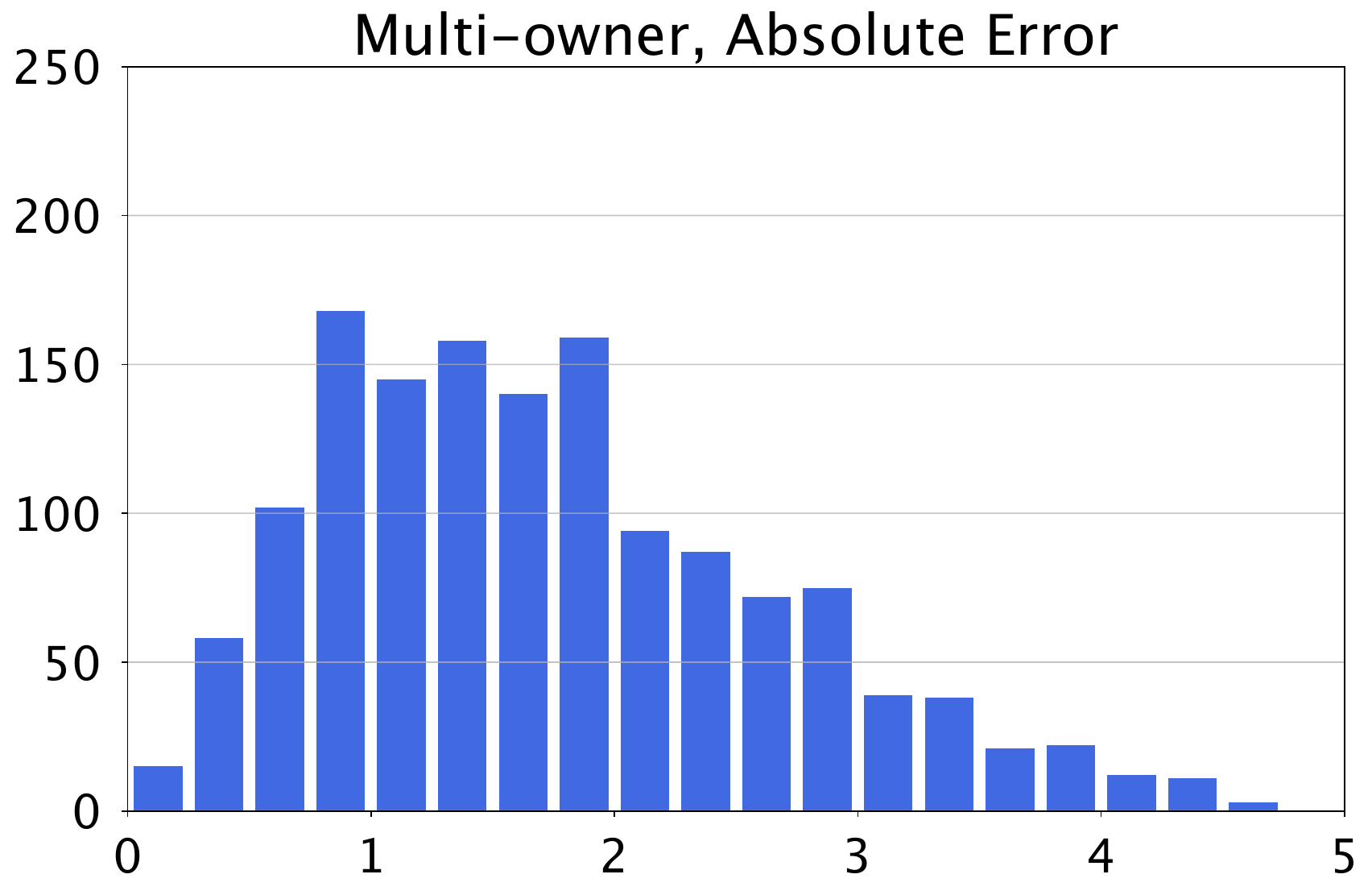}
            \end{subfigure}
    \caption{Difference between cumulative distributions of proxy errors for isotonic and raw scores. Left panel: At $x$-axis value $x$, the $y$-axis value represents $|\{i: (\hat{y}_i^{\textnormal{Iso}} - y_i')^2 \leq x \}| - |\{i: (\hat{y}_i^{\textnormal{Ave}} - y_i')^2 \leq x \}|$, where $|\{i: (\hat{y}_i^{\textnormal{Iso}} - y_i')^2 \leq x \}|$ denotes the number of submissions with isotonic scores having proxy MSE less than or equal to $x$. Right panel: At $x$-axis value $x$, the $y$-axis value represents $|\{i: |\hat{y}_i^{\textnormal{Iso}} - y_i'| \leq x \}| - |\{i: |\hat{y}_i^{\textnormal{Ave}} - y_i'| \leq x \}|$. The consistently positive difference demonstrates that isotonic scores generally yield smaller proxy errors than raw scores in distribution.}
    \label{fig:histogram_residual_difference}
\end{figure}

\begin{figure}[!htp]
  \centering
  \includegraphics[width=0.48\textwidth]{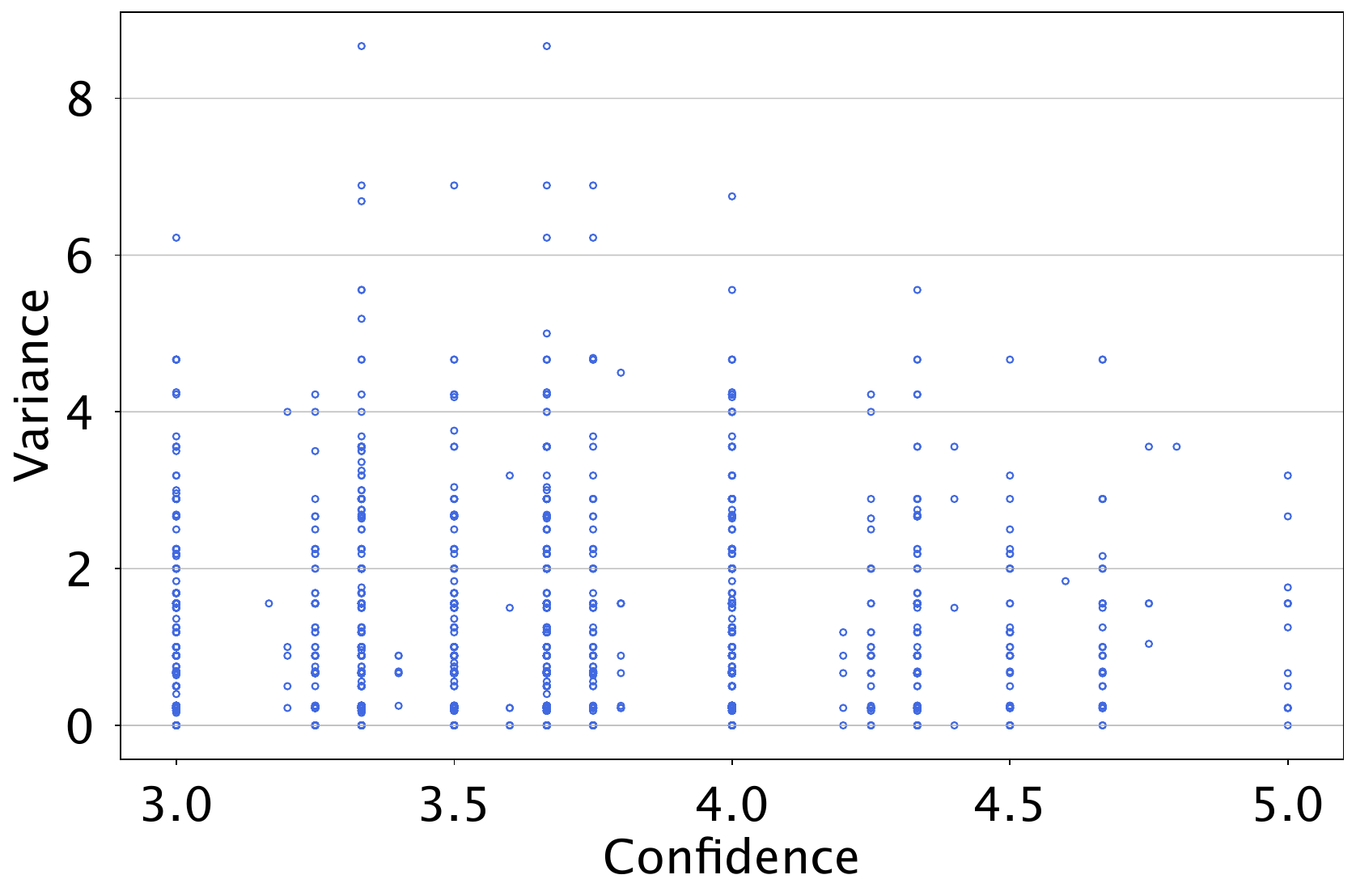}
  \caption{Scatter plots of review confidence against raw score variance. The correlation between raw score variance and review confidence is small.
  }
  \label{fig:confidence}
\end{figure}

Figure~\ref{fig:histogram_residual_difference} shows the difference between the cumulative distributions of proxy errors for the isotonic and raw scores, supporting the findings presented in 
Figure 4.
Table \ref{table:disagreement_coauthor} 
indicates that coauthors reach agreement more easily when there are substantial score gaps between their evaluations. 
Conversely, when the quality of papers is closely matched, resulting in small score gaps, coauthors are more likely to experience disagreements.
Figure \ref{fig:confidence} 
illustrates the correlation between review confidence and score variance. 
Notably, reviewers with high confidence do not always agree with one another or assign consistent scores to identical submissions.
This observation suggests that the residual scores contain the information pertaining to both variance and confidence.

\begin{figure*}[!htp]
    \centering
    \includegraphics[width=0.48\textwidth]{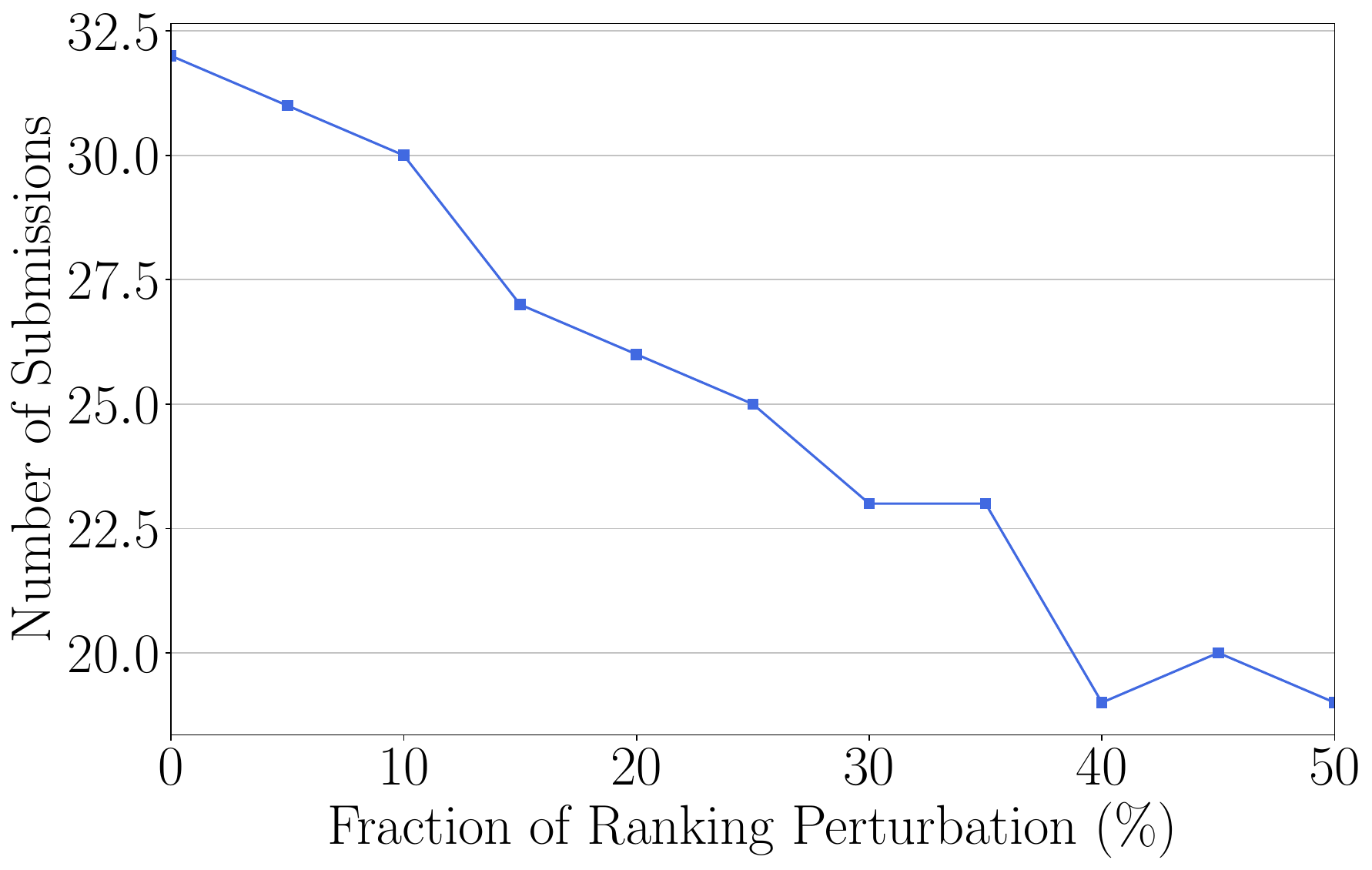} 
    \caption{Number of top-isotonic-scored submissions being ``Accepted as Oral'' under different fractions of ranking perturbation.}
    \label{appfig:overlap_perturbation}
\end{figure*}

Figure \ref{appfig:overlap_perturbation} demonstrates the high quality of authors' rankings and shows that the collected rankings are locally optimal in the sense that any perturbation would decrease their effectiveness. To show this, we randomly select a subset of authors, varying the fraction from 0\% to 50\%, and reverse the order of their rankings. For instance, if an author originally ranks their submissions as $A>B>C$, their perturbed ranking becomes $C>B>A$. Using these perturbed rankings, we recompute the isotonic scores and identify the top 81 submissions with the highest isotonic scores, which we then compare against the 81 submissions ``Accepted as Oral'' at ICML 2023. Under the original, unperturbed rankings, 32 of the top isotonic-scored submissions were ``Accepted as Oral''. However, when 20\% of the rankings are randomly perturbed, this overlap decreases to 26 submissions.

This analysis also provides evidence for using isotonic scores in the oversight of ACs’ recommendations in 
Section 4. Our results indicate that 32 of the top isotonic-scored submissions were ``Accepted as Oral,'' compared to 31 for the top raw-scored submissions. Furthermore, the Isotonic Mechanism uniquely identifies 8 ``Accepted as Oral'' submissions that were not among the top raw-scored submissions. This result underscores the value of isotonic scores in identifying high-quality submissions. More importantly, by leveraging isotonic scores to flag submissions, SACs can more effectively identify those that may require further review.

Table \ref{apptab:filtered_mse} presents a screening procedure to identify informative or reliable rankings before applying them in the final analysis. We filter out 481 authors who were neither reviewers nor area chairs at ICML 2023, considering rankings only from the remaining authors. Since reviewers and ACs are generally more experienced researchers, their rankings are expected to be of higher quality. This filtering results in a modest improvement in the Isotonic Mechanism’s performance, with the MSE reduction increasing from 21.30\% to 22.52\% (greedy strategy), as shown in Table~\ref{apptab:filtered_mse}.

Note that 
Table 2
uses a randomly selected raw-score estimator. We repeat this random selection process six times and aggregate the results in Tables \ref{apptab:greedy_multi_isotonic_mse_rep} and \ref{apptab:greedy_multi_isotonic_mse_rep2}, which remain highly consistent with those shown in 
Table 2, 
suggesting that the randomness in the selection process has little impact on the observed improvements.

\begin{figure}[!htp]
    \centering
    \resizebox{0.3\textwidth}{!}{\begin{minipage}[b]{0.40\textwidth}
        \stackunder[0pt]{ \includegraphics[width=\textwidth]{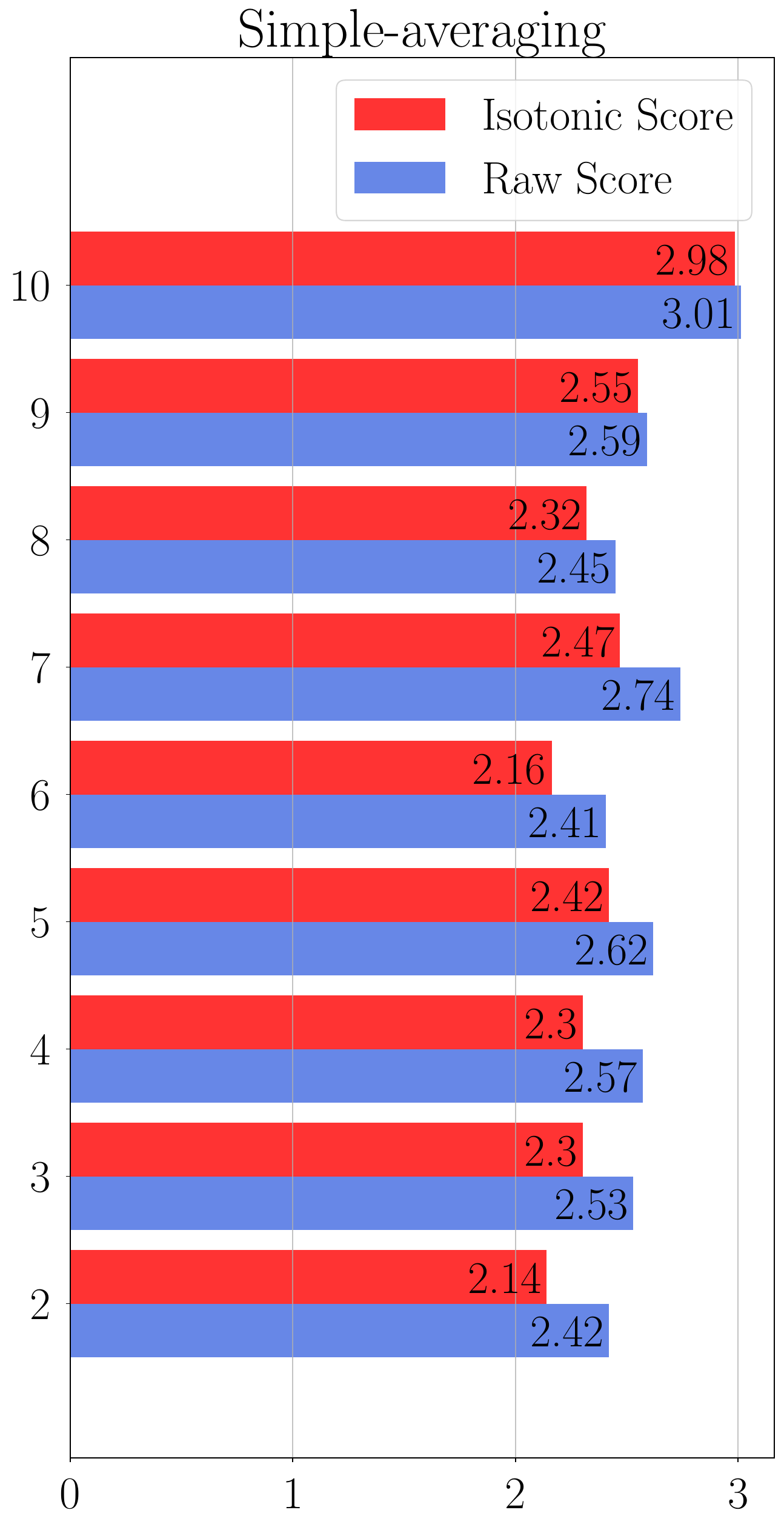}
        }{Mean Squared Error}
    \end{minipage}}
    \resizebox{0.3\textwidth}{!}{\begin{minipage}[b]{0.40\textwidth}
        \stackunder[0pt]{ \includegraphics[width=\textwidth]{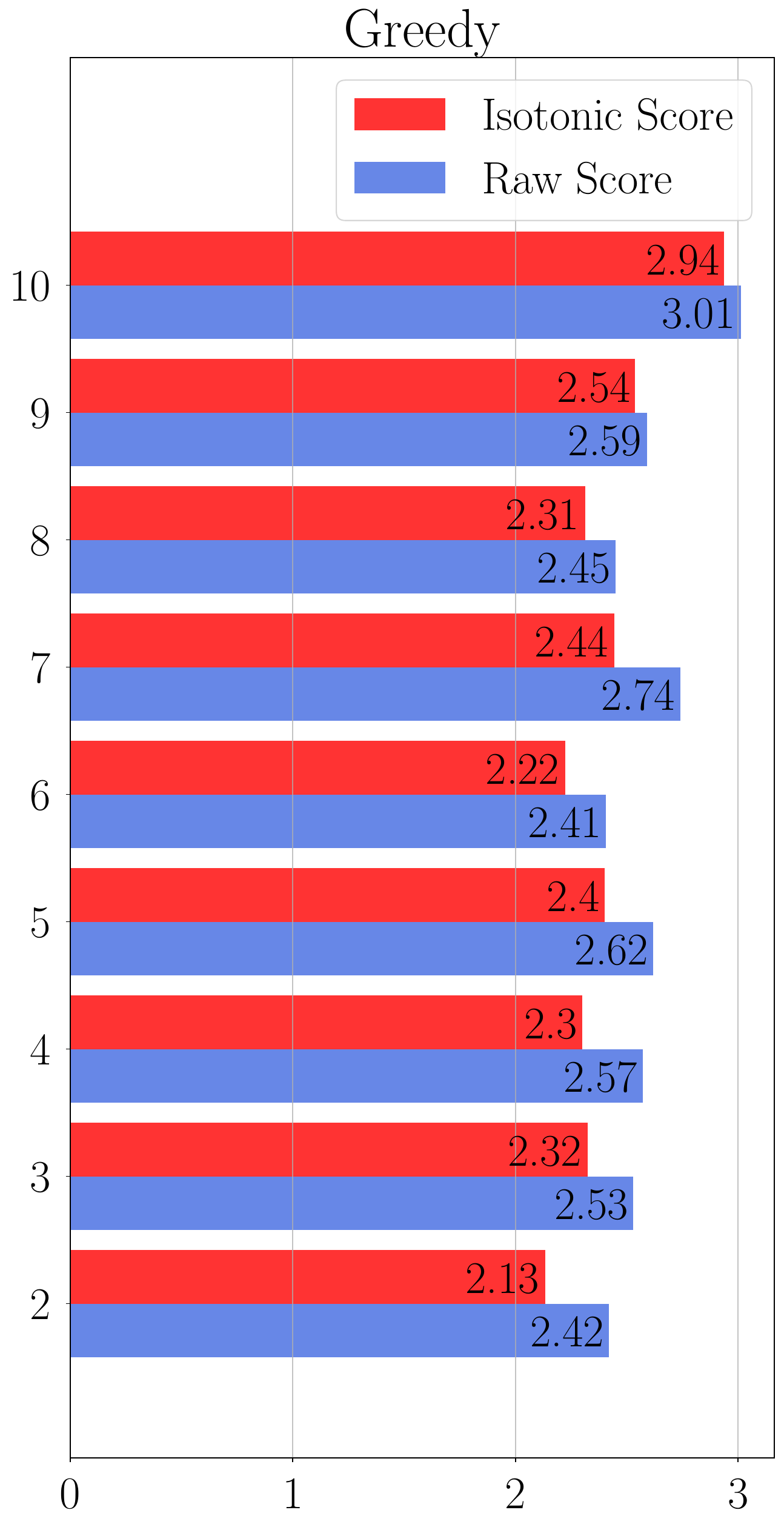}
        }{Mean Squared Error}
    \end{minipage}}
    \resizebox{0.3\textwidth}{!}{\begin{minipage}[b]{0.40\textwidth}
        \stackunder[0pt]{ \includegraphics[width=\textwidth]{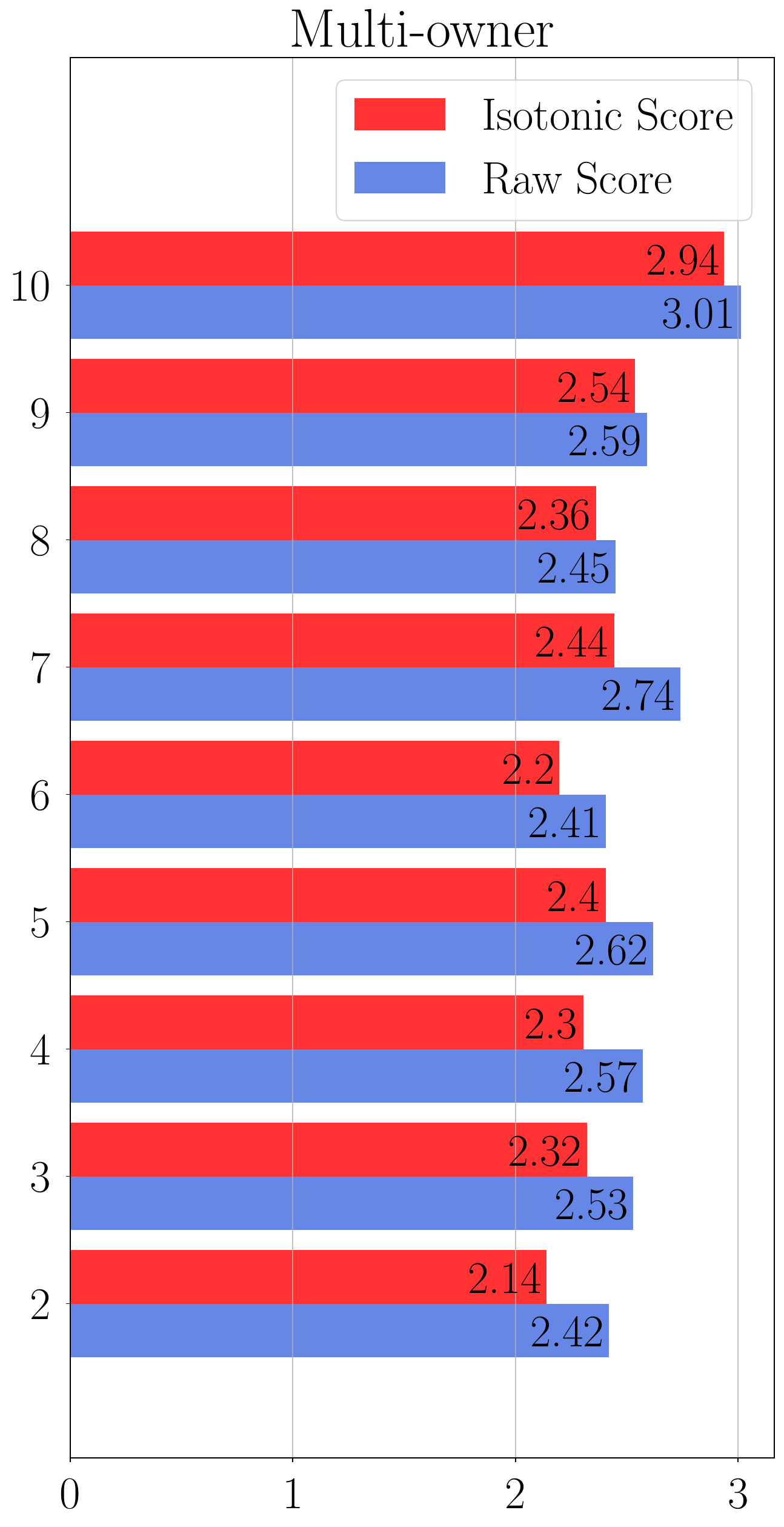}
        }{Mean Squared Error}
    \end{minipage}}
    \resizebox{0.3\textwidth}{!}{\begin{minipage}[b]{0.40\textwidth}
        \stackunder[0pt]{ \includegraphics[width=\textwidth]{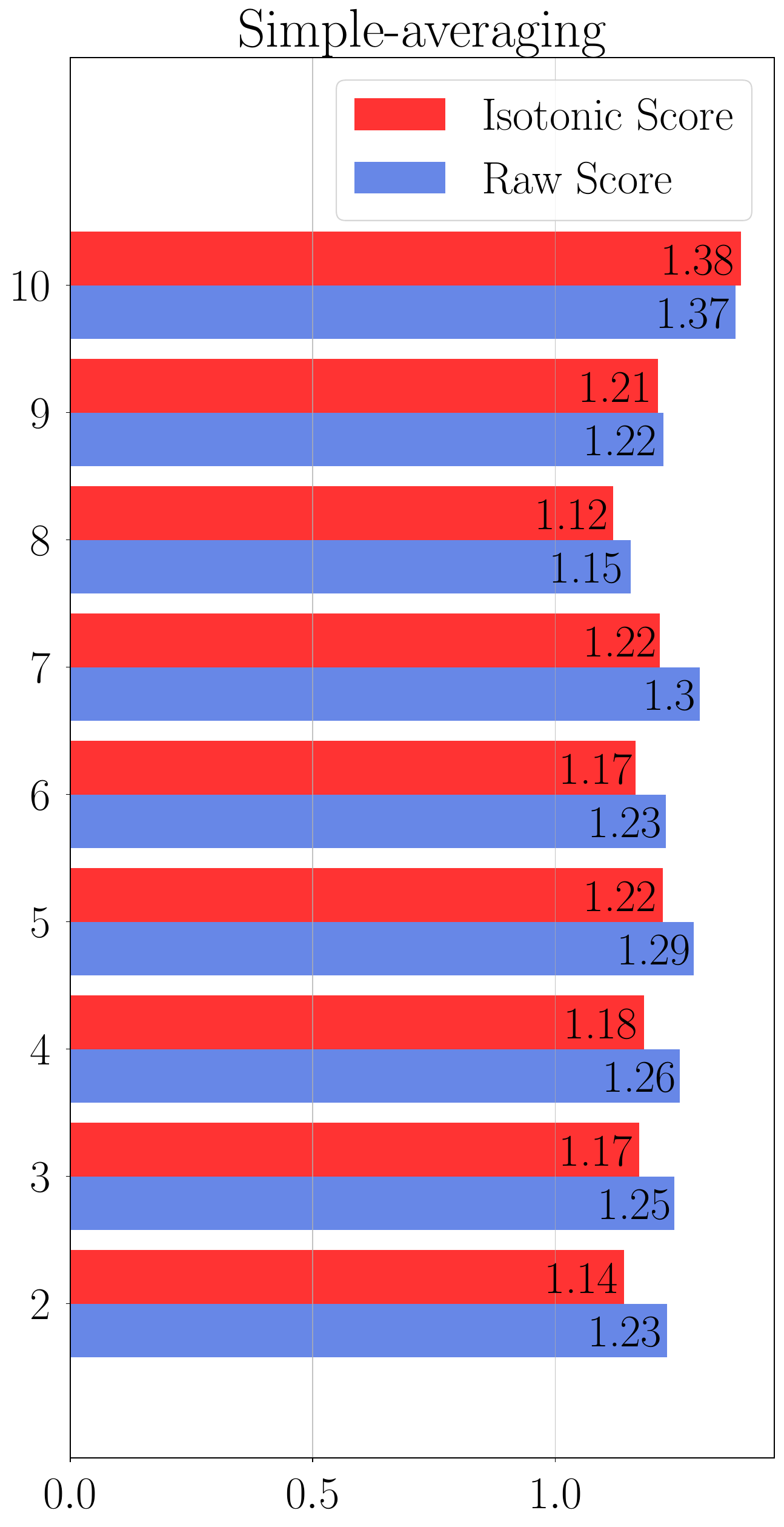}
        }{Mean Absolute Error}
    \end{minipage}}
    \resizebox{0.3\textwidth}{!}{\begin{minipage}[b]{0.40\textwidth}
        \stackunder[0pt]{ \includegraphics[width=\textwidth]{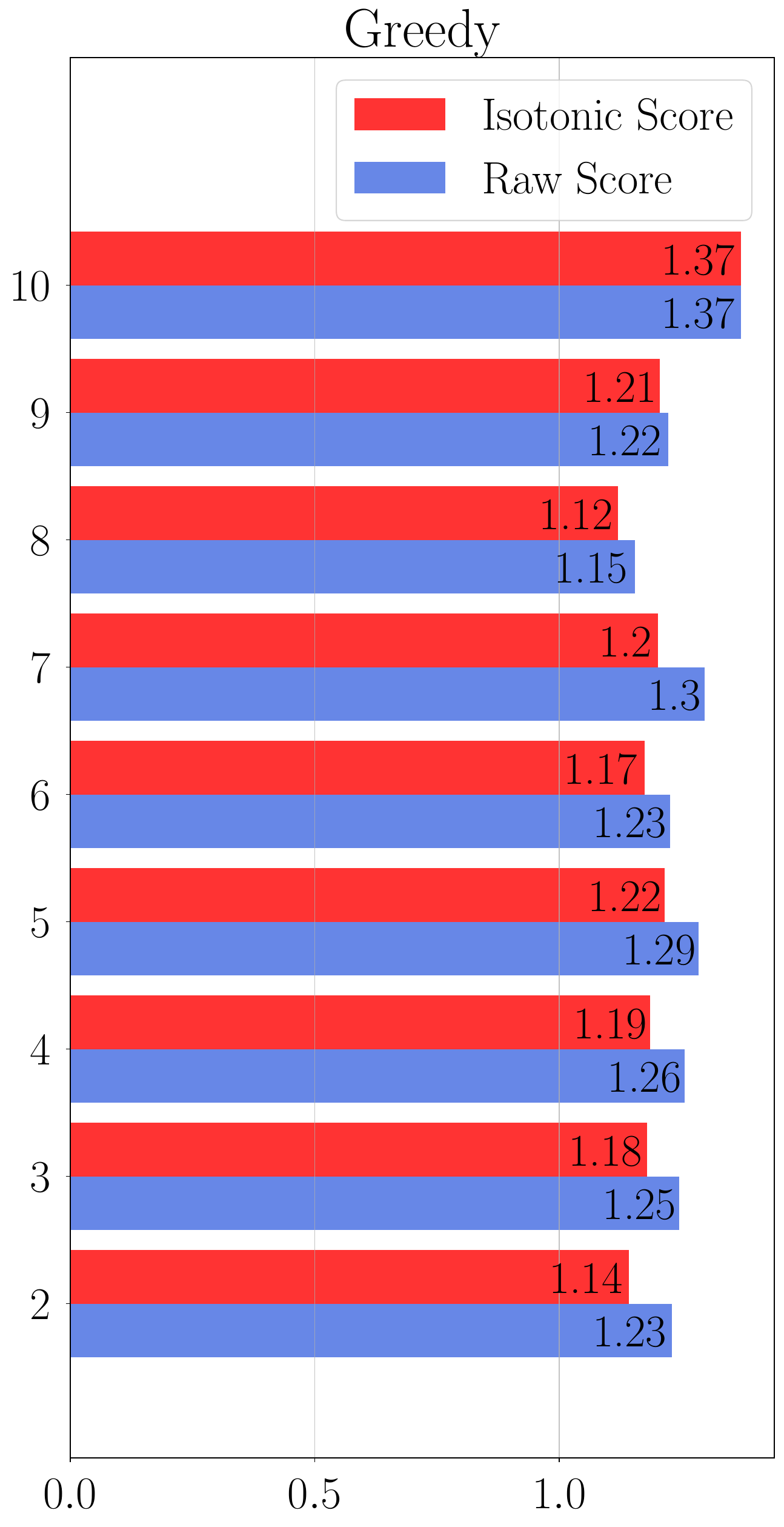}
        }{Mean Absolute Error}
    \end{minipage}}
    \resizebox{0.3\textwidth}{!}{\begin{minipage}[b]{0.40\textwidth}
        \stackunder[0pt]{ \includegraphics[width=\textwidth]{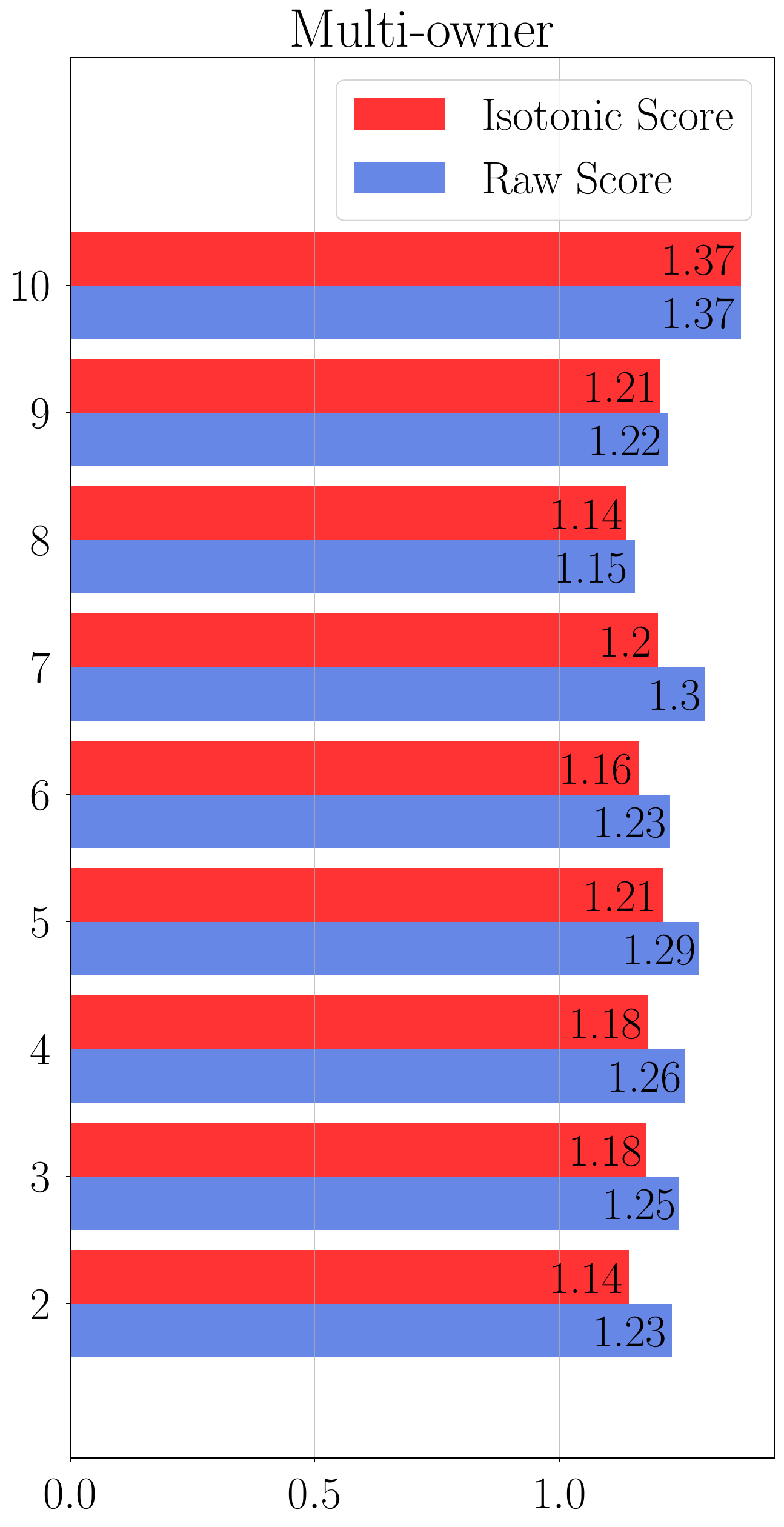}
        }{Mean Absolute Error}
    \end{minipage}}
    \caption{
    Comparison between isotonic and raw scores depending on the number of authors a submission has. The vertical axis denotes the number of authors, ranging from 2 to 10, and the horizontal axis represents the proxy MSE (top) or MAE (bottom). Isotonic scores consistently achieve lower estimation errors than raw scores across all author group sizes.}
    \label{appfig:coauthor_mse}
\end{figure}

Additionally, Figure \ref{appfig:coauthor_mse} examines the relationship between the number of coauthors and the improvement achieved by the Isotonic Mechanism. We categorize submissions into nine groups based on the number of coauthors, ranging from 2 to 10. Within each group, we exclude authors with only a single submission. We then compare the MSE and MAE of isotonic scores with those of raw scores, computed using the proxy ground truth. 
Figure \ref{appfig:coauthor_mse} shows that the Isotonic Mechanism achieves a relatively smaller reduction in MSE and MAE for submissions with a larger number of coauthors. A possible explanation is that as the number of coauthors increases, disagreements over submission quality become more frequent, potentially adding noise to the ranking process.

Furthermore, to account for scenarios where each submission typically receives more than four review scores, Table \ref{apptab:fuzzy_proxy_mse} presents an additional analysis by randomly selecting two review scores per submission as the raw-score-estimator as defined in the ``Evaluation Metrics''. The average of the remaining scores serves as a ``fuzzy'' proxy for the ground truth. We compare the MSE and MAE of isotonic scores against the raw-score-estimator, computed using the ``fuzzy'' proxy ground truth. Table \ref{apptab:fuzzy_proxy_mse} indicates that the Isotonic Mechanism continues to reduce error compared to raw scores.

\begin{table}[H]
\centering
\renewcommand{\arraystretch}{1.2}
\begin{tabular}{l||c|c}

\hline\hline
& \begin{tabular}{@{}c@{}} Score Gap $\leq 1.0$ \end{tabular} & \begin{tabular}{@{}c@{}} Score Gap $> 1.0$ \end{tabular} \\ 

\hline 
\begin{tabular}{@{}c@{}} Sample Size \end{tabular} & $433$ & $291$ \\ 

\hline
\begin{tabular}{@{}c@{}} Disagreement  \end{tabular} & $30.7 \%$ & $26.1 \%$ \\ 

\hline
\end{tabular}
\caption{The fraction of disagreements between coauthors in ranking two submissions depends on the gap between the average review scores of the two submissions. 
}
\label{table:disagreement_coauthor}
\end{table}

\begin{table}[H]
\centering
\renewcommand{\arraystretch}{1.05}
\resizebox{\textwidth}{!}{
\begin{tabular}{l||c|c|c|c|c|c}

\hline \hline
\multirow{2}{*}{} & \multicolumn{3}{c|}{Proxy MSE } & \multicolumn{3}{c}{Proxy MAE } \\

\cline{2-7}
& \begin{tabular}{@{}c@{}} Error \end{tabular} & \begin{tabular}{@{}c@{}} Improvement \end{tabular} & 
\begin{tabular}{@{}c@{}} $p$-value \end{tabular} & \begin{tabular}{@{}c@{}} Error \end{tabular} & \begin{tabular}{@{}c@{}} Improvement \end{tabular} & 
\begin{tabular}{@{}c@{}} $p$-value \end{tabular} 
\\

\hline
\begin{tabular}{@{}c@{}} Raw Score \end{tabular} & $2.54$ & \texttt{NA} & \texttt{NA} & $1.25$ & \texttt{NA} & \texttt{NA}\\ 

\hline
\begin{tabular}{@{}c@{}} Simple-averaging Strategy \end{tabular} & $1.93$ & $23.90 \%$ & $2.51 \times 10^{-34}$ & $1.08$ & $13.10 \%$ & $1.65 \times 10^{-29}$\\ 

\hline
\begin{tabular}{@{}c@{}} Greedy Strategy \end{tabular} & $1.97$ & $22.52 \%$ & $3.81 \times 10^{-30}$ & $1.09$ & $12.38 \%$ & $5.70 \times 10^{-26}$\\ 

\hline
\begin{tabular}{@{}c@{}} Multi-owner Strategy \end{tabular} & $2.03$ & $20.09 \%$ & $1.47 \times 10^{-27}$ & $1.11$ & $10.90 \%$ & $2.83 \times 10^{-23}$\\ 

\hline
\end{tabular}}
\caption{Reduction of MSE and MAE using the Isotonic Mechanism with various strategies, using only rankings from reviewers or ACs. A paired one-sided $t$-test shows that the reduction in proxy errors is statistically highly significant.}
\label{apptab:filtered_mse}
\end{table}

\begin{table}[H]
\centering
\renewcommand{\arraystretch}{1.05}
\resizebox{\textwidth}{!}{
\begin{tabular}{l||c|c|c|c|c|c}

\hline \hline
\multirow{2}{*}{} & \multicolumn{3}{c|}{Proxy MSE } & \multicolumn{3}{c}{Proxy MAE } \\

\cline{2-7}
& \begin{tabular}{@{}c@{}} Error \end{tabular} & \begin{tabular}{@{}c@{}} Improvement \end{tabular} & 
\begin{tabular}{@{}c@{}} $p$-value \end{tabular} & \begin{tabular}{@{}c@{}} Error \end{tabular} & \begin{tabular}{@{}c@{}} Improvement \end{tabular} & 
\begin{tabular}{@{}c@{}} $p$-value \end{tabular} 
\\

\hline
\begin{tabular}{@{}c@{}} Raw Score \end{tabular} & $2.61$ & \texttt{NA} & \texttt{NA} & $1.27$ & \texttt{NA} & \texttt{NA}\\ 

\hline
\begin{tabular}{@{}c@{}} Simple-averaging Strategy \end{tabular} & $1.99$ & $23.66 \%$ & $7.86 \times 10^{-86}$ & $1.11$ & $12.99 \%$ & $1.16 \times 10^{-76}$\\ 

\hline
\begin{tabular}{@{}c@{}} Greedy Strategy \end{tabular} & $2.07$ & $20.85 \%$ & $7.06 \times 10^{-66}$ & $1.12$ & $11.75 \%$ & $1.93 \times 10^{-59}$\\ 

\hline
\begin{tabular}{@{}c@{}} Multi-owner Strategy \end{tabular} & $2.11$ & $19.35 \%$ & $6.54 \times 10^{-70}$ & $1.14$ & $10.77 \%$ & $4.80 \times 10^{-61}$\\ 

\hline
\end{tabular}}
\caption{Reduction of proxy MSE and MAE using the Isotonic Mechanism with various strategies. A paired one-sided $t$-test shows that the reduction in proxy errors is statistically highly significant. Note that the results depend on the randomly selected raw-score-estimator. We repeat this random process three times and aggregate the results here.}
\label{apptab:greedy_multi_isotonic_mse_rep}
\end{table}

\begin{table}[H]
\centering
\renewcommand{\arraystretch}{1.05}
\resizebox{\textwidth}{!}{
\begin{tabular}{l||c|c|c|c|c|c}

\hline \hline
\multirow{2}{*}{} & \multicolumn{3}{c|}{Proxy MSE } & \multicolumn{3}{c}{Proxy MAE } \\

\cline{2-7}
& \begin{tabular}{@{}c@{}} Error \end{tabular} & \begin{tabular}{@{}c@{}} Improvement \end{tabular} & 
\begin{tabular}{@{}c@{}} $p$-value \end{tabular} & \begin{tabular}{@{}c@{}} Error \end{tabular} & \begin{tabular}{@{}c@{}} Improvement \end{tabular} & 
\begin{tabular}{@{}c@{}} $p$-value \end{tabular} 
\\

\hline
\begin{tabular}{@{}c@{}} Raw Score \end{tabular} & $2.67$ & \texttt{NA} & \texttt{NA} & $1.29$ & \texttt{NA} & \texttt{NA}\\ 

\hline
\begin{tabular}{@{}c@{}} Simple-averaging Strategy \end{tabular} & $2.03$ & $24.08 \%$ & $2.61 \times 10^{-83}$ & $1.12$ & $12.88 \%$ & $1.14 \times 10^{-75}$\\ 

\hline
\begin{tabular}{@{}c@{}} Greedy Strategy \end{tabular} & $2.09$ & $21.84 \%$ & $3.49 \times 10^{-66}$ & $1.14$ & $11.75 \%$ & $1.60 \times 10^{-59}$\\ 

\hline
\begin{tabular}{@{}c@{}} Multi-owner Strategy \end{tabular} & $2.15$ & $19.34 \%$ & $6.15 \times 10^{-66}$ & $1.15$ & $10.51 \%$ & $2.80 \times 10^{-59}$\\ 

\hline
\end{tabular}}
\caption{Reduction of proxy MSE and MAE using the Isotonic Mechanism with various strategies. A paired one-sided $t$-test shows that the reduction in proxy errors is statistically highly significant. Note that the results depend on the randomly selected raw-score-estimator. We repeat this random process three more times and aggregate the results here.}
\label{apptab:greedy_multi_isotonic_mse_rep2}
\end{table}

\begin{table}[H]
\centering
\renewcommand{\arraystretch}{1.05}
\resizebox{\textwidth}{!}{
\begin{tabular}{l||c|c|c|c|c|c}

\hline \hline
\multirow{2}{*}{} & \multicolumn{3}{c|}{Proxy MSE } & \multicolumn{3}{c}{Proxy MAE } \\

\cline{2-7}
& \begin{tabular}{@{}c@{}} Error \end{tabular} & \begin{tabular}{@{}c@{}} Improvement \end{tabular} & 
\begin{tabular}{@{}c@{}} $p$-value \end{tabular} & \begin{tabular}{@{}c@{}} Error \end{tabular} & \begin{tabular}{@{}c@{}} Improvement \end{tabular} & 
\begin{tabular}{@{}c@{}} $p$-value \end{tabular} 
\\

\hline
\begin{tabular}{@{}c@{}} Raw Score \end{tabular} & $2.66$ & \texttt{NA} & \texttt{NA} & $1.26$ & \texttt{NA} & \texttt{NA}\\ 

\hline
\begin{tabular}{@{}c@{}} Simple-averaging Strategy \end{tabular} & $2.31$ & $9.38 \%$ & $3.40 \times 10^{-15}$ & $1.22$ & $3.36 \%$ & $1.96 \times 10^{-6}$\\ 

\hline
\begin{tabular}{@{}c@{}} Greedy Strategy \end{tabular} & $2.34$ & $8.28 \%$ & $1.79 \times 10^{-11}$ & $1.22$ & $3.22 \%$ & $9.91\times 10^{-6}$\\ 

\hline
\begin{tabular}{@{}c@{}} Multi-owner Strategy \end{tabular} & $2.36$ & $7.63 \%$ & $3.80 \times 10^{-11}$ & $1.22$ & $2.75 \%$ & $4.04 \times 10^{-5}$\\ 

\hline
\end{tabular}}
\caption{Reduction of proxy MSE and MAE using the Isotonic Mechanism with various strategies, computed using the ``fuzzy'' proxy ground truth. A paired one-sided $t$-test shows that the reduction in proxy errors is statistically highly significant.}
\label{apptab:fuzzy_proxy_mse}
\end{table}

We finally explain a claim made in the second paragraph of 
Section 3.2.
The $21.3\%$ reduction in the overall proxy MSE shows that 
\[
\begin{aligned}
1 - 21.3\% &= \frac{\sum_{i=1}^{2530} (\hat{y}_i^{\textnormal{Iso}} - y'_i)^2}{\sum_{i=1}^{2530} (\hat{y}_i^{\textnormal{Ave}} - y'_i)^2} \\
&\approx \frac{\sum_{i=1}^{2530} \mathbb{E} (\hat{y}_i^{\textnormal{Iso}} - y'_i)^2}{\sum_{i=1}^{2530} \mathbb{E}(\hat{y}_i^{\textnormal{Ave}} - y'_i)^2 }\\
& = \frac{\sum_{i=1}^{2530} \left[ \mathrm{MSE}(\hat{y}_i^{\textnormal{Iso}}) + \mathrm{Var}(y'_i) \right]}{\sum_{i=1}^{2530} \left[ \mathrm{MSE}(\hat{y}_i^{\textnormal{Ave}}) + \mathrm{Var}(y'_i) \right]}.\\
\end{aligned}
\]
Recognizing the fact that $\frac{a + c}{b + c} > \frac{a}{b}$ for any $b > a > 0$ and $c > 0$, we get
\[
\frac{\sum_{i=1}^{2530} \left[ \mathrm{MSE}(\hat{y}_i^{\textnormal{Iso}}) + \mathrm{Var}(y'_i) \right]}{\sum_{i=1}^{2530} \left[ \mathrm{MSE}(\hat{y}_i^{\textnormal{Ave}}) + \mathrm{Var}(y'_i) \right]} > \frac{\sum_{i=1}^{2530} \mathrm{MSE}(\hat{y}_i^{\textnormal{Iso}})}{\sum_{i=1}^{2530} \mathrm{MSE}(\hat{y}_i^{\textnormal{Ave}})}.
\]
Thus, we have
\[
\frac{\sum_{i=1}^{2530} \mathrm{MSE}(\hat{y}_i^{\textnormal{Ave}}) - \sum_{i=1}^{2530} \mathrm{MSE}(\hat{y}_i^{\textnormal{Iso}})}{\sum_{i=1}^{2530} \mathrm{MSE}(\hat{y}_i^{\textnormal{Ave}})} \ge 21.3\%.
\]

\section{Additional Synthetic Experiments}
\label{app:syn_exp}

Table \ref{apptab:synthetic_noisy_rank_mse} presents an additional analysis to assess the robustness of the Isotonic Mechanism under noisy rankings. Our results demonstrate that, despite ranking noise, the Isotonic Mechanism improves the raw scores by approximately 10\% in terms of MSE when compared to the synthetic ground truth. Moreover, we find that the simple-averaging Isotonic Mechanism outperforms the other two mechanisms in handling noisy rankings.

To evaluate this, we first generate a synthetic ground truth score for each submission in ICML 2023, sampling from \( R = \max\{\min\{R_0, 10\}, 0\} \) with \( R_0 \sim \mathcal{N}(5, 1.25) \). Reviewer scores are then generated as unbiased estimates of the ground truth, expressed as \( y = \max\{\min\{y_0, 10\}, 0\} \) with \( y_0 = \text{round} \left\{ R + \mathcal{N}(0, 1.25) \right\} \), where the noise is independent of \( R \). Each author subsequently ranks their submissions according to the Plackett-Luce model, providing a total ranking of \( M \) submissions \( \{i_1, \dots, i_M\} \). The probability of observing a ranking \( i_1 \succ \dots \succ i_M \) follows:
\begin{align*}
    \mathbb{P}(i_1 \succ \dots \succ i_M) = \prod_{j=1}^{M-1} \bigg[\frac{e^{R_{i_j}}}{\sum_{k=j}^{M} e^{R_{i_k}}}\bigg],
\end{align*}
where \( R_{i_j} \) and \( R_{i_k} \) are the synthetic ground truth scores of submissions \( i_j \) and \( i_k \), respectively. 
Finally, we evaluate the performance of the Isotonic Mechanism by comparing the MSE and MAE of the isotonic scores and the raw scores, computed using the synthetic ground truth. Table \ref{apptab:synthetic_noisy_rank_mse} suggests that the Isotonic Mechanism, especially the simple-averaging Isotonic Mechanism, remains effective even when the ranking input is imperfect or noisy.

\begin{table}[H]
\centering
\renewcommand{\arraystretch}{1.05}
\resizebox{\textwidth}{!}{
\begin{tabular}{l||c|c|c|c|c|c}

\hline \hline
\multirow{2}{*}{} & \multicolumn{3}{c|}{ MSE } & \multicolumn{3}{c}{ MAE } \\

\cline{2-7}
& \begin{tabular}{@{}c@{}} Error \end{tabular} & \begin{tabular}{@{}c@{}} Improvement \end{tabular} & 
\begin{tabular}{@{}c@{}} $p$-value \end{tabular} & \begin{tabular}{@{}c@{}} Error \end{tabular} & \begin{tabular}{@{}c@{}} Improvement \end{tabular} & 
\begin{tabular}{@{}c@{}} $p$-value \end{tabular} 
\\

\hline
\begin{tabular}{@{}c@{}} Raw Score \end{tabular} & $0.55$ & \texttt{NA} & \texttt{NA} & $0.59$ & \texttt{NA} & \texttt{NA}\\ 

\hline
\begin{tabular}{@{}c@{}} Simple-averaging Strategy \end{tabular} & $0.47$ & $14.03\%$ & $4.64 \times 10^{-10}$ & $0.54$ & $7.65 \%$ & $1.93 \times 10^{-10}$\\ 

\hline
\begin{tabular}{@{}c@{}} Greedy Strategy \end{tabular} & $0.50$ & $9.52 \%$ & $4.05 \times 10^{-5}$ & $0.56$ & $4.75 \%$ & $9.94 \times 10^{-5}$\\ 

\hline
\begin{tabular}{@{}c@{}} Multi-owner Strategy \end{tabular} & $0.50$ & $8.79 \%$ & $3.87 \times 10^{-5}$ & $0.56$ & $4.89 \%$ & $1.60 \times 10^{-5}$\\ 

\hline
\end{tabular}}
\caption{Reduction of MSE and MAE using the Isotonic Mechanism with various strategies, computed using noisy rankings. A paired one-sided $t$-test shows that the reduction in proxy errors is statistically highly significant. Note that the errors reported in this table are substantially smaller than those in 
Table 2, as this table presents MSE and MAE computed using the ground truth rather than proxy ground truth.}
\label{apptab:synthetic_noisy_rank_mse}
\end{table}

Table \ref{apptab:synthetic_bias_mse} suggests that the Isotonic Mechanism is more useful in cases where some reviewers are biased, such as when emergency reviewers are needed. 
To simulate a setting with biased raw scores, we generate a synthetic ground truth for each submission at ICML 2023 by sampling from \( R = \max\{\min\{R_0, 10\}, 0\} \) with \( R_0 \sim \mathcal{N}(5, 1.25) \). We then introduce reviewer biases in the following manner. Each review score \( y \) is modeled as \( y = R + z(R) \), where the noise term \( z(R) \) depends on the ground truth score \( R \). Specifically, we consider two types of reviewers:

\begin{itemize}
    \item ``Bold'' reviewers tend to overestimate high-quality submissions ($R > 5$) and underestimate lower-quality ones ($R \leq 5$). Their scores follow the distribution $y = \max\{\min\{y_0, 10\}, 0\}$ with
  \begin{align*}
      y_0 = \text{round}\left\{ R + \mathcal{N}(0, 1.25) + \left[ \text{sigmoid}(R - 5) - 0.5 \right] \right\}.
  \end{align*}

  \item ``Conservative'' reviewers exhibit the opposite bias. Their scores follow the distribution $y = \max\{\min\{y_0, 10\}, 0\}$ with
  \begin{align*}
      y_0 = \text{round}\left\{ R + \mathcal{N}(0, 1.25) - \left[ \text{sigmoid}(R - 5) - 0.5 \right] \right\}.
  \end{align*}
\end{itemize}

Each submission is assigned three reviewers, with each reviewer independently categorized as either ``bold'' or ``conservative'' with equal probability. We assume that each author ranks their submissions based on the ground truth scores.

To evaluate the impact of reviewer biases, we compare the MSE and MAE of the Isotonic Mechanism against the raw scores in Table \ref{apptab:synthetic_bias_mse}, computed using the synthetic ground truth. Our results show that the Isotonic Mechanism improves estimation accuracy by a comparable percentage even when reviewers are biased.

\begin{table}[!htp]
\centering
\renewcommand{\arraystretch}{1.05}
\resizebox{\textwidth}{!}{
\begin{tabular}{l||c|c|c|c|c|c}

\hline \hline
\multirow{2}{*}{} & \multicolumn{3}{c|}{ MSE } & \multicolumn{3}{c}{ MAE } \\

\cline{2-7}
& \begin{tabular}{@{}c@{}} Error \end{tabular} & \begin{tabular}{@{}c@{}} Improvement \end{tabular} & 
\begin{tabular}{@{}c@{}} $p$-value \end{tabular} & \begin{tabular}{@{}c@{}} Error \end{tabular} & \begin{tabular}{@{}c@{}} Improvement \end{tabular} & 
\begin{tabular}{@{}c@{}} $p$-value \end{tabular} 
\\

\hline
\begin{tabular}{@{}c@{}} Raw Score \end{tabular} & $0.57$ & \texttt{NA} & \texttt{NA} & $0.60$ & \texttt{NA} & \texttt{NA}\\ 

\hline
\begin{tabular}{@{}c@{}} Simple-averaging Strategy \end{tabular} & $0.44$ & $23.54 \%$ & $1.48 \times 10^{-49}$ & $0.52$ & $13.45 \%$ & $3.67\times 10^{-64}$\\ 

\hline
\begin{tabular}{@{}c@{}} Greedy Strategy \end{tabular} & $0.44$ & $22.71 \%$ & $7.56 \times 10^{-45}$ & $0.52$ & $13.01 \%$ & $9.01 \times 10^{-58}$\\ 

\hline
\begin{tabular}{@{}c@{}} Multi-owner Strategy \end{tabular} & $0.47$ & $18.49 \%$ & $2.38 \times 10^{-36}$ & $0.54$ & $6.13 \%$ & $5.11 \times 10^{-46}$\\ 

\hline
\end{tabular}}
\caption{Reduction of MSE and MAE using the Isotonic Mechanism with various strategies, computed using biased review scores. A paired one-sided $t$-test shows that the reduction in errors is statistically highly significant.}
\label{apptab:synthetic_bias_mse}
\end{table}

Table \ref{apptab:synthetic_outlier_mse} examines how the Isotonic Mechanism could potentially address issues arising from ``outlier'' scores. We design an experiment where synthetic ``outlier'' scores are introduced. Specifically, for each submission, we generate an additional ``outlier'' score by sampling uniformly from \( \{1,2,\dots,10\} \). 
Instead of randomly selecting a single score as before, we modify the raw-score-estimator to take the average of one randomly selected review score and the generated ``outlier'' score. The remaining review scores are averaged as the proxy ground truth. Table \ref{apptab:synthetic_outlier_mse} demonstrates that the Isotonic Mechanism effectively reduces the MSE and MAE introduced by ``outlier'' scores.

\begin{table}[H]
\centering
\renewcommand{\arraystretch}{1.05}
\resizebox{\textwidth}{!}{
\begin{tabular}{l||c|c|c|c|c|c}

\hline \hline
\multirow{2}{*}{} & \multicolumn{3}{c|}{Proxy MSE } & \multicolumn{3}{c}{Proxy MAE } \\

\cline{2-7}
& \begin{tabular}{@{}c@{}} Error \end{tabular} & \begin{tabular}{@{}c@{}} Improvement \end{tabular} & 
\begin{tabular}{@{}c@{}} $p$-value \end{tabular} & \begin{tabular}{@{}c@{}} Error \end{tabular} & \begin{tabular}{@{}c@{}} Improvement \end{tabular} & 
\begin{tabular}{@{}c@{}} $p$-value \end{tabular} 
\\

\hline
\begin{tabular}{@{}c@{}} Raw Score \end{tabular} & $3.95$ & \texttt{NA} & \texttt{NA} & $1.62$ & \texttt{NA} & \texttt{NA}\\ 

\hline
\begin{tabular}{@{}c@{}} Simple-averaging Strategy \end{tabular} & $2.76$ & $30.05\%$ & $5.49 \times 10^{-74}$ & $1.34$ & $17.19 \%$ & $9.29 \times 10^{-69}$\\ 

\hline
\begin{tabular}{@{}c@{}} Greedy Strategy \end{tabular} & $2.82$ & $28.50 \%$ & $1.86 \times 10^{-64}$ & $1.34$ & $16.86 \%$ & $2.30 \times 10^{-61}$\\ 

\hline
\begin{tabular}{@{}c@{}} Multi-owner Strategy \end{tabular} & $2.94$ & $25.65 \%$ & $2.51 \times 10^{-60}$ & $1.38$ & $14.88 \%$ & $6.59 \times 10^{-56}$\\ 

\hline
\end{tabular}}
\caption{Reduction of proxy MSE and MAE using the Isotonic Mechanism with various strategies, computed using review scores that include synthetic ``outlier'' scores. A paired one-sided $t$-test shows that the reduction in proxy errors is statistically highly significant.}
\label{apptab:synthetic_outlier_mse}
\end{table}

\end{document}